\documentclass[twocolumn]{aastex702} 
\usepackage{natbib}
\usepackage{caption}
\usepackage{pifont}

\graphicspath{{./}{figures/}}
\usepackage{color,subfigure,lineno,booktabs} 

\definecolor{firebrick}{rgb}{0.7, 0.13, 0.13}

\newcommand{\Hb}{H$\beta$}
\newcommand{\Ha}{H$\alpha$}

\def\HeI{He\,{\sc i}}

\def\MgII{Mg\,{\sc ii}}

\def\FeII{Fe\,{\sc ii}}
\def\FeIII{Fe\,{\sc iii}}
\def\OIII{[O\,{\sc iii}]}

\def\CaII{Ca\,{\sc ii}}
\def\SiII{Si\,{\sc ii}}
\def\SII{S\,{\sc ii}}

\def\OII{[O\,{\sc ii}]}
\def\OI{[O\,{\sc i}]}

\newcommand{\cmark}{\ding{51}}
\newcommand{\xmark}{\ding{55}}
\newcommand{\namark}{$\bigcirc$}

\def\proj{\texttt{NEXUS}}

\shorttitle{NEXUS Transient Survey} %
\shortauthors{Zhuang et~al.}

\begin{document}

\title{NEXUS: Transient Searches and First Results from Year One Observations}


\author[0000-0001-5105-2837]{Ming-Yang Zhuang}
\affiliation{Department of Astronomy, University of Illinois Urbana-Champaign, Urbana, IL 61801, USA}
\email[show]{mingyang@illinois.edu}

\author[0000-0001-7201-1938]{Lei Hu}
\affiliation{Department of Physics and Astronomy, University of Pennsylvania, Philadelphia, PA 19104, USA}
\affiliation{McWilliams Center for Cosmology and Astrophysics, Department of Physics, Carnegie Mellon University, 5000 Forbes Avenue, Pittsburgh, PA 15213, USA}
\email{leihu@sas.upenn.edu}

\author[0000-0002-2361-7201]{Justin D. R. Pierel}
\affiliation{Space Telescope Science Institute, Baltimore, MD 21218, USA}
\altaffiliation{NASA Einstein Fellow}
\email{justin.pierel@gmail.com}

\author[0000-0003-0230-6436]{Zhiwei Pan}
\affiliation{Department of Astronomy, University of Illinois Urbana-Champaign, Urbana, IL 61801, USA}
\email{zhiweip@illinois.edu}

\author[0000-0003-1659-7035]{Yue Shen}
\affiliation{Department of Astronomy, University of Illinois Urbana-Champaign, Urbana, IL 61801, USA}
\affiliation{National Center for Supercomputing Applications, University of Illinois Urbana-Champaign, Urbana, IL 61801, USA}
\email{shenyue@illinois.edu}

\author[0000-0002-6523-9536]{Adam J.\ Burgasser}
\affiliation{Department of Astronomy \& Astrophysics, UC San Diego, La Jolla, CA 92093, USA}
\email{aburgasser@ucsd.edu}

\author[0000-0003-3310-0131]{Xiaohui Fan}
\affiliation{Steward Observatory, University of Arizona, 933 N. Cherry Ave., Tucson, AZ 85721, USA}
\email{xfan@arizona.edu}

\author[0000-0001-6947-5846]{Luis C. Ho}
\affiliation{Kavli Institute for Astronomy and Astrophysics, Peking University, Beijing, 100871, Beijing, People's Republic of China}
\affiliation{Department of Astronomy, School of Physics, Peking University, Beijing, 100871, Beijing, People's Republic of China}
\email{lho.pku@gmail.com}

\author[0000-0002-1605-915X]{Junyao Li}
\affiliation{Department of Astronomy, University of Illinois at Urbana-Champaign, Urbana, IL 61801, USA}
\email{junyaoli@illinois.edu}

\author[0000-0003-3509-4855]{Alice E. Shapley}
\affiliation{Department of Physics \& Astronomy, University of California, Los Angeles, 430 Portola Plaza, Los Angeles, CA 90095, USA}
\email{aes@astro.ucla.edu}

\author[0000-0002-8501-3518]{Zachary Stone}
\affiliation{Department of Astronomy, University of Illinois at Urbana-Champaign, Urbana, IL 61801, USA}
\email{stone28@illinois.edu}

\author[0000-0002-7633-431X]{Feige Wang}
\affiliation{Department of Astronomy, University of Michigan, 1085 S. University Ave., Ann Arbor, MI 48109, USA}
\email{fgwang@umich.edu}

\author[0000-0001-7092-9374]{Lifan Wang}
\affiliation{The George P. and Cynthia Woods Mitchell Institute for Fundamental Physics and Astronomy and Physics and Astronomy Department Texas A\&M University College Station, TX 77843, USA}
\email{lifan@tamu.edu}

\begin{abstract}
We describe ongoing efforts of high-redshift transient searches using multi-epoch NIRCam imaging (F200W+F444W) and NIRSpec MSA/PRISM spectroscopy from the NEXUS JWST multi-cycle Treasury program targeting the north ecliptic pole region. The transient search area covers $\sim 61.5\,{\rm arcmin^2}$ between the reference epoch and each subsequent NEXUS-Deep epoch at a cadence of $\sim2$~months. In the first year of observations from NEXUS, we detect {68} robust transients, with the host photometric redshift distribution declining rapidly at $z>2$ but extending to $z_{\rm phot}\approx 6$. In addition, we obtained secure spectroscopic redshifts for {37} transients {($\sim54\%$) from NIRSpec/PRISM and NIRCam/WFSS}, with seventeen at $1 < z < 2$, eight at $2 < z < 3$, two at $3 < z < 4$, and one tentative host association at z = 6.151. {Overall, the NEXUS program recovers observed supernova (SN) rates broadly consistent with other SN search programs with JWST. While NEXUS achieves the highest transient detection efficiency, 6.6 SNe per imaging hr ($2.6\times$ COSMOS-SN and $26\times$ JADES-SN), the limited filter coverage (F200W+F444W only) limits robust identification and classification of high-$z$ SNe for deep spectroscopic follow-up.} We describe the details of the data reduction and transient detection pipeline, and report the Year-1 transient sample along with their light curves, available MSA spectroscopy, host association, and the raw detection rate. {We also describe and release a new PSF photometry package that properly accounts for correlated pixel noise from combining drizzled images. }     
\end{abstract}

\section{Introduction}\label{sec:introduction}

Supernovae (SNe) provide a unique window into stellar evolution, galaxy growth, and cosmology across cosmic time. Core-collapse supernovae (CCSNe), the explosive deaths of massive stars, both trace recent star formation and play a central role in regulating the evolution of galaxies through the injection of energy and newly synthesized elements into their environments \citep{Gentry_SNe_momentum_2020,Gelli_sne_quench_2024,Ibrahim_MZR_evolution_2024}. Measurements of the CCSN population as a function of redshift therefore provide an independent probe of the cosmic star-formation history, while their rates and relative subtype fractions can constrain changes in the initial mass function (IMF), progenitor populations, and stellar evolution across cosmic time \citep{Madau_SFH_2014,Dahlen2004_high_z_SN_rates,Dahlen2012_CCSNe_rates,Strolger2015_CCSNe_rates}. These measurements become particularly valuable at high redshift, where the lower metallicities and different star-forming environments of young galaxies may produce SN progenitors and explosions that differ systematically from those observed in the local Universe.

Type Ia supernovae (SNe\,Ia) provide a complementary view of the high-redshift Universe. Their use as standardized luminosity-distance indicators enabled the discovery of cosmic acceleration and they remain one of the primary tools for constraining the expansion history of the Universe \citep{riess_observational_1998,perlmutter_measurements_1999}. However, increasingly precise cosmological measurements require understanding whether the properties and standardization of SNe\,Ia remain unchanged across the dramatically different stellar populations encountered over cosmic history. The highest-redshift SNe\,Ia provide a particularly powerful test. At $z\gtrsim1.5$, the contribution of dark energy to the expansion rate becomes small, and the expected distance--redshift relation is relatively insensitive to the detailed properties of dark energy. Deviations from this relation can therefore provide a relatively clean test for redshift-dependent SN~Ia luminosity biases or evolution \citep{riess_first_2006}. Building a statistically meaningful population of SNe\,Ia in this regime is thus important both for understanding their progenitors and for validating their use in future precision cosmology.

Until recently, observational studies of both populations were strongly limited at high redshift. The combination of cosmological dimming and the redshifting of rest-frame optical emission into the infrared makes SNe at $z\gtrsim1$ exceptionally challenging for previous facilities. The Hubble Space Telescope created the first sample of such SNe with a few dozen discoveries to $z\sim 2$, but most were photometrically classified and had limited data coverage \citep{rodney_type_2014,strolger_rate_2015}. The James Webb Space Telescope (JWST) fundamentally changed this landscape. Its infrared sensitivity, wavelength coverage, aperture, and angular resolution enable the discovery and characterization of SNe at redshifts that were previously accessible only in exceptional cases. Early JWST observations have already demonstrated this potential, producing substantial samples of SNe beyond $z=1$, including spectroscopically confirmed CCSNe and SNe\,Ia at redshifts well beyond the reach of most previous SN surveys \citep{2023ApJS..269...43Y,JADES_transients,siebert_discovery_2024,coulter_discovery_2025,pierel_discovery_2024,pierel_testing_2025,siebert_sn_2025,2026ApJ...998..115Y,2026arXiv260820781P}. These first results have begun to probe whether CCSN explosion properties and subtype populations evolve with redshift, while simultaneously extending tests of SN~Ia standardization into an entirely new cosmological regime. However, these samples were generally created from ad hoc surveys \citep[e.g.,][]{JADES_transients, COSMOS-Web_transients}, whereas repeated deep JWST imaging with long baselines and spectroscopic follow-up would offer an opportunity not simply to extend existing SN surveys to greater distances, but to begin measuring the transient populations during substantially earlier epochs of stellar and galaxy evolution.

\proj\ \citep{NEXUS} is a multi-cycle (Cycles 3-5) JWST imaging and spectroscopic survey around the North Ecliptic Pole, with a key science component on systematic discovery of high-redshift SNe. It features two overlapping tiers. The \textbf{Wide tier} covers a footprint of $\sim 400\,{\rm arcmin^2}$, and performs NIRCam imaging in six filters (F090W, F115W, F150W, F200W, F356W, F444W) and WFSS slitless spectroscopy in F322W2 and F444W. The Wide tier is revisited annually over three cycles with a $\Delta t$ of about 17 months (PA orientation about $\pm 150$~deg). However, due to additional scheduling constraints, the first Wide epoch (Wide Ep01) was split in two, with the central $\sim 100\,{\rm arcmin^2}$ observed in September 2024 \citep{NEXUS_EDR} and the rest of Wide Ep01 observed in June 2025. The \textbf{Deep tier} covers the central $\sim 50\ {\rm arcmin^2}$ within the Wide tier, and performs NIRSpec MSA/PRISM spectroscopy and F200W+F444W NIRCam imaging on a 2-month cadence through early 2028. For both Wide and Deep observations, there are coordinated parallel observations with MIRI imaging and NIRCam imaging in additional filters, but these parallel observations have non-uniform and smaller coverage than the primary observations. The schedule for the multi-epoch Wide and Deep observations is now fixed, as shown in {Table~1} of \citet{NEXUS_QDR}. The full technical details of the program design and science cases are presented in the NEXUS overview paper \citep{NEXUS}, and the relevant Quick Release data (e.g., multi-epoch imaging and spectra in the Deep area) are detailed in \citet{NEXUS_QDR}. 

One of the main science goals of NEXUS is to discover and characterize the high-redshift transient population detectable with JWST. In this work we describe the effort of transient searches within the Deep tier of NEXUS, and present the results from the first-year observations. The scheduling of each year's observations of NEXUS does not follow exactly JWST cycle definitions. For year one observations, we refer to observations that include the first Wide (reference) epoch and six Deep epochs, i.e., the first third of the full NEXUS data. 

This paper is organized as follows. {In Section~\ref{sec:data}, we outline the NEXUS survey design, including NIRCam imaging and NIRSpec/MOS spectroscopic follow-up observations. Section~\ref{sec:detection} details our transient selection algorithm and astrometric refinement pipeline. In Section~\ref{sec:PSF_photometry}, we present a new Python package developed for high-precision point spread function (PSF) photometry on JWST imaging. Section~\ref{sec:results} describes the Year~1 NEXUS transient sample, transient light curves, host-galaxy association, and spectroscopic/photometric redshift determination. In Section~\ref{sec:discussion}, we discuss non-SN variable sources, evaluate JWST transient survey strategies and efficiency metrics, present raw detection rates, and examine the individual properties of our highest-redshift ($z > 3$) SN candidates. Main conclusions and forward-looking remarks are presented in Section~\ref{sec:con}. In this paper, we define high redshift as $z>3$.} Throughout the paper, we adopt a cosmology with $H_0=70$ km s$^{-1}$ Mpc$^{-1}$, $\Omega_{\rm m}=0.3$, and $\Omega_{\Lambda}=0.7$; all magnitudes are in the AB system.

\begin{deluxetable}{cccc}
\tablecaption{NEXUS Year 1 Observations\label{table:obs_date}}
\tablehead{
\colhead{Epoch Name} & \colhead{APT Obs No} & \colhead{Date (UTC)} & \colhead{Median MJD}}
\startdata
\multicolumn{4}{c}{\textbf{Reference NIRCam imaging}} \\
Wide Ep01 & 1 & 2024-09-12 & 60565.96\\
\hline
\multicolumn{4}{c}{\textbf{NIRCam imaging F200W + F444W}} \\
Deep Ep01 & 2 & 2025-06-01 & 60827.79\\
Deep Ep02 & 4 & 2025-07-30 & 60886.09\\
Deep Ep03 & 6 & 2025-09-28 & 60946.81\\
Deep Ep04 & 8 & 2025-11-28 & 61007.26\\
Deep Ep05 & 10 & 2026-01-28 & 61068.48\\
Deep Ep06 & 12 & 2026-03-28 & 61127.42\\
\hline
\multicolumn{4}{c}{\textbf{NIRSpec/MOS PRISM}} \\
Deep Ep01 & 3 & 2025-06-01 & 60827.96\\
Deep Ep02 & 5 & 2025-07-30 & 60886.26\\
Deep Ep03 & 7 & 2025-09-28 & 60946.98\\
Deep Ep04 & 9 & 2025-11-28 & 61007.43\\
Deep Ep05 & 11 & 2026-01-28 & 61068.65\\
Deep Ep06 & 13 & 2026-03-28 & 61127.60\\
\enddata
\end{deluxetable}

\begin{figure*}[t]
    \centering
    \includegraphics[width=\linewidth]{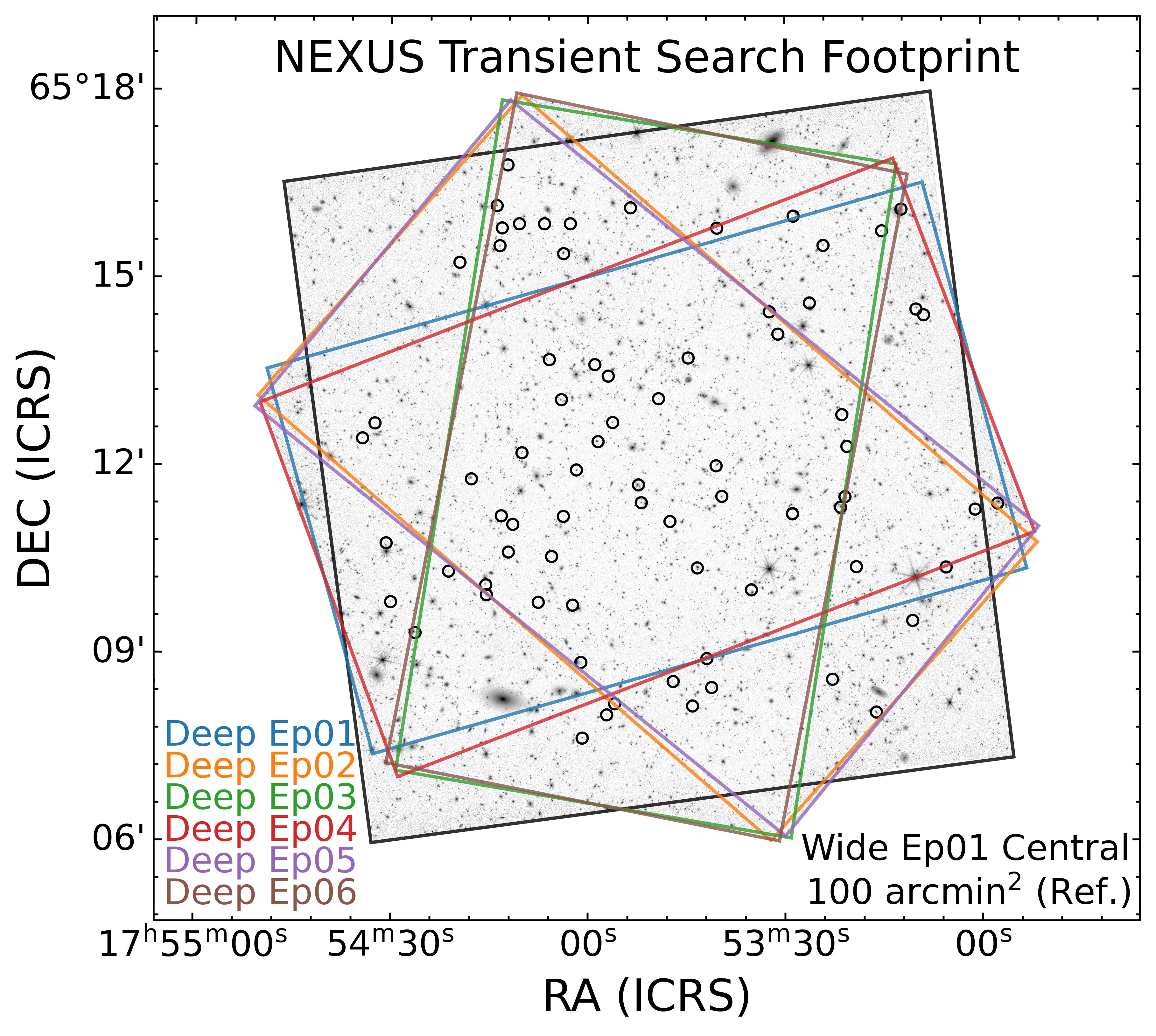}
    \caption{NIRCam imaging footprints of NEXUS Deep epoch transient search area. Black, blue, orange, green, red, purple, and brown curves represent the footprints of the central $100\, {\rm arcmin^2}$ Wide Ep01 area (used as reference image), and  coverage of Deep Ep01 through Ep06, respectively. The locations of the discovered transients are marked with open black circles. The background image shows the inverse weight averaged F444W image using data from Wide Ep01 and Deep Ep01-Ep06. {Each Deep epoch covers $\sim 61.5\,{\rm arcmin^2}$ within the reference Wide Ep01 footprint, and the central $\sim 35.8\,{\rm arcmin^2}$ area is covered by all Deep epochs.} }
    \label{fig:NEXUS_deep_FoV}
\end{figure*}

\begin{figure*}[t]
    \centering
    \includegraphics[width=0.99\linewidth]{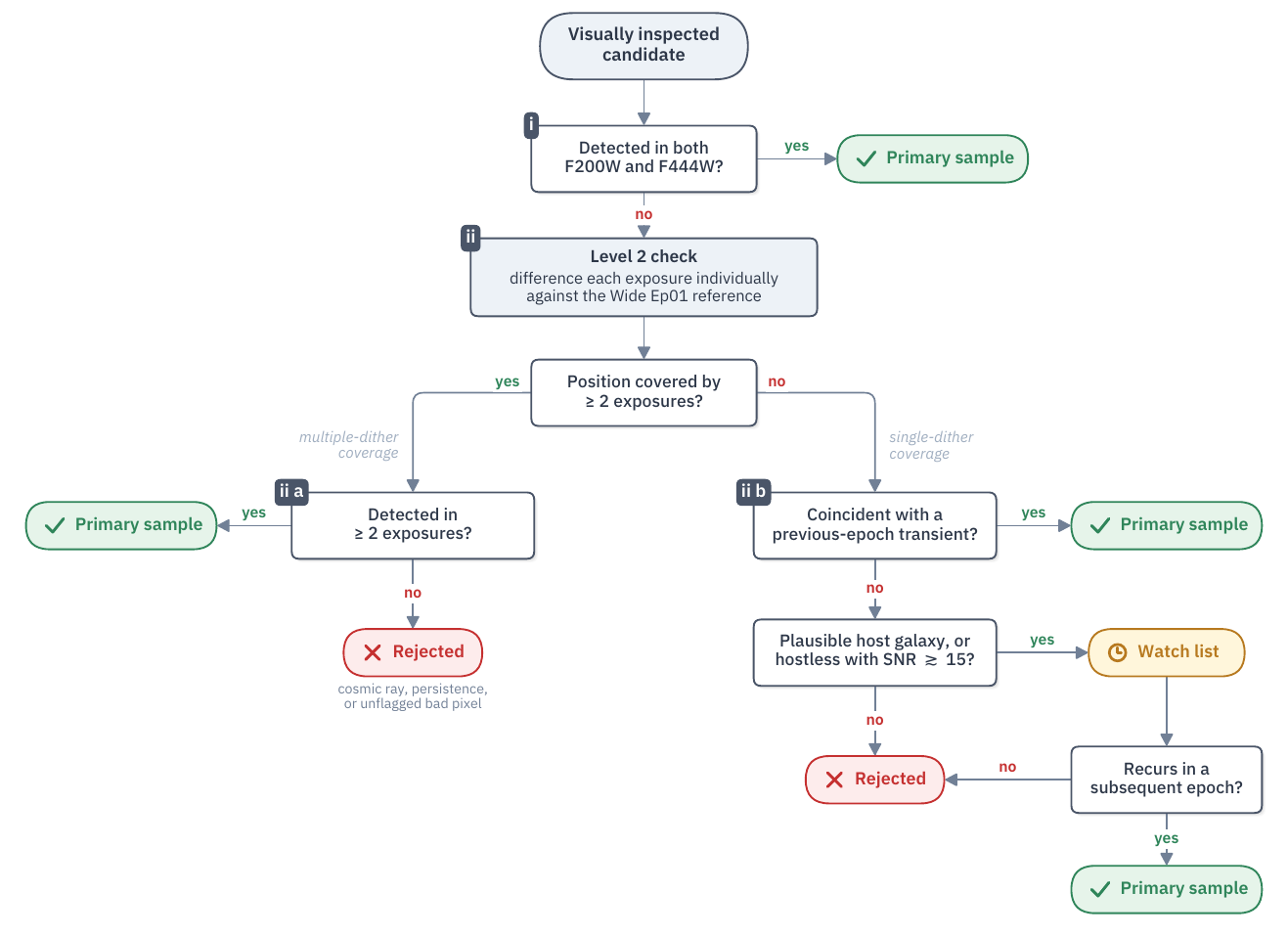}
    \caption{{Flowchart of the transient detection pipeline adopted in this work.}}
    \label{fig:flowchart}
\end{figure*}

\begin{figure*}[t]
    \centering
    \includegraphics[width=0.8\linewidth]{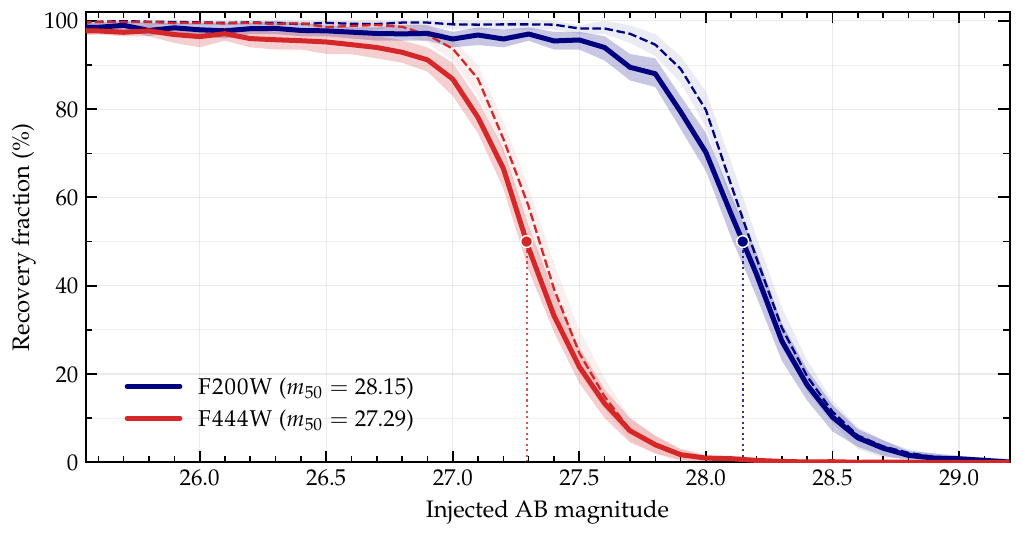}
    \caption{{Recovery fraction as a function of injected SN magnitude for the NEXUS deep-epoch transient search. F200W and F444W are shown in blue and red, respectively. The thick solid curves show the mean recovery fractions for SNe Ia and CCSNe using the host-offset prescriptions of \cite{JADES_II_SNRate}, while the thin dashed curves show the random-position baseline. A simulated source is considered recovered if it is matched to a pipeline detection with $\rm SNR \geq5$. The shaded regions show the 16th--84th percentile confidence intervals. The circles and dotted vertical lines mark the 50\% recovery limits, $m_{50}=28.15$~mag in F200W and $27.29$~mag in F444W.}}
    \label{fig:transient_completeness}
\end{figure*}

\begin{figure*}[t]
    \centering
    \includegraphics[width=\linewidth]{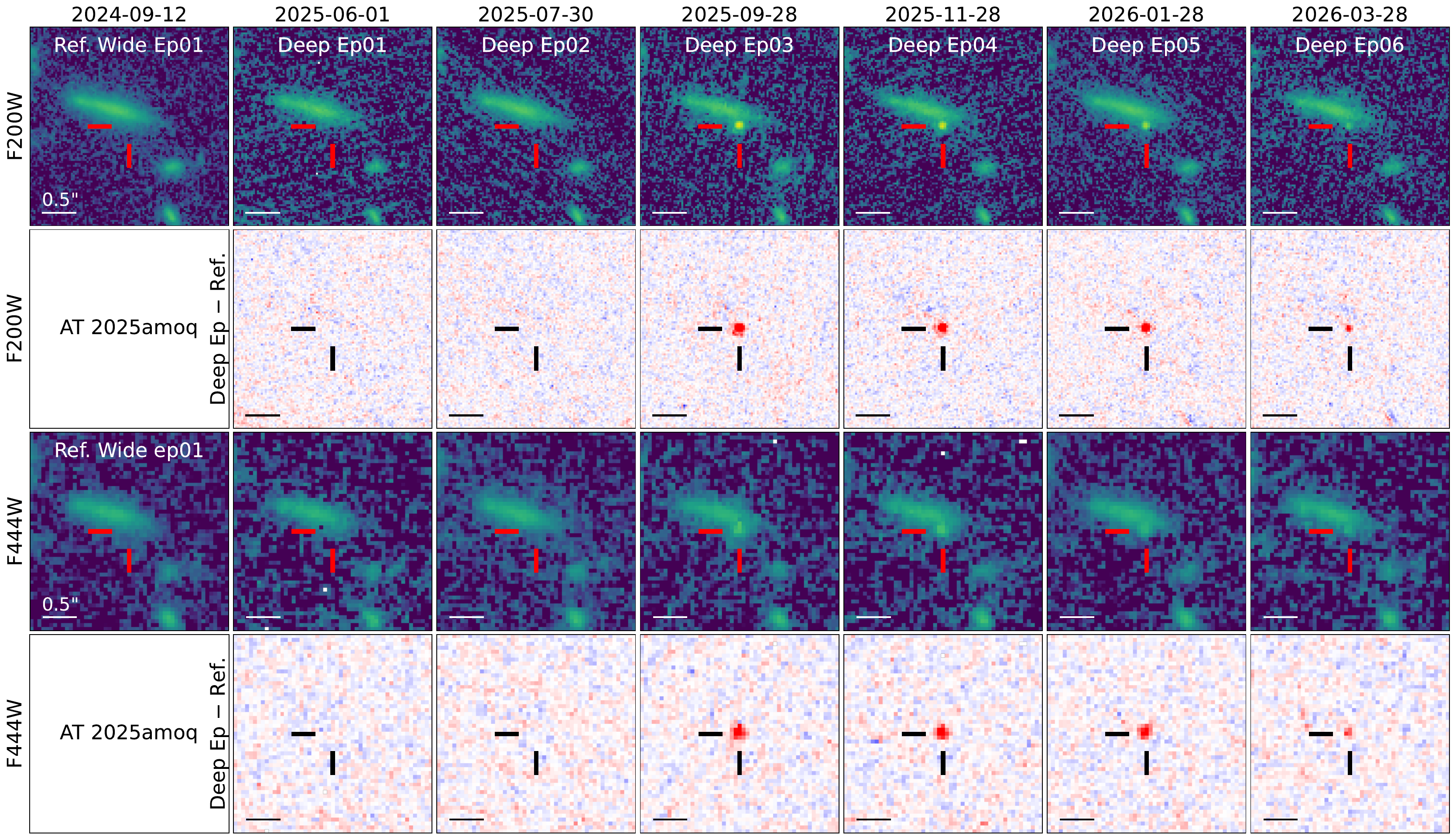}
    \caption{F200W (top two rows) and F444W (bottom two rows) image stamps of AT~2025amoq (indicated by red and black crosshairs) at $z=0.562$ in Wide Ep01 (reference epoch) and six Deep epochs (Ep01 to Ep06). The second and fourth rows show the difference images in F200W and F444W, respectively.}
    \label{fig:SN_diff_img}
\end{figure*}

\section{Observation and Data Reduction}\label{sec:data}

\subsection{Imaging Data}\label{sec:2.1}

To search for transients, we use the central $\sim100\, {\rm arcmin^2}$ of the NEXUS-Wide Ep01 NIRCam imaging as the reference and utilize the subsequent cadenced NIRCam observations in NEXUS Deep. Transient searches based on repeated Wide-area NIRCam imaging will be reported in a separate paper. The reference NIRCam observations were acquired between UT 2024 September 12 and 13. The follow-up NIRCam imaging observations began approximately 8.5 months later with a cadence of $\sim$60 days, extending through spring 2028. Table~\ref{table:obs_date} summarizes the Year 1 observations of the NEXUS program. 

Wide Ep01 NIRCam imaging observations include six filters (F090W, F115W, F150W, F200W, F356W, and F444W). The vast majority of this footprint has an exposure time of $\sim$20 min for F200W and $\sim$15 min for F444W, reaching $5\sigma$ depths of $\sim$28.2 and 27.9 mag, respectively (measured using randomly placed 0\farcs3-diameter apertures after aperture correction). The Wide Ep01 data are included in the early data release and are described in details in \citet{NEXUS_EDR}.

Deep epoch imaging observations utilize two filters (F200W and F444W), and largely overlap with the central Wide Ep01 NIRCam field. To maximize the sky coverage, we adopted the \texttt{FULLBOX} primary dither pattern with a \texttt{2TIGHTGAPS} step and no secondary dither. This setup results in a sky coverage of $\sim70\, {\rm arcmin^2}$, with $\sim$46\% and 54\% of the total area covered by one and two dithered exposures per individual epoch, respectively. Given the relatively short exposure time for individual exposure ($\sim$5 min), the $5\sigma$ depths are $\sim$27.2--27.6 for F200W and 27.3--27.6~mag for F444W measured using randomly placed $r=0\farcs15$ apertures without aperture correction. The footprint of the transient searching area analyzed in this paper is shown in Figure~\ref{fig:NEXUS_deep_FoV}. {The overlap region between the reference Wide Ep01 and each Deep epochs is on average $61.5\, {\rm arcmin^2}$ ($\sim88\%$ of the individual Deep epoch footprint), and the central $\sim35.8\,{\rm arcmin^2}$ area is covered by all Deep epochs.}

All images were reduced following the same procedures described in \citet{NEXUS_EDR} and were resampled to pixel scales of 30 mas for F200W and 60 mas for F444W. Because the NEXUS Deep epoch NIRCam imaging observations employ only two dither positions, these pixel scale choices closely approximate the native detector scales ($\sim31.5$ mas for F200W and 63 mas for F444W) to mitigate sub-sampling artifacts and the resulting ``pixelated'' structure in the point spread function (PSF). All epochs are pixel aligned on a common world coordinate system. 

We construct dedicated PSF models for F200W and F444W in each epoch using the same set of bright, isolated stars following the methods presented in \citet{Zhuang_PSF_2024}. These PSF models are oversampled by a factor of 4 and have a stamp size of $\sim4\arcsec\times4\arcsec$ ($133\times133$ pixel for F200W and $67\times67$ pixel for F444W). All magnitudes reported in this paper include aperture corrections derived from $2' \times 2'$ model PSFs generated using \texttt{stpsf} \citep{stpsf}.

\subsection{Spectroscopic Data}\label{sec:2.2}

For each Deep epoch, four pointings of NIRSpec/MOS PRISM spectroscopy were conducted immediately following the NIRCam imaging observations in the same epoch, covering the central $\sim50\, {\rm arcmin^2}$ region. PRISM spectroscopy covers a wide wavelength range of 0.55--5.5 \micron\ with low spectral resolution $R=\lambda/\Delta\lambda\sim30-300$. With a two-shutter slitlet nodding pattern for each pointing, each target received an effective exposure time of $\sim21.4$ min, reaching $3\sigma$ depths of $\sim27.2$~mag at 1.5 \micron\ and 25.9~mag at 4 \micron. 

For each Deep epoch, $\sim750-950$ objects were selected for Micro-Shutter Array (MSA) spectroscopy. The majority of the targets are bright galaxies with F444W brighter than 26~mag. Transient candidates were assigned higher weights to prioritize slit allocation. Candidates will remain in the target pool unless their fluxes fall below the detection threshold. For transient candidates located close to their host galaxies and thus contaminated by host galaxy emission, a final spectrum is scheduled to be obtained after the transient has faded to secure host properties and enable subsequent host subtraction if prior spectra of the transient were taken. Details regarding target weight assignments, spectroscopic data reduction, and Deep epoch data releases are provided in \citet{NEXUS_QDR}. Across the first six Deep epochs, we have acquired 41 spectra for {31 unique} transient candidates.

\section{Transient Detection and Astrometry Refinement}\label{sec:detection}

\subsection{Transient Selection Algorithm}

Transient candidates are identified in each Deep epoch by differencing its F200W and F444W mosaics against the corresponding Wide~Ep01 reference mosaics. As described in Section~\ref{sec:2.1}, all mosaics are registered to a common pixel grid and share a uniform photometric calibration. Image subtraction follows the methodology of \citet{Hu2024}, which was also adopted by \citet{Stone2026} for the NEXUS variability search. We first cross-convolve the reference and science mosaics with each other's PSF models generated using \texttt{stpsf} \citep{stpsf} before subtraction. The resulting difference image is then convolved with a noise-decorrelation kernel, which whitens the correlated background noise while recovering a compact effective PSF for transients. To process the large mosaics efficiently, the subtraction is executed in parallel over smaller image tiles.

For each epoch, we extract transient candidates from the decorrelated difference images using \textsc{SExtractor} \citep{SExtractor}, with \texttt{DETECT\_THRESH} $=1.5$ and \texttt{DETECT\_MINAREA} $=3$, and associate detections across images using a matching radius of $0\farcs2$. All candidates are visually inspected and classified according to the following selection criteria:
\begin{enumerate}
    \item[(i)] \textit{Dual-band detections.} 
    A candidate detected in both the F200W and F444W difference mosaics is directly included in the primary sample for that epoch.

    \item[(ii)] \textit{Single-band detections.} 
    A candidate detected in only one band is further examined using the Level~2 products, with each contributing exposure differenced individually against the reference image. 
    \textit{(a) Multiple-dither coverage:} A candidate detected in at least two individual exposures is included in the primary sample. If multiple exposures cover its position but the signal appears in only one, the candidate is rejected as a likely cosmic ray or instrumental artifact, such as a persistence residual or an unflagged bad pixel.
    \textit{(b) Single-dither coverage:} A candidate coincident with a transient identified in a previous epoch is included in the primary sample. Otherwise, a candidate associated with a plausible host galaxy, or a high-significance ``hostless'' candidate (signal-to-noise ratio SNR$\gtrsim15$), is placed on a watch list and promoted to the primary sample if it recurs in a subsequent epoch.
\end{enumerate}
Throughout the vetting procedure, detections exhibiting anomalous PSF morphologies are excluded from the candidate sample. {Figure~\ref{fig:flowchart} displays the flowchart of our candidate vetting procedure.}

We measured the transient detection completeness by injecting simulated supernovae with \(25.0\leq m_{\rm AB}\leq30.5\) into the difference images and rerunning our detection pipeline. The injected PSF models were generated with \texttt{stpsf} and propagated to the corresponding difference images. For host-associated injections, we followed the prescription of \cite{JADES_II_SNRate}: host galaxies were drawn at random, with SN offsets following an exponential distribution with \(\lambda=1.5\) half-light radii for CCSNe and an \(n=2\) Sérsic profile for SNe~Ia, both truncated at six half-light radii. Figure~\ref{fig:transient_completeness} shows the resulting completeness curves averaged over the SNe~Ia and CCSNe injections, yielding 50\% completeness depths of 28.15 and 27.29~mag in F200W and F444W, respectively. {We note that our selection criteria favor dual-band detections; for single-band detection, restricting to candidates with prior detection history or higher SNR thresholds effectively raises the operational detection threshold and slightly reduces overall sample completeness near the limit.}

Figure~\ref{fig:SN_diff_img} shows an example of a transient identified by our selection algorithm. As illustrated in the image cutouts from the reference and science epochs, the transient, AT~2025amoq appeared in Deep Ep03 on 2025 September 28 and remained detected for approximately six months through Deep Ep06 on 2026 March 28.

\begin{figure*}[t]
    \centering
    \includegraphics[width=0.8\linewidth]{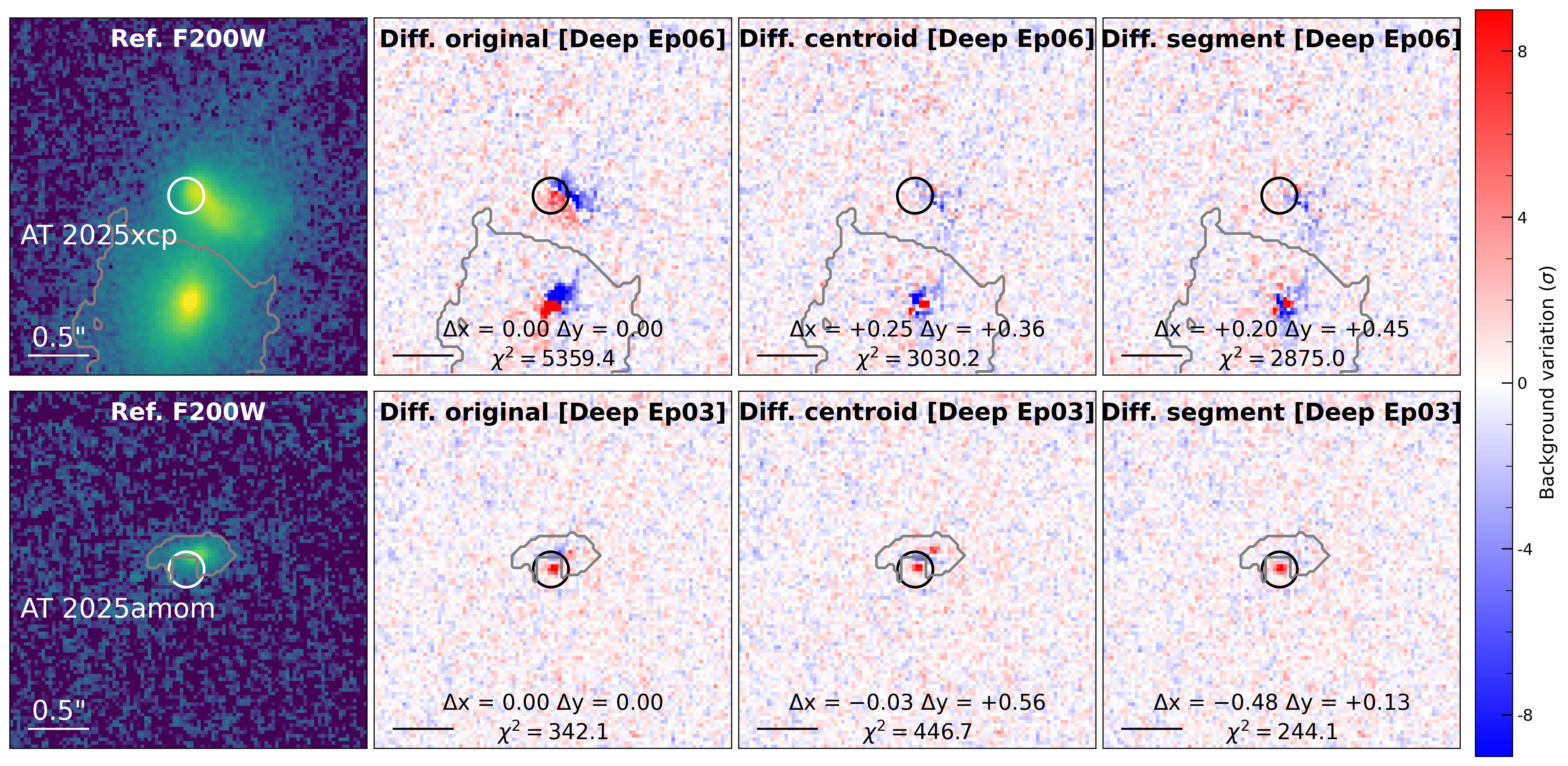}
    \caption{Comparison of F200W difference images before (Diff. original; second column) and after astrometry refinement using centroid only (Diff. centroid; third column) and all the pixel within the galaxy segment (Diff. segment; last column). First row shows false positive SN detection from AT~2025xcp host galaxy, and the second row shows the refined SN morphology and thus improved flux measurement for the genuine transient AT~2025amom. The first column shows F200W reference images. Offsets in X and Y directions and the normalized $\chi^2$ within the segment are shown at the bottom of each panel.}
    \label{fig:astrometry_refine_demo}
\end{figure*}

\subsection{Astrometry Refinement}

Although the global astrometric calibration in our pipeline achieves high precision ($0\farcs01$ rms) across the full field of view, local geometric distortion residuals can still introduce significant offsets between epochs. Sub-pixel offset as small as $\sim0.5$ pixels ($0\farcs015$) can produce substantial dipole residuals in difference images, severely hindering accurate transient photometry.

To address this issue, the JADES transient survey \citep{JADES_transients} implemented a local correction by spatially oversampling pixels by a factor of 8 via bilinear interpolation and performing an integer grid-search shift on the resampled scale, followed by visual inspection to select the optimal subtraction. While effective, this approach relies on a coarse, discrete spatial grid (equivalent to a $0.125$-pixel resolution limit) and introduces potential smoothing artefacts during the interpolation process. Furthermore, relying on visual inspection lacks formal statistical optimization and becomes impractical for automated pipelines. 

To overcome these limitations and achieve higher accuracy, we extend the framework of \citet{JADES_transients} by developing an automated, two-stage local astrometric refinement routine, leveraging Fourier-domain sub-pixel shifts and pixel-to-pixel intensity $\chi^2$ optimization. 

For each transient, we extract a local $12\arcsec\times 12\arcsec$ cutout ($401\times401$ in F200W and $201\times201$ in F444W) from each epoch. We first generate source segmentation to identify the closest bright galaxy (SNR $>100$) unblended with the transient (separation $>0\farcs12$). In the case where the transient is located in close proximity to the galaxy center, we mask the transient with a $5\times5$ pixel box to prevent self-subtraction bias. 

Rather than utilizing real-space interpolation, which alters noise properties and degrades high-frequency structural details, we apply shifts in the Fourier domain. Alignment accuracy is evaluated by calculating the reduced $\chi^2$ of the unmasked galaxy residual pixels: 
\begin{equation}
\chi^2 = \sum \frac{|f_{\text{ref}} - f_{\text{sci}}|^2}{\sigma_{\text{ref}}^2 + \sigma_{\text{sci}}^2},
\end{equation}
where $f$ and $\sigma$ denote the pixel flux values and corresponding uncertainties.

The optimization proceeds in two iterations: 
1. A coarse $21 \times 21$ grid search covering $\pm1.0$ pixel in steps of $0.1$ pixels.
2. A refined $21 \times 21$ grid centered on the $\chi^2$ minimum from the first pass, narrowing the search window to $\pm0.1$ pixels with a step size of $0.01$ pixels (0.3--0.6 mas). 

The final, optimal shift is determined by the absolute global minimum of this localized $\chi^2$ surface. This approach ensures that the spatial registration is entirely determined by the physical morphology of a close high SNR galaxy. By standardizing and automating sub-pixel alignment down to the $1/100\text{th}$ pixel level, this method eliminates subjective visual bias, preserves native pixel noise characteristics, and significantly minimizes dipole subtraction artifacts to enable reliable PSF photometry. 

Figure~\ref{fig:astrometry_refine_demo} demonstrates the performance of our approach compared to the default pipeline and centroid-based method. Our full pixel-matching method yields substantial improvements in subtraction residuals and $\chi^2$. Specifically, this refined astrometry eliminates a false-positive detection for AT~2025xcp in Deep Ep06, while restoring the correct morphology and flux for AT~2025amom in Deep Ep03.

Astrometry refinement was performed to 23 transients that exhibited noticeable offsets and were located close to their host galaxies. While offset can be as large as $\sim$0.8 pixel for one object, the median offset across the sample is $\sim$5 mas (1/6 pixel) in F200W and 4 mas (1/15 pixel) in F444W, with 95th percentile offsets of $\sim$14 mas (0.5 pixel) in F200W and 9 mas (0.15 pixel) in F444W. 

\begin{figure*}[t]
    \centering
    \includegraphics[width=0.4\linewidth]{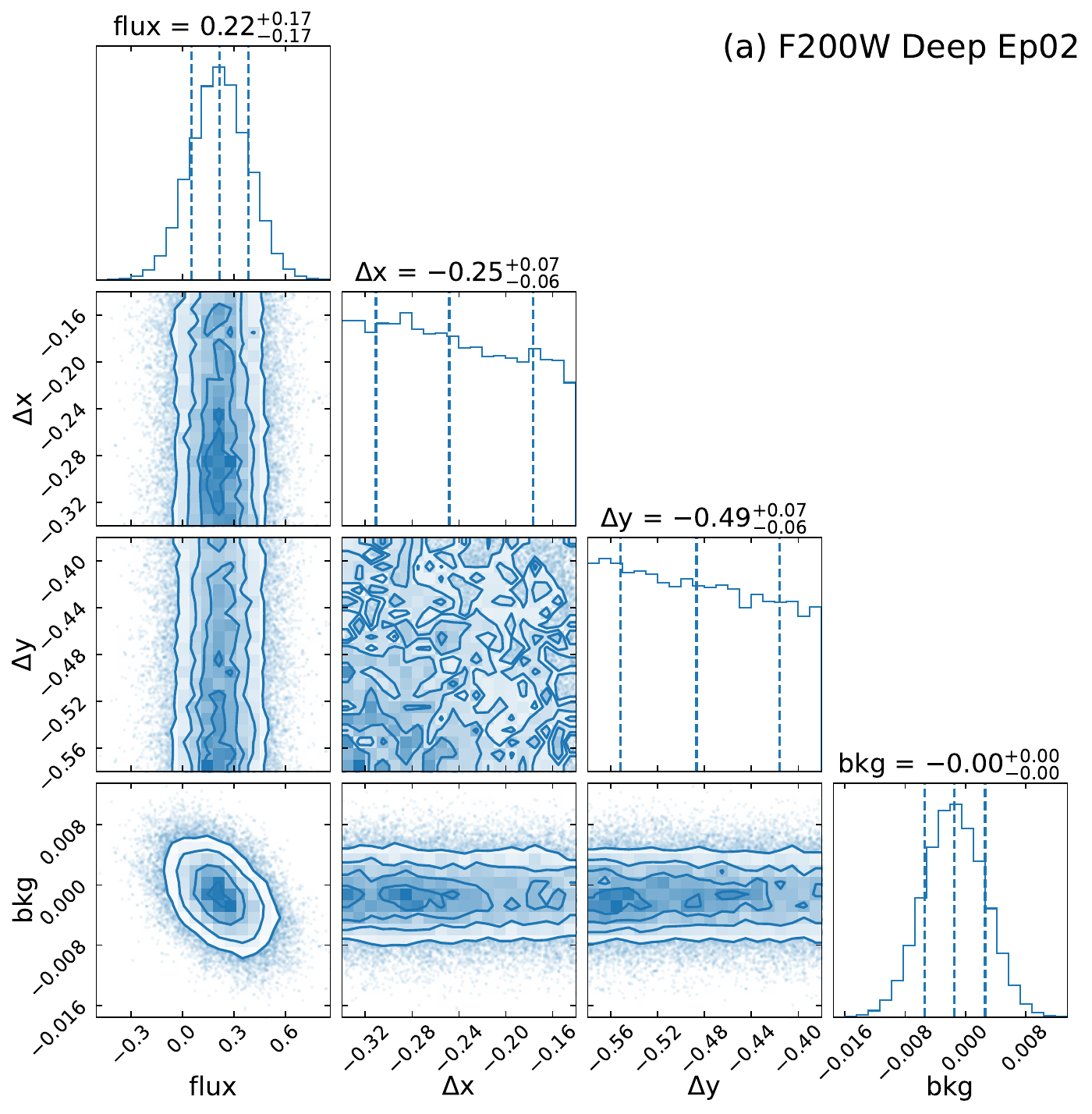}
    \includegraphics[width=0.4\linewidth]{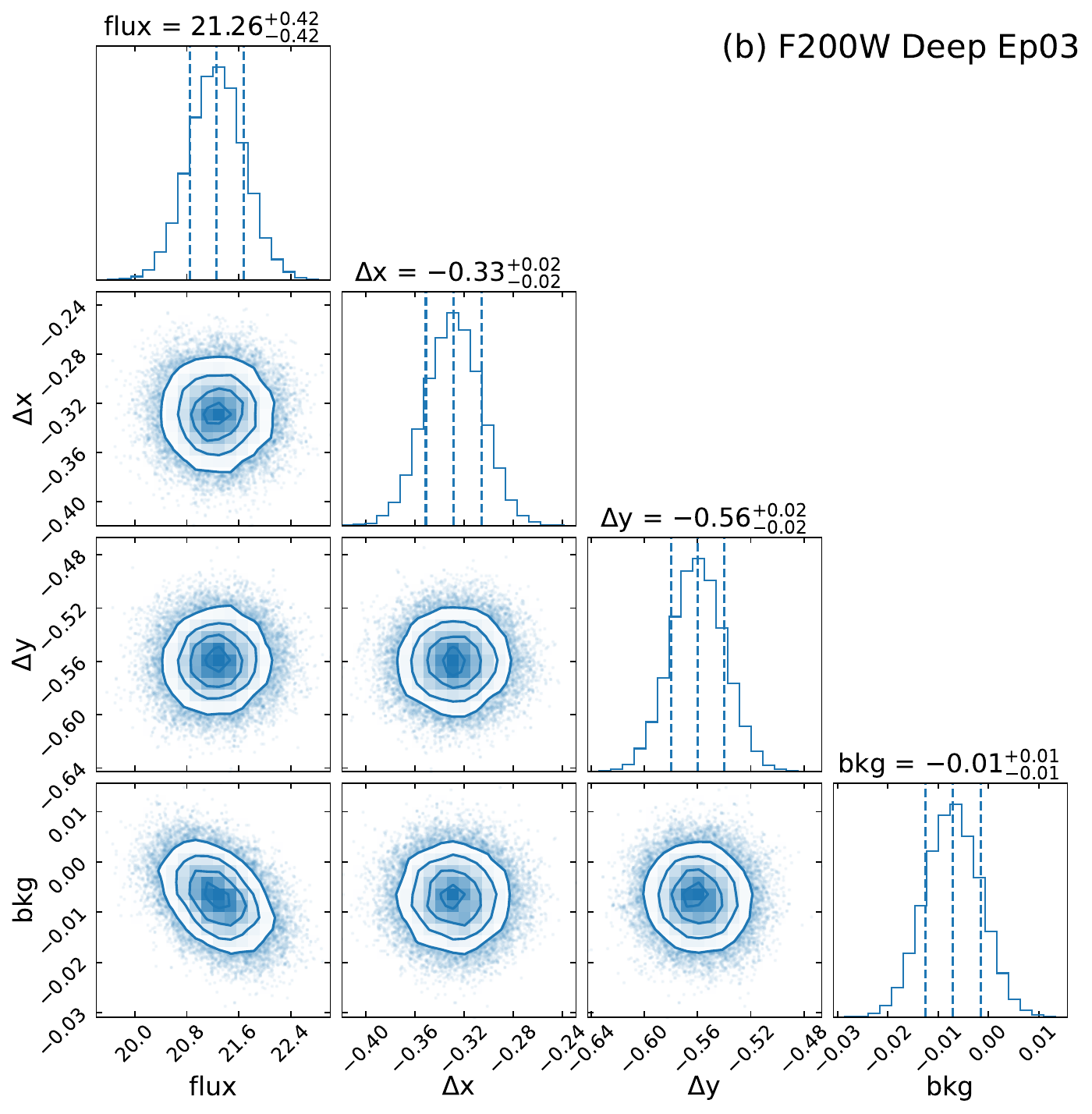}
    \includegraphics[width=0.4\linewidth]{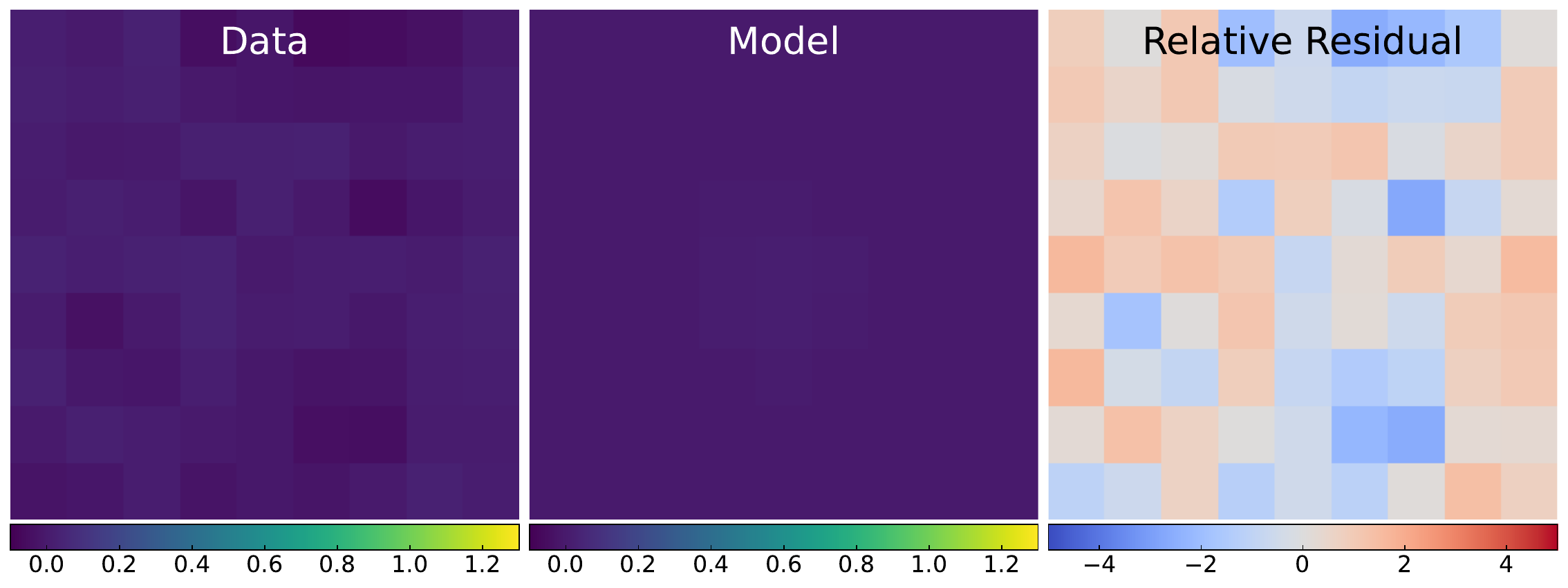}
    \includegraphics[width=0.4\linewidth]{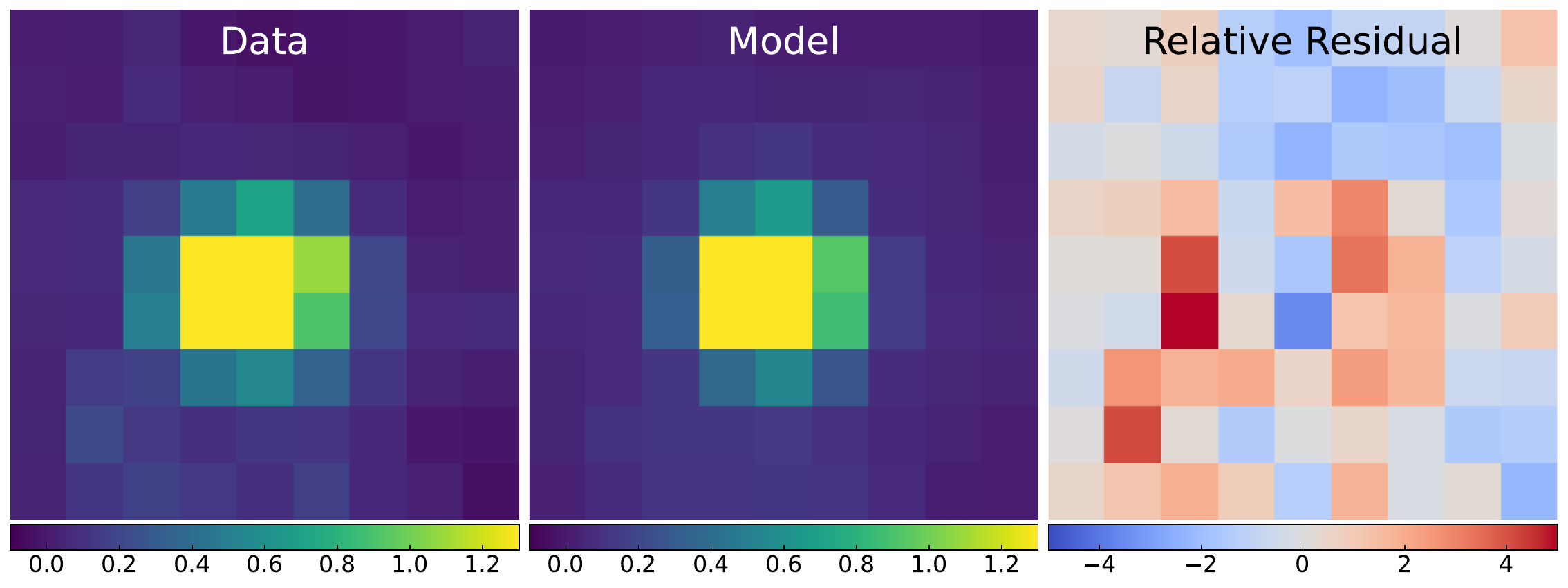}
    \caption{PSF photometry results of AT~2025amoq (the same object in Figure~\ref{fig:SN_diff_img}) for F200W in (a) Deep Ep02 and (b) Deep Ep03 using our method (Section~\ref{sec:PSF_photometry}). Top panels show posterior distributions of the fitting parameters, while the bottom panels display input data (difference image), best-fit model, and the relative residual ($\rm \frac{Data-Model}{Error}$). Vertical dashed lines indicate 16th, 50th, and 84th percentiles of the posterior distributions with values shown at the top of each subpanel. AT~2025amoq is not detected in Deep Ep02 (significance $=$ flux/error $\approx1.3\sigma$) but robustly detected in Deep Ep03 (significance $\approx50\sigma$).}
    \label{fig:SN_PSF_modeling}
\end{figure*}

\begin{figure*}[t]
    \centering
    \includegraphics[width=0.4\linewidth]{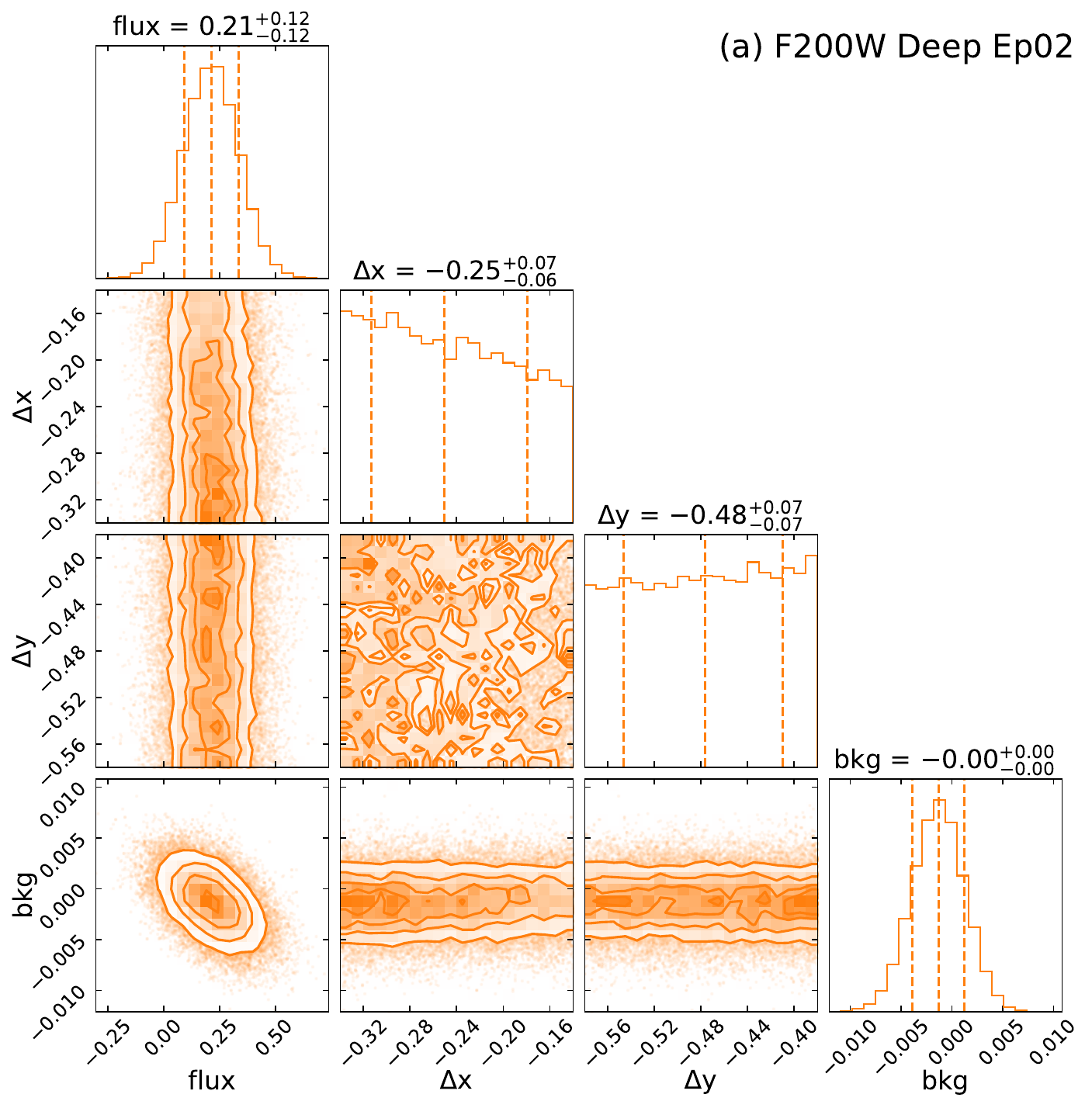}
    \includegraphics[width=0.4\linewidth]{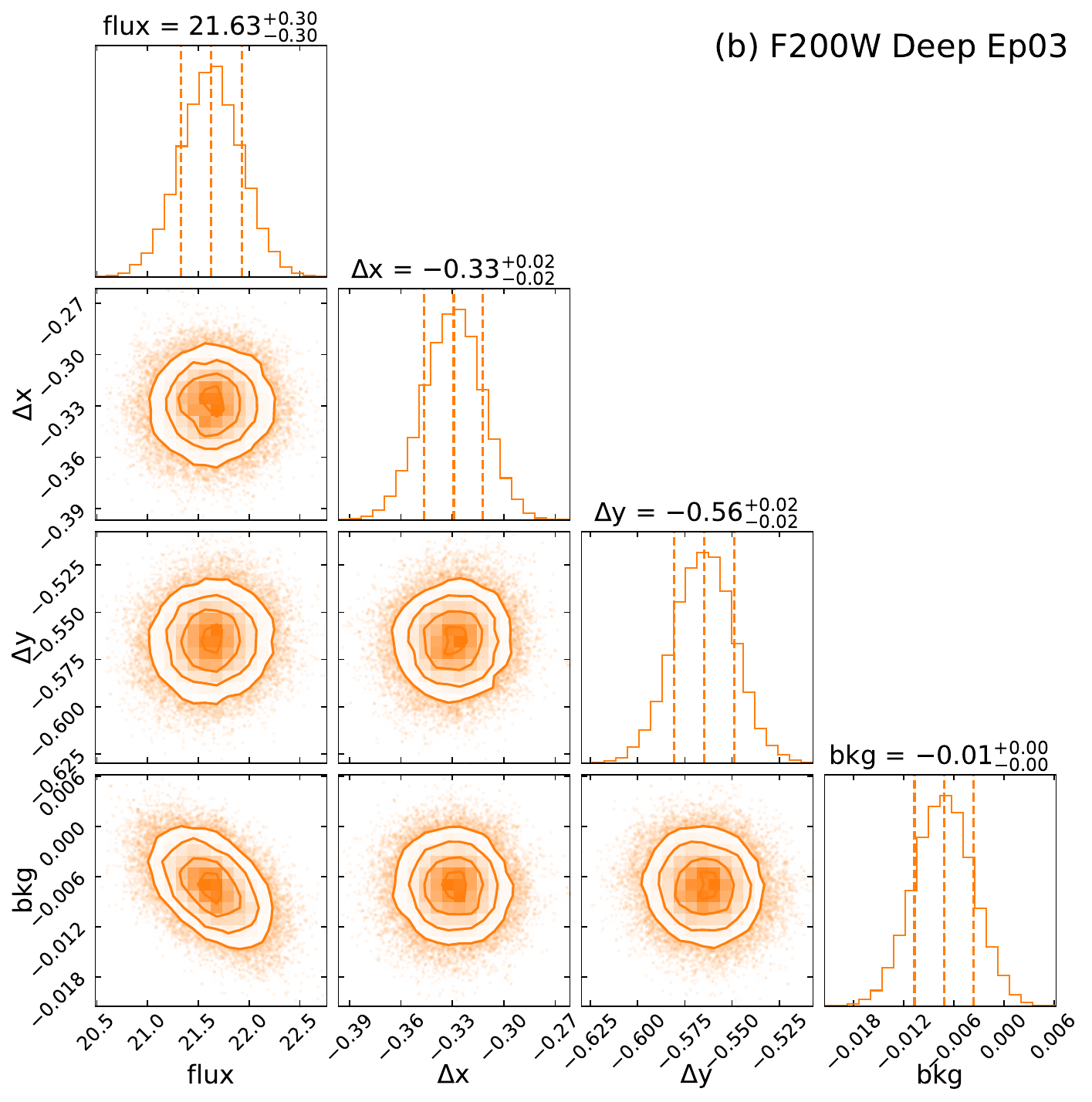}
    \caption{Same as upper panels of Figure~\ref{fig:SN_PSF_modeling} but without correlated pixel noise correction. The flux uncertainties are artificially reduced, compared with our fiducial approach in Section~\ref{sec:PSF_photometry}. }
    \label{fig:SN_PSF_modeling_no_covariance_correction}
\end{figure*}

\section{A New Package for Point Spread Function Photometry}\label{sec:PSF_photometry}

Precise photometric measurements in deep JWST imaging require careful treatment of correlated noise introduced by the resampling process. Standard PSF-fitting approaches typically assume spatially independent Gaussian noise, corresponding to a diagonal covariance matrix in pixel space. However, this assumption is not strictly valid for drizzled images, where neighboring pixels exhibit significant covariance affected by the drizzle parameters including the pixel fraction and the output pixel scale. Ignoring these correlations can distort the likelihood surface and lead to substantially underestimated photometric uncertainties, particularly for faint sources near the detection limit.

To address this issue, we have developed a new Markov chain Monte Carlo (MCMC)-based PSF photometry framework \texttt{jwst\_psfmc}\footnote{\url{https://github.com/mingyangzhuang/jwst_psfmc}} that explicitly models the correlated noise properties of the drizzled images. Instead of constructing the full covariance matrix directly in pixel space, which is computationally prohibitive for large image stamps, we estimate the empirical covariance kernel from blank-sky regions and evaluate the likelihood function in Fourier space, where the covariance becomes diagonal under the assumption of spatial homogeneity of the background noise. This approach enables an efficient and numerically stable treatment of correlated noise while preserving the full covariance information. The details of our method are presented in Appendix~\ref{Appendix_A}.

Our PSF photometry is performed on difference images with  Wide Ep01 as the reference epoch and produced by subtracting Wide Ep01 stamps from Deep epochs. However, Wide Ep01 F200W and F444W do not cover AT~2026rgb and AT~2026rge, respectively. Therefore, we use Deep Ep01 stamps as the reference epochs.

We first obtain the initial guess of transient centroids using F200W image in the discovery epoch and constrain their coordinates to be $\pm1$ pixel (flat prior) of this initial guess. For simplicity, we choose the relative distance in X and Y direction with respective to the central pixel ($\Delta$x and $\Delta$y) as fitting parameters. We adopt conservative constraints for the total integrated flux of the PSF model and the constant residual sky background with flat priors. These two parameters never reach boundaries. We adopt a fitting region of $9\times9$ pixels to avoid potential sky variation and speed up the modeling. In total, we have four free parameters. 

We directly evaluate the likelihood function in Fourier-space after correcting the correlated noise, and use \texttt{emcee} \citep{emcee} to perform MCMC sampling to explore the posterior distributions of the input parameters. We run the MCMC sampling with 48 workers and a chain length of 3000 steps ($N=3000$). 

With fitting results in hands, we examine posterior distributions of coordinates and the significance of flux to determine whether we need to adjust the boundary of coordinates. For transients with skewed Gaussian-like coordinate posterior distributions, we use the peak of their coordinate posterior distribution as the new initial guess and its $\pm1$ pixel values as the boundary. While for transients with relatively flat coordinate posterior distribution, we shrink the prior from $\pm1$ pixel to $\pm0.1$ pixel to obtain a more reliable flux upper limit at exactly the coordinates of the detected epoch(s). We repeat the fitting and obtain the final results for these objects.

For our sample, the integrated autocorrelation time ($\tau$) is $\sim40\text{--}50$ steps, well above the standard threshold of $N/\tau=50$, indicating that our MCMC sampling is converged and the parameter estimates are statistically reliable. Figure~\ref{fig:SN_PSF_modeling} shows two examples of our PSF photometry for a detection case and a non-detection case. 

PSF photometry can also reach better depths than aperture photometry. For example, AT~2025amoq is not detected in Deep Ep02 (Figure~\ref{fig:SN_PSF_modeling}a), with a $3\sigma$ upper limit of 28.8~mag from our PSF photometry. This upper limit is $\sim$0.8~mag deeper than that estimated from aperture photometry with a $3\sigma$ upper limit of 28.0~mag in a $0\farcs3$-diameter aperture after aperture correction.

For traditional PSF photometry methods, where correlated pixel noise is not corrected, Figure~\ref{fig:SN_PSF_modeling_no_covariance_correction} demonstrate that while the median values are generally consistent with each other, flux uncertainties would be underestimated by $\sim$30\% using the same examples. This may lead to systematics when directly using these ``incorrectly precise'' light curves to interpret supernova properties such as types and peak phase.

\begin{deluxetable*}{lcccclcccc}[t]
\tablecaption{NEXUS Year 1 Supernovae Sample\label{table:sample}}
\tablehead{
\colhead{SN ID} & \colhead{R.A.} & \colhead{Decl.} & \colhead{Detection} & \colhead{Discovery} &
\colhead{SN ID} & \colhead{R.A.} & \colhead{Decl.} & \colhead{Detection} & \colhead{Discovery}
}
\startdata
AT 2025xcm & 268.56441 & 65.16542 & \cmark\cmark\cmark\cmark\cmark\xmark & Deep Ep01 & AT 2025amor & 268.56472 & 65.16812 & \xmark\xmark\cmark\cmark\cmark\cmark & Deep Ep03 \\
AT 2025xcn & 268.43058 & 65.17262 & \cmark\cmark\cmark\xmark\xmark\xmark & Deep Ep01 & AT 2025amos & 268.58837 & 65.17169 & \xmark\namark\cmark\cmark\namark\cmark & Deep Ep03 \\
AT 2025xco & 268.46800 & 65.19471 & \cmark\cmark\cmark\xmark\xmark\xmark & Deep Ep01 & AT 2025amot & 268.54774 & 65.18417 & \xmark\xmark\cmark\cmark\xmark\xmark & Deep Ep03 \\
AT 2025xcp & 268.41505 & 65.19169 & \cmark\cmark\cmark\xmark\xmark\xmark & Deep Ep01 & AT 2025amou & 268.33975 & 65.18878 & \xmark\xmark\cmark\cmark\cmark\cmark & Deep Ep03 \\
AT 2025xcq & 268.41858 & 65.19983 & \cmark\cmark\xmark\xmark\xmark\xmark & Deep Ep01 & AT 2025amov & 268.33553 & 65.20494 & \xmark\xmark\cmark\cmark\cmark\cmark & Deep Ep03 \\
AT 2025xcr & 268.43634 & 65.22851 & \cmark\cmark\cmark\cmark\xmark\xmark & Deep Ep01 & AT 2025amow & 268.54373 & 65.26428 & \namark\xmark\cmark\namark\cmark\cmark & Deep Ep03 \\
AT 2025xcs & 268.51694 & 65.21741 & \cmark\cmark\cmark\cmark\xmark\xmark & Deep Ep01 & AT 2025amox & 268.55105 & 65.28000 & \namark\xmark\cmark\namark\cmark\cmark & Deep Ep03 \\
AT 2025xct & 268.53149 & 65.16343 & \cmark\cmark\xmark\xmark\xmark\xmark & Deep Ep01 & AT 2025amoy\tablenotemark{a} & 268.50449 & 65.14747 & \xmark\xmark\xmark\cmark\xmark\xmark & Deep Ep04 \\
AT 2025xcu & 268.49572 & 65.22676 & \cmark\cmark\cmark\cmark\xmark\cmark & Deep Ep01 & AT 2025amoz & 268.62518 & 65.16353 & \xmark\namark\namark\cmark\namark\namark & Deep Ep04 \\
AT 2025xcv & 268.35921 & 65.24322 & \cmark\namark\xmark\cmark\namark\cmark & Deep Ep01 & AT 2025ampa\tablenotemark{a} & 268.32962 & 65.17288 & \xmark\xmark\namark\cmark\xmark\namark & Deep Ep04 \\
AT 2025xcw & 268.37011 & 65.18701 & \cmark\cmark\cmark\xmark\xmark\xmark & Deep Ep01 & AT 2025ampb & 268.62808 & 65.17919 & \xmark\namark\namark\cmark\namark\namark & Deep Ep04 \\
AT 2025xcx\tablenotemark{a} & 268.28657 & 65.24001 & \cmark\namark\namark\xmark\namark\namark & Deep Ep01 & AT 2025ampc & 268.44791 & 65.18498 & \xmark\xmark\xmark\cmark\xmark\xmark & Deep Ep04 \\
AT 2025xcy & 268.29142 & 65.24148 & \cmark\namark\namark\xmark\namark\namark & Deep Ep01 & AT 2025ampd\tablenotemark{a} & 268.55498 & 65.18647 & \xmark\xmark\xmark\cmark\xmark\xmark & Deep Ep04 \\
AT 2025xcz & 268.42458 & 65.14847 & \cmark\cmark\xmark\xmark\xmark\xmark & Deep Ep01 & AT 2025ampe & 268.33678 & 65.19156 & \xmark\xmark\xmark\cmark\cmark\cmark & Deep Ep04 \\
AT 2025xda & 268.55057 & 65.17682 & \cmark\cmark\cmark\cmark\cmark\xmark & Deep Ep01 & AT 2025ampf & 268.64317 & 65.20714 & \xmark\xmark\namark\cmark\cmark\namark & Deep Ep04 \\
AT 2025xdb & 268.48707 & 65.22368 & \cmark\cmark\cmark\cmark\xmark\xmark & Deep Ep01 & AT 2025ampg & 268.48432 & 65.21133 & \xmark\xmark\xmark\cmark\cmark\xmark & Deep Ep04 \\
AT 2025xdc & 268.46610 & 65.18997 & \cmark\cmark\cmark\xmark\xmark\xmark & Deep Ep01 & AT 2025amph & 268.52453 & 65.22811 & \xmark\xmark\xmark\cmark\cmark\cmark & Deep Ep04 \\
AT 2025amnx & 268.50727 & 65.19869 & \cmark\cmark\cmark\cmark\cmark\xmark & Deep Ep01 & AT 2025ampi & 268.35056 & 65.25850 & \xmark\namark\xmark\cmark\namark\cmark & Deep Ep04 \\
AT 2025amny & 268.54182 & 65.20325 & \cmark\cmark\cmark\xmark\xmark\xmark & Deep Ep01 & AT 2026rfz & 268.44593 & 65.14233 & \namark\xmark\xmark\namark\cmark\cmark & Deep Ep05 \\
AT 2025amnz & 268.63525 & 65.21120 & \cmark\cmark\namark\cmark\xmark\namark & Deep Ep01 & AT 2026rga & 268.39636 & 65.16672 & \xmark\xmark\xmark\xmark\cmark\cmark & Deep Ep05 \\
AT 2025amoa & 268.33868 & 65.21341 & \cmark\cmark\cmark\cmark\cmark\cmark & Deep Ep01 & AT 2026rgb & 268.27265 & 65.17276 & \xmark\xmark\namark\xmark\cmark\namark & Deep Ep05 \\
AT 2025amob & 268.36958 & 65.26635 & \cmark\namark\cmark\cmark\namark\cmark & Deep Ep01 & AT 2026rgc & 268.52298 & 65.17564 & \xmark\xmark\xmark\xmark\cmark\cmark & Deep Ep05 \\
AT 2025amoc & 268.31725 & 65.13412 & \namark\cmark\namark\namark\xmark\namark & Deep Ep02 & AT 2026rgd & 268.25415 & 65.18814 & \xmark\xmark\namark\xmark\cmark\namark & Deep Ep05 \\
AT 2025amod & 268.43375 & 65.13583 & \namark\cmark\cmark\namark\cmark\cmark & Deep Ep02 & AT 2026rge & 268.23995 & 65.18970 & \xmark\namark\namark\xmark\cmark\namark & Deep Ep05 \\
AT 2025amoe & 268.34494 & 65.14289 & \namark\cmark\namark\namark\xmark\namark & Deep Ep02 & AT 2026rgf\tablenotemark{a} & 268.57409 & 65.19630 & \xmark\xmark\xmark\xmark\cmark\xmark & Deep Ep05 \\
AT 2025amof & 268.29406 & 65.15853 & \namark\cmark\namark\namark\cmark\namark & Deep Ep02 & AT 2026rgg & 268.37934 & 65.23489 & \xmark\xmark\xmark\xmark\cmark\cmark & Deep Ep05 \\
AT 2025amog & 268.51563 & 65.18636 & \xmark\cmark\cmark\cmark\cmark\cmark & Deep Ep02 & AT 2026rgh & 268.55471 & 65.26320 & \namark\xmark\xmark\namark\cmark\cmark & Deep Ep05 \\
AT 2025amoh & 268.49353 & 65.20624 & \xmark\cmark\cmark\xmark\xmark\xmark & Deep Ep02 & AT 2026rgi & 268.47298 & 65.26856 & \namark\xmark\xmark\namark\cmark\cmark & Deep Ep05 \\
AT 2025amoi & 268.58161 & 65.25402 & \namark\cmark\namark\namark\xmark\namark & Deep Ep02 & AT 2026rgj & 268.48813 & 65.13345 & \namark\namark\xmark\namark\namark\cmark & Deep Ep06 \\
AT 2025amoj & 268.51545 & 65.25634 & \namark\cmark\cmark\namark\xmark\xmark & Deep Ep02 & AT 2026rgk & 268.42161 & 65.14071 & \namark\xmark\xmark\namark\xmark\cmark & Deep Ep06 \\
AT 2025amok & 268.55607 & 65.25846 & \namark\cmark\cmark\namark\cmark\cmark & Deep Ep02 & AT 2026rgl & 268.60954 & 65.15535 & \xmark\namark\xmark\xmark\namark\cmark & Deep Ep06 \\
AT 2025amol & 268.52772 & 65.26428 & \namark\cmark\cmark\namark\cmark\xmark & Deep Ep02 & AT 2026rgm & 268.45512 & 65.21774 & \xmark\xmark\xmark\xmark\xmark\cmark & Deep Ep06 \\
AT 2025amom & 268.51121 & 65.26433 & \namark\cmark\cmark\namark\cmark\cmark & Deep Ep02 & AT 2026rgn\tablenotemark{a} & 268.38473 & 65.24082 & \xmark\xmark\xmark\xmark\xmark\cmark & Deep Ep06 \\
AT 2025amon & 268.55789 & 65.26916 & \namark\cmark\cmark\namark\cmark\xmark & Deep Ep02 & AT 2026rgo & 268.31321 & 65.26238 & \xmark\namark\xmark\xmark\namark\cmark & Deep Ep06 \\
AT 2025amoo & 268.50353 & 65.12727 & \namark\namark\cmark\namark\namark\xmark & Deep Ep03 & AT 2026rgp & 268.41821 & 65.26315 & \namark\namark\xmark\xmark\namark\cmark & Deep Ep06 \\
AT 2025amop & 268.48297 & 65.13640 & \namark\xmark\cmark\namark\namark\cmark & Deep Ep03 & AT 2026rgq & 268.30074 & 65.26813 & \xmark\namark\namark\xmark\namark\cmark & Deep Ep06 \\
AT 2025amoq & 268.50975 & 65.16271 & \xmark\xmark\cmark\cmark\cmark\cmark & Deep Ep03 & AT 2026rhb & 268.37018 & 65.18708 & \xmark\xmark\xmark\xmark\xmark\cmark & Deep Ep06 \\
\enddata
\tablecomments{Detection indicates the detection status of SN in NIRCam images of six NEXUS Deep epochs (Ep01-Ep06). \cmark: detected; \xmark: not detected; \namark: not covered. The ``Discovery'' column marks the epoch where SNe first appeared.}
\tablenotetext{a}{Tentative transient sources with only one detection in one band, and covered by only one dither point. }
\end{deluxetable*}

\begin{figure*}[t]
    \centering
    \includegraphics[width=0.9\linewidth]{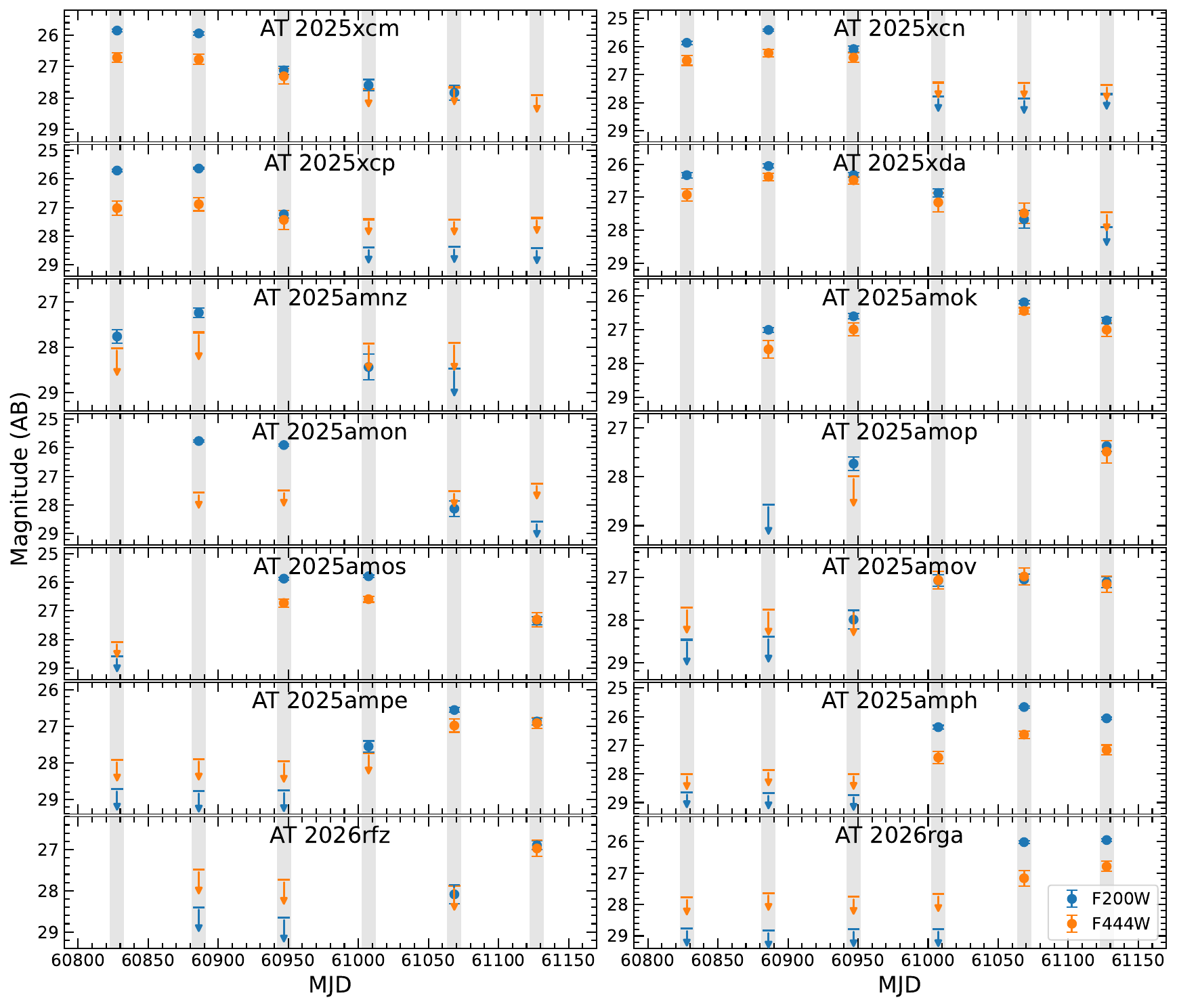}
    \caption{Light curves of 14 transients with their early rising phases covered by NEXUS-Deep epochs. Blue and orange dots represent F200W and F444W, respectively. Errorbars indicate $1\sigma$ uncertainties, while downward arrows indicate $3\sigma$ upper limits. Vertical gray stripes indicate the date of six NEXUS Deep epochs.}
    \label{fig:lightcurves}
\end{figure*}

\begin{figure*}[t]
    \centering
    \includegraphics[width=0.8\linewidth]{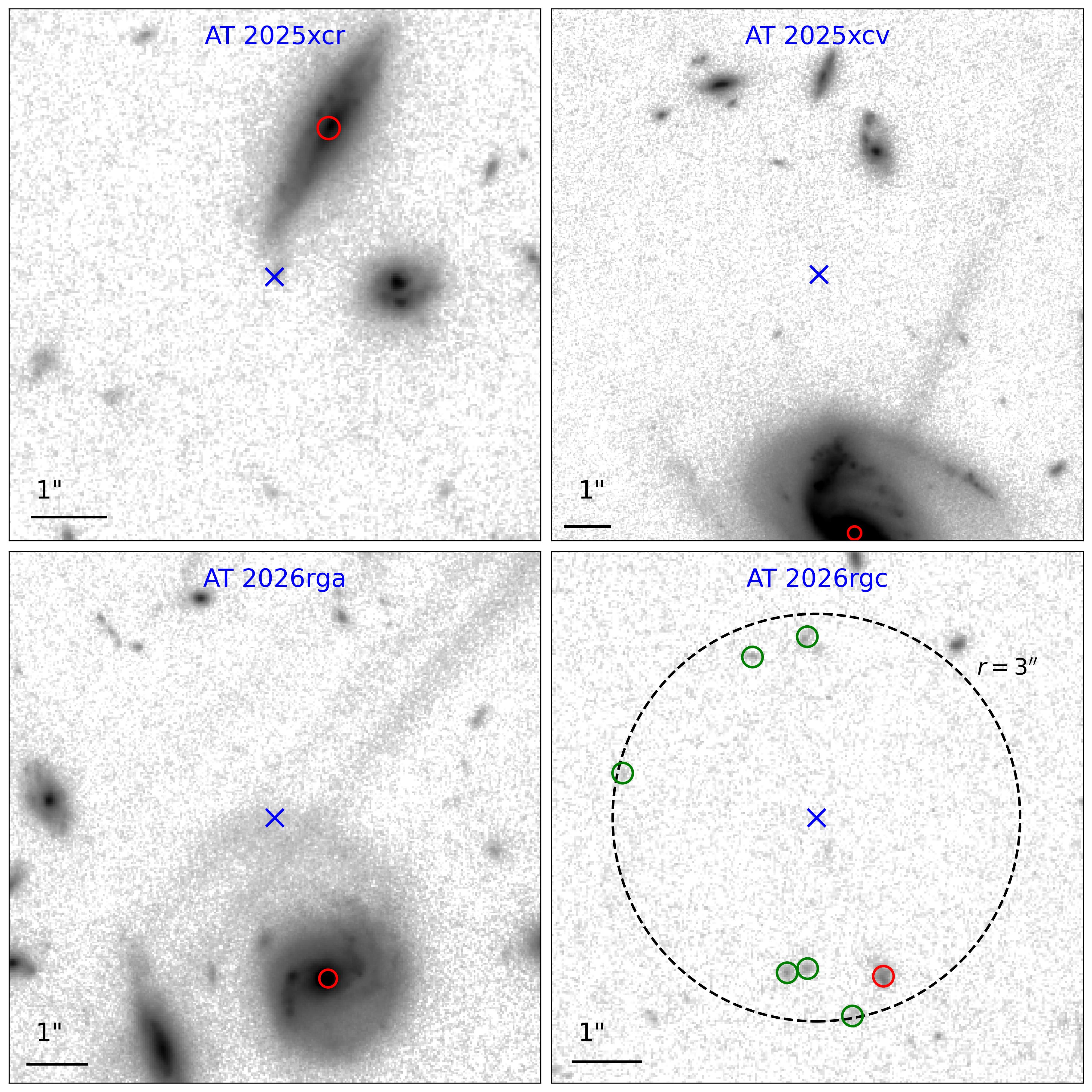}
    \caption{F200W cutouts of four transients (blue cross) and their assigned host galaxies (red circle) with separation greater than 2\arcsec. All other candidate host galaxies of AT~2026rgc within 3\arcsec\ are marked with green circles.}
    \label{fig:transient_hosts}
\end{figure*}

\section{Results}\label{sec:results}

\subsection{The NEXUS Year 1 Transient Sample}

We have identified 74 transients across all six Deep epochs in the NEXUS-Deep tier. The basic properties of the sample are summarized in Table~\ref{table:sample}. Multi-epoch F200W and F444W NIRCam reference, science, and difference image cutouts for all sources are provided in Appendix~\ref{sec:stamp_gallery}. Among the 74 transients, 68 are robust transients with detections in more than one epoch/exposure or detected simultaneously in both the F200W and F444W filters in at least one epoch. The remaining six objects (AT~2025xcx, AT~2025amoy, AT~2025ampa, and AT~2025ampd, AT~2026rgf, AT~2026rgn) are classified as tentative candidates, as each was only detected in a single epoch, within a single band, and covered by a single dither position. These six candidates are reported here for completeness but are excluded from subsequent analysis.

\subsection{Transient Light Curves}\label{sec5.2}

The cadenced NIRCam imaging and dedicated PSF photometry of the NEXUS survey enable us to construct long term, moderate cadenced light curves. Appendix~\ref{Appendix_B} presents the six-epoch F200W and F444W photometry for our sample. In total, 33 sources ($\sim$47\%) are detected in at least three epochs and 21 sources ($\sim31\%$) have at least six photometric measurements. Due to the generally higher brightness for SNe at shorter wavelengths in the infrared, more sources have F200W detections. As a result, 21 sources ($\sim31\%$) have F444W detections spanning at least three Deep epochs. 

Notably, the NEXUS light curves constrain the pre-peak or peak epochs for 14 transients (Figure~\ref{fig:lightcurves}). In particular, several targets exhibit two early epochs of similar flux followed by decaying brightness in subsequent observations, indicating that the observational window straddled their light curve peaks. Comprehensive light-curve template fitting and modeling for the full NEXUS transient sample will be presented in a companion paper.

\subsection{Host Galaxy Association}\label{sec:host_assoc}

To identify the host galaxy of our transient sample, we adapt a method based on a quantity termed the ``directional light radius'' \citep[DLR;][]{2006ApJ...648..868S, 2016AJ....152..154G, 2018PASP..130f4002S}. We calculate a dimensionless quantity $d_{\rm Kron}$, defined as the angular distance between a galaxy and the transient normalized by the Kron radius (square root of the product of the semi-major and semi-minor axes) of that galaxy, for each galaxy located within 3\arcsec\ of the transient. The galaxy with the minimum $d_{\rm Kron}$ is assigned as the host galaxy of a transient after visual inspection. 

The basic properties of the host galaxies are presented in Table~\ref{table:SN_host}. Among the 68 robust transients, 56 are close to their host galaxies with separations $\leq 1\arcsec$. Transients lacking an obvious host galaxy candidate {within 1\arcsec\ radius} are classified as ``hostless'' transients, though we still assign a host galaxy based on the $d_{\rm Kron}$ metric. Whether or not these transients with large host galaxy offsets are true associations will be investigated in future work. As noted in Section~\ref{sec:data}, because the Deep epochs are shallower than the Wide Ep01 reference epoch, we can exclude the possibility that these apparent ``hostless'' transients reside in faint, underlying galaxies undetected in the reference epoch. Most ``hostless'' transients are associated with large, extended foreground galaxies at $z<1$ (see Section~\ref{sec:redshifts}). Notably, four transients lie $>2\arcsec$ from their assigned host galaxies (Figure~\ref{fig:transient_hosts}):
\begin{itemize}
    \item \textbf{AT~2025xcr:} The association is clear, as the transient is located just beyond the stellar disk of the a large, extended galaxy. 
    \item \textbf{AT~2025xcv:} It has the largest angular separation from its candidate host galaxy in our sample ($\sim5\farcs9$). The assigned host is a giant, low-$z$ spiral galaxy at $z_{\rm spec}=0.359$. Despite the large offset, this galaxy has a significantly smaller $d_{\rm Kron}=2.6$ than all other neighbors within 3\arcsec, which have $d_{\rm Kron}\gtrsim4.5$.
    \item \textbf{AT~2026rga:} Spectroscopic follow-up yields $z_{\rm spec}=0.759$ for the transient, consistent with the redshift of the assigned host galaxy located $\sim2\farcs9$ away. The transient resides within asymmetric stellar streams/tidal tails, tracing the interaction-induced  \textit{in situ} star formation. 
    \item \textbf{AT~2026rgc:} Association remains ambiguous, as there are seven candidates located within a $2\farcs5\text{--}3\farcs0$ annulus. Our assigned host is the brightest and  largest candidate in this group. Follow-up spectroscopy of these host candidates, as well as the transient itself, is required to confirm the association. 
\end{itemize}

In addition to AT~2025xcv and AT~2026rgc, AT~2025xct is the only other transient located at a separation exceeding twice the host galaxy Kron radius ($d_{\rm Kron} > 2$). Its assigned host galaxy, located at an angular offset of $\sim 1\farcs6$, is the only detected source within $2\farcs4$ and the largest galaxy (Kron radius $\sim 0\farcs5$) within $3\arcsec$.

We highlight the discovery of two distinct transients (AT~2025xcw and AT~2026rhb) associated with the same host galaxy (ID~94163 at $z_{\rm spec}=1.946$). The two transients are separated by 0\farcs26, which corresponds to a projected distance of $\sim$2.2 kpc. AT~2025xcw was discovered in Deep Ep01 with comparable brightness in F200W and F444W ($\sim$27.5 mag) before fading below the detection threshold $\sim$180 days later in Deep Ep04. In contrast, AT~2026rhb remained undetected until Deep Ep06 ($\sim$180 days after AT~2025xcw completely faded) and exhibits a distinctly red color ($\text{F200W} - \text{F444W}\approx0.3$ mag). Further analysis of this multiple-transient host will be presented in future work after collecting more data of AT~2026rhb in later Deep epochs.

\begin{figure*}[t]
    \centering
    \includegraphics[width=\linewidth]{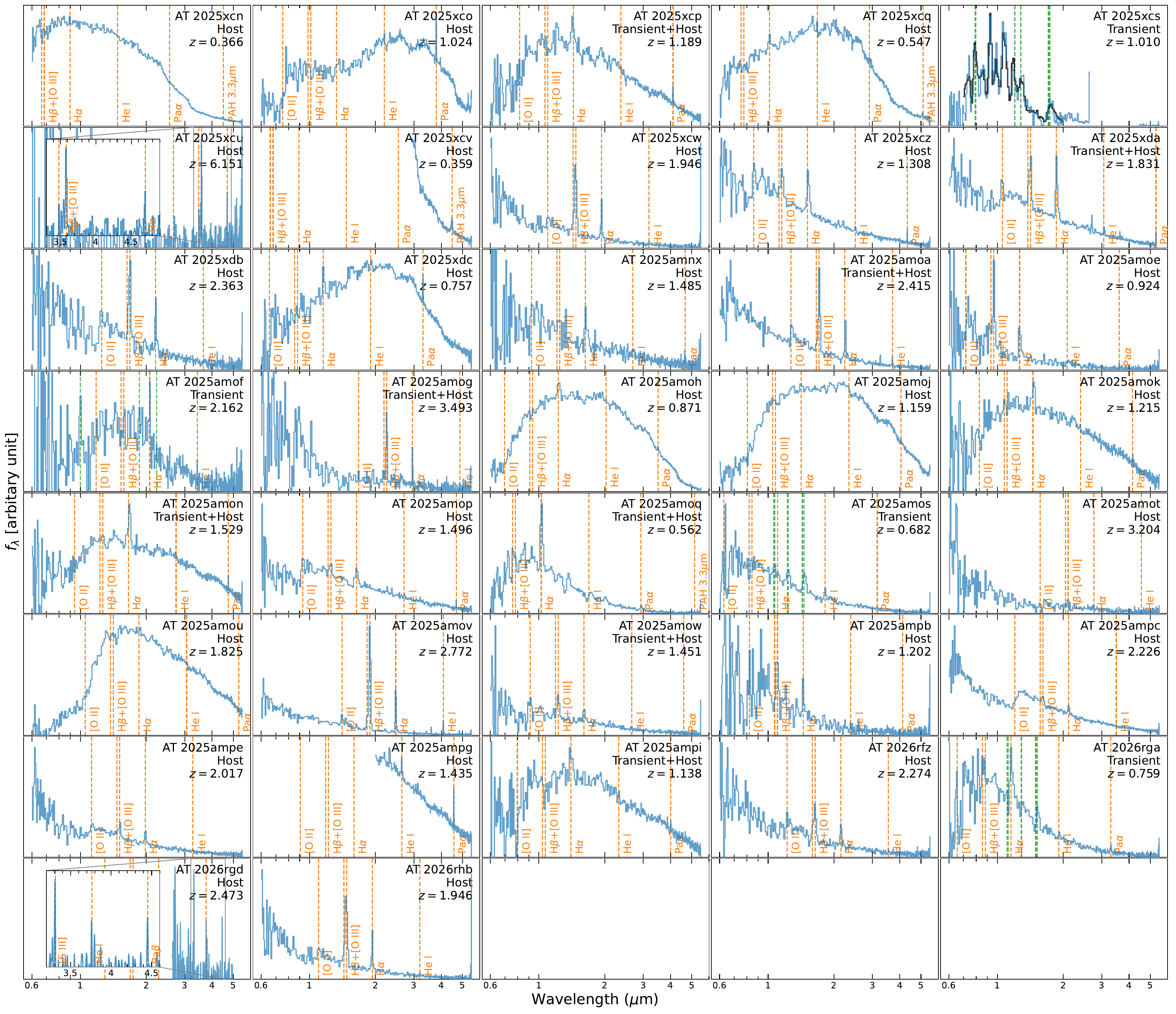}
    \caption{NEXUS NIRSpec PRISM spectra used for spectroscopic redshift ($z_{\rm spec}$) determination. Prominent host galaxy emission lines are marked with orange vertical dashed lines. The name of the transient, target of the spectrum (host: host-only, transient+host: transient+underlying host (dominated), or transient: transient-dominated), and $z_{\rm spec}$ are listed in the upper-right corner. {Vertical green dashed lines highlight spectral features for specific objects: \CaII\ H+K, \SiII\ $\lambda 5972\ \text{\AA}$, \SiII\ $\lambda 6355\ \text{\AA}$, and \CaII\ triplet $\lambda\lambda\lambda 8498, 8542, 8662\ \text{\AA}$ in AT~2025xcs; O\,{\sc iii} $\lambda3133\,\text{\AA}$, \HeI\ ($\lambda5876, \lambda7065$ for AT~2025amof; \OI\,$\lambda\lambda$ 6300, 6364 doublet, [\CaII]\,$\lambda\lambda$7291, 7324\, doublet, \CaII\,$\lambda\lambda\lambda$8498, 8542, 8662\, triplet for AT~2025amos and AT~2026rga. The spectrum for AT~2026rgd is obtained from NIRCam F322W2+F444W  WFSS and smoothed for better visualization. Insets in the AT~2025xcu and AT~2026rgd panels show zoomed-in views of key emission lines. The thick black curve in the panel of AT~2025xcs represents the best-fit SN~Ia template from \texttt{salt3-nir} using \texttt{SNCosmo} \citep{sncosmo}.}}
    \label{fig:Transient_host_spec}
\end{figure*}

\begin{figure}[t]
    \centering
    \includegraphics[width=\linewidth]{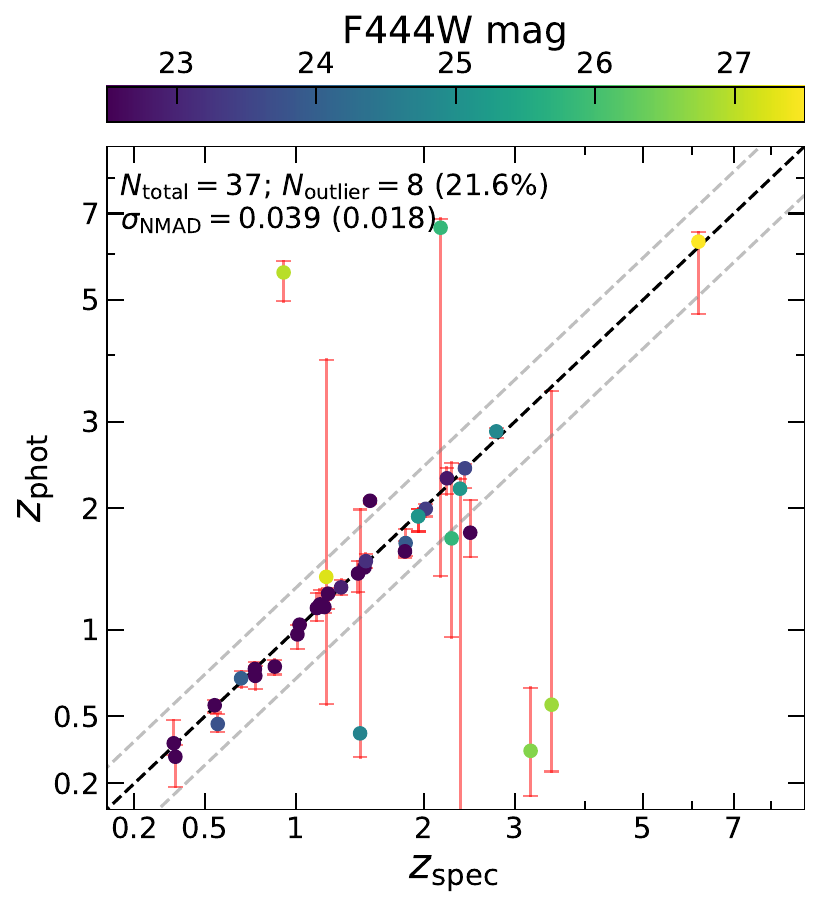}
    \caption{Comparison between photometric redshift $z_{\rm phot}$ and $z_{\rm spec}$ for 37 transient host galaxies, with colors indicating their F444W magnitudes. The black dashed line indicates the 1:1 relation ($z_{\rm phot}=z_{\rm spec}$), while gray dashed lines represent the boundary for outlier selection, defined as relative redshift deviation $|\Delta z| \equiv \frac{|z_{\rm phot}-z_{\rm spec}|}{1+z_{\rm spec}}=0.15$. The upper-left corner shows the total number of sources, the number of outliers ($|\Delta z|>0.15$) and its fraction, and the normalized median absolute deviation ($\sigma_{\rm NMAD}$) for the full sample and the non-outlier sample (in the parentheses). Error bars correspond to the 16th--84th confidence interval of the $z_{\rm phot}$ probability density distributions.}
    \label{fig:zspec_zphot}
\end{figure}

\subsection{Transient Redshifts}\label{sec:redshifts}

Spectroscopic redshifts ($z_{\rm spec}$) are obtained through follow-up NEXUS NIRSpec/MOS PRISM spectroscopy of the host galaxies and transients themselves by utilizing all available spectra up to Deep Epo08. As an augmentation, we also utilize NIRCam WFSS available for all the sources within the NEXUS field to search for emission lines in sources without PRISM spectra. For sources without spectroscopic coverage, we adopt photometric redshifts ($z_{\rm phot}$) estimated using \texttt{EAZY} \citep{EAZY} with the \texttt{agn\_blue\_sfhz\_13} template set. We supplement near-infrared coverage from NIRCam with optical $grizy$ broad-band and NB816 and NB921 narrow-band photometry from the Subaru Hyper Suprime-Cam HEROES survey \citep{HSC_HEROES}, {with $5\sigma$ point-source depths of 27.2, 26.4, 26.0, 25.4, 24.2, 24.5, and 24.6 mag, respectively.} For NIRCam host photometry, we use combined Wide Ep01 (F090W--F444W) and Deep Ep01-Ep04 (F200W, F210M, F360M, and F444W). However, for transients identified in Deep Ep01--Ep04 and lying close to their host centers, host photometry is extracted exclusively from Wide Ep01 to avoid transient contamination. Moreover, we also utilize NIRCam F322W2 and F444W WFSS available to all the sources in the NEXUS field. 

Redshifts of the NEXUS transient sample are presented in Table~\ref{table:SN_host}. We have obtained high-quality $z_{\rm spec}$ for {36} transients from PRISM spectra and for AT~2026rgd from NIRCam WFSS spectra (Figure~\ref{fig:Transient_host_spec}), including seventeen at $1<z<2$, eight at $2<z<3$, two at $3<z<4$, and one high-$z$ candidate (AT~2025xcu) at $z=6.151$. Of these {37} $z_{\rm spec}$ measurements, 25 are derived from host galaxy spectra and 12 are from spectra targeted at the transient location. Among the 12 spectra targeting the transients, eight are heavily contaminated by their host galaxies, while {four} exhibit distinctive SN features, as summarized below:
\begin{itemize}
    \item \textbf{AT~2025xcs at $z_{\rm spec}=1.010$:} At rest-frame $\lesssim 30$ days post-peak, the spectrum exhibits prominent broad, blueshifted P-Cygni absorption features, including \CaII\ H+K, \SiII\ $\lambda5972$ and $\lambda6355$, the \CaII\ triplet ($\lambda\lambda\lambda 8498, 8542, 8662\ \text{\AA}$), as well as \SiII\ $\lambda4130$ and $\lambda5640$ and a complex blend of \MgII, \FeII/\FeIII, \SiII\, and \SII\ features, characteristic of an SN~Ia. The best-fit SN~Ia template from \texttt{salt3-nir} shows an excellent match to the observed spectrum, yielding a rest-frame phase of $\sim22$ days post-peak and providing strong spectroscopic confirmation of its SN~Ia classification.

    \item {\textbf{AT~2025amof at $z_{\rm spec}=2.162$:} Observed in Deep Ep03 (one epoch after its discovery in Deep Ep02 at ${\text{F200W}} \approx 25.7$ mag), the spectrum displays \Ha\ and three \HeI\ lines ($\lambda5876, \lambda7065, \lambda10830$). Notably, we identify a strong O\,{\sc iii} $\lambda3133\,\text{\AA}$ Bowen fluorescence line, likely produced in dense circumstellar medium by shocks (as in an SN~IIn) or by intense photoionization from shock breakout flash during the early phase of a CCSN \citep[e.g.,][]{2002ApJ...572..350F, 2008Sci...321..223S, 2014ApJ...797..118F}. Given its high absolute magnitude ($M_V \lesssim -20.5$ from F200W in Deep Ep02), which exceeds that of typical CCSN, AT~2025amof thus is qualified as a superluminous SN~II (SLSN-II).}

    \item \textbf{AT~2025amos at $z_{\rm spec}=0.682$:} At rest-frame $\sim 100$ days post-peak (Epoch Ep06), the spectrum of AT~2025amos ($z_{\rm spec}=0.682$) displays prominent emission features including the \CaII\ triplet, [\CaII] $\lambda\lambda7291, 7324$, \HeI, and \OI $\lambda\lambda6300, 6364$. The complete absence of \OIII\ confirms that the weak \Ha\ emission is intrinsic to the SN rather than from its host galaxy. Combined with the strong \CaII\ and \HeI\ lines, these features suggest that AT~2025amos is an SN~IIb transitioning from photospheric to nebular phase.

    \item \textbf{AT~2026rga at $z_{\rm spec}=0.759$:} At rest-frame $\lesssim 30$ days post-peak (Deep Ep06), the spectrum of AT~2026rga exhibits prominent broad, blueshifted P-Cygni profiles in \Ha\ and the \CaII\ triplet, characteristic of an SN~IIP in its early photospheric plateau phase. Combined with the characteristically weak [\CaII] doublet, which has not yet developed due to high optical depth and electron density, these signatures confirm AT~2026rga as an early, optically thick SN~IIP before the onset of the nebular transition.
\end{itemize}

{All except one (AT~2025amou) of the host-dominated spectra show a collection of prominent emission features for reliable redshift determination.} The host of AT~2025amou displays a continuum-dominated spectrum with no prominent emission lines. We obtain its $z_{\rm spec}$ by performing spectral fitting to its continuum using \texttt{bagpipes} \citep{bagpipes}. 
Finally, we note that the redshift of AT~2025xcu is tentative due to host association ambiguities (see Section~\ref{sec:highz-SN}).

{To assess the reliability of our $z_{\rm phot}$ for the full transient sample, we compare it with $z_{\rm spec}$ in Figure~\ref{fig:zspec_zphot} for the spectroscopic subset. We find an excellent overall agreement, with a normalized median absolute deviation (NMAD) of $\sigma_{\rm NMAD}=0.039$ for the normalized difference $\Delta z \equiv \frac{z_{\rm phot}-z_{\rm spec}}{1+z_{\rm spec}}$. For all but three sources, the 16th--84th percentile $z_{\text{phot}}$ confidence intervals overlap the $|\Delta z| < 0.15$ region. Eight sources ($\sim21.6\%$) are classified as outliers ($|\Delta z|>0.15$), with four ($\sim10.8\%$) exhibiting catastrophic failures ($|\Delta z|>0.50$). These four catastrophic outliers are faint host galaxies with F444W in the range of 25.8--27.0~mag that lack optical detections. The median relative uncertainty of the photometric redshifts is $\sigma_{z} / (1 + z_{\rm phot}) = 0.066$. Nevertheless, several other faint host galaxies, such as those of AT~2025xct, AT~2025amob, AT~2025amol, AT~2025amph, AT~2026rgc, and AT~2026rgg, also suffer from poor optical constraints, resulting in substantially larger $z_{\rm phot}$ uncertainties.}

\begin{figure}[t]
    \centering
    \includegraphics[width=\linewidth]{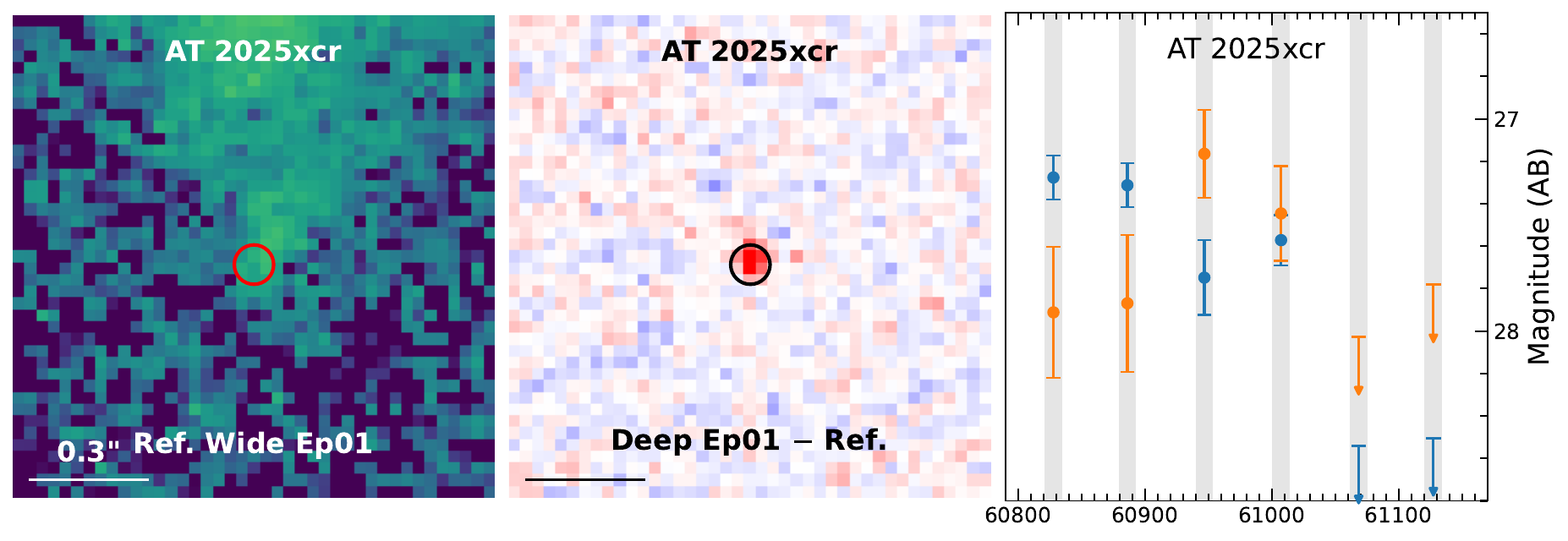}
    \includegraphics[width=\linewidth]{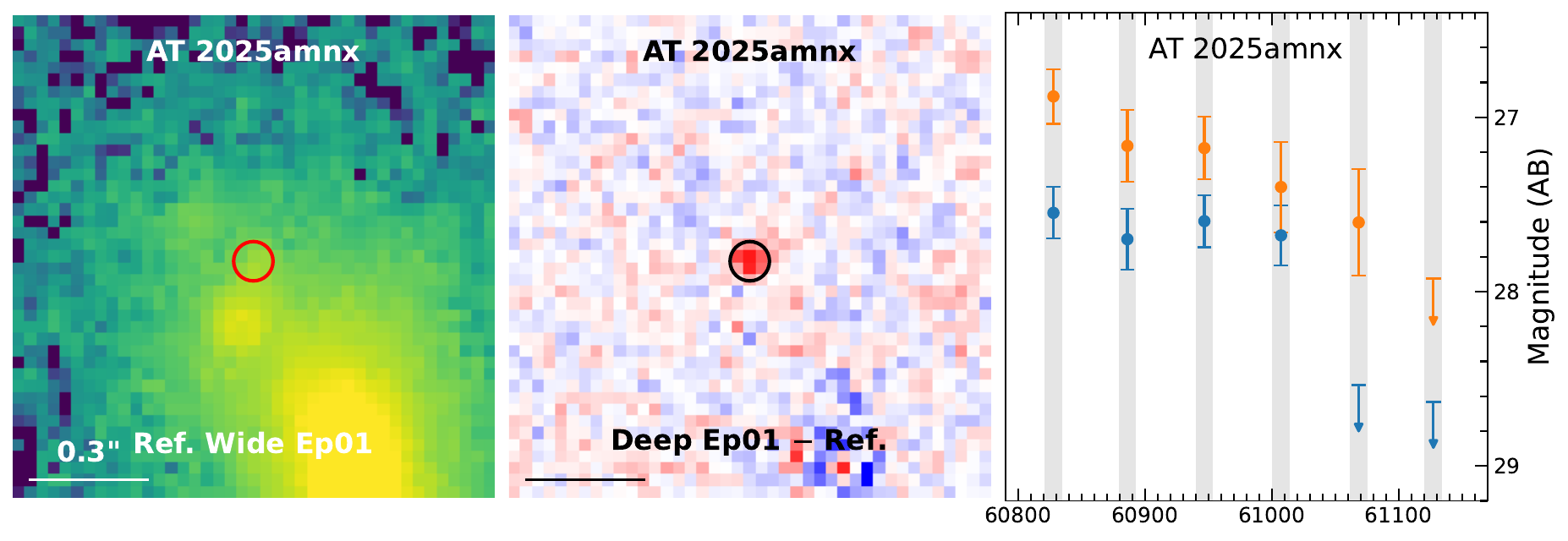}
    \includegraphics[width=\linewidth]{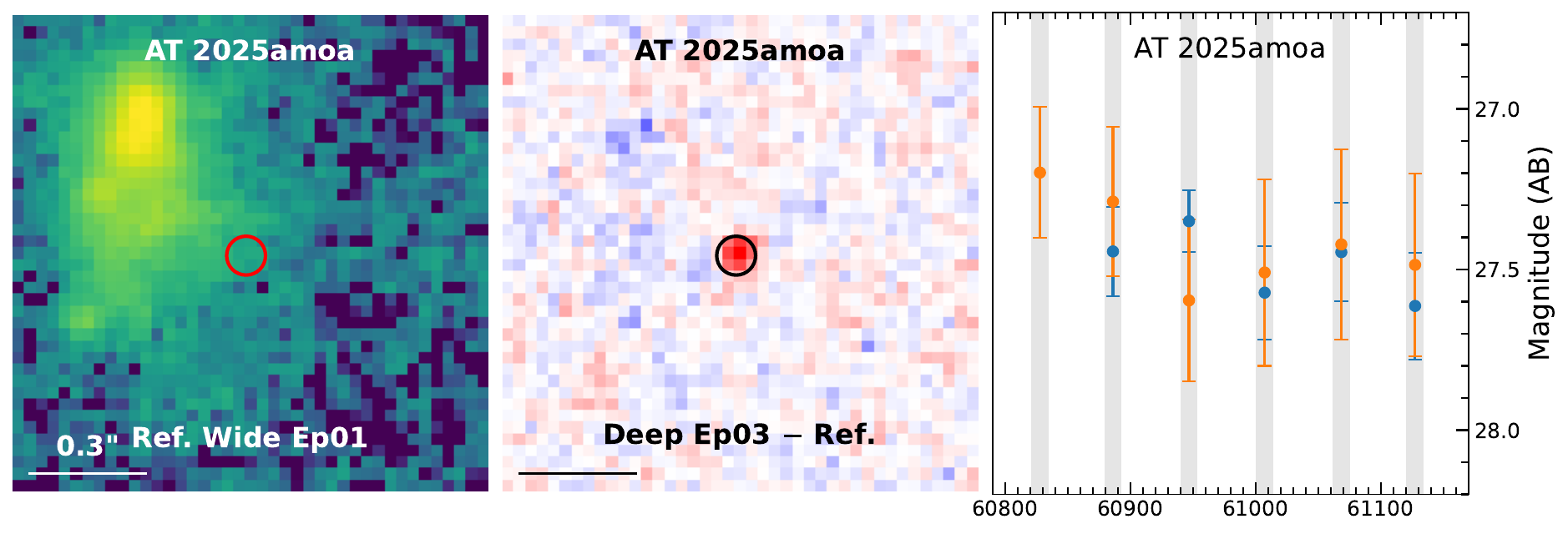}
    \includegraphics[width=\linewidth]{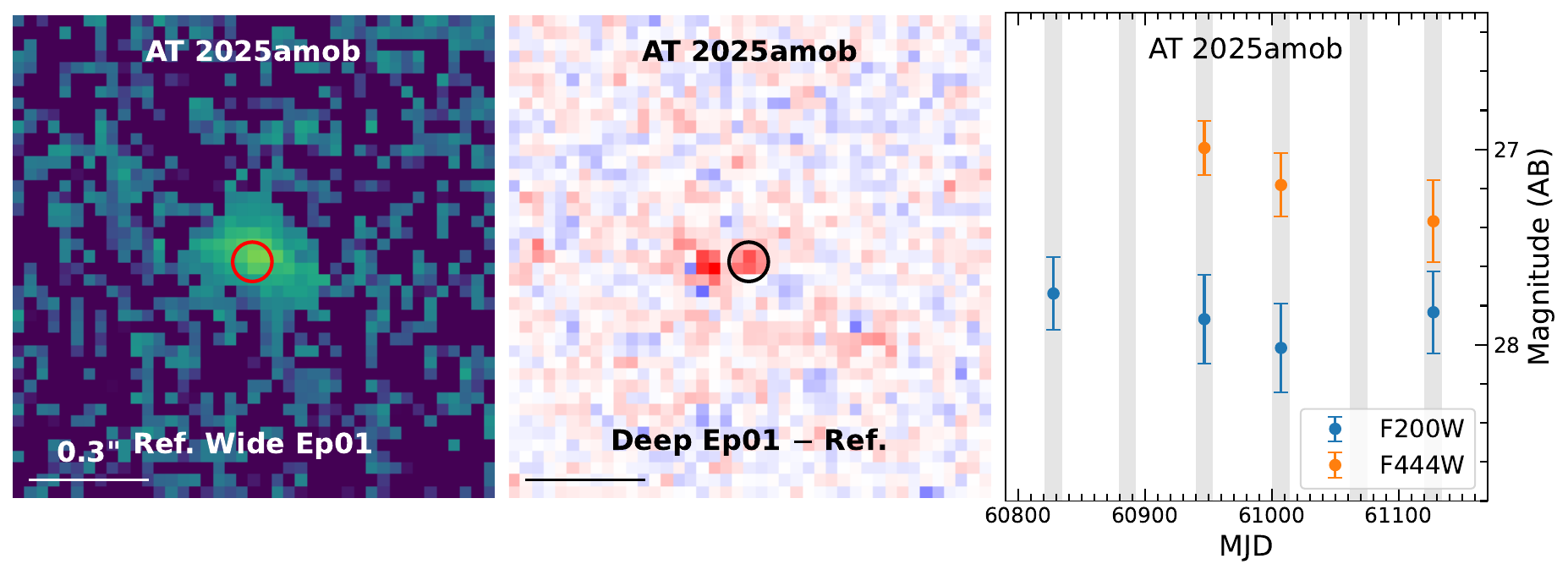}
    \caption{NIRCam F200W cutout stamps and multi-band light curves for four non-SN candidate variables. The first column shows the reference image with no emission from the transient/variable. The second column displays representative difference images, with the target location marked with an open circle. Light curves in the F200W (blue) and F444W (orange) bands are shown in the third column. AT~2025amob is a nuclear variable candidate likely associated with AGN activity, while the other three exhibit significant spatial offsets from their host centers, likely candidates for dwarf AGNs in satellite galaxies, wandering black holes, or recoiling black holes. We note that the red spot located left to AT~2025amob is an artifact.}
    \label{fig:non_SN}
\end{figure}

\section{Discussion}\label{sec:discussion}

\subsection{Supernovae and Other Variable Sources}\label{sec:non-sn}

To isolate a clean supernova sample from other variable sources such as active galactic nuclei (AGNs) and tidal disruption events (TDEs), we evaluate both the light curves and the projected offsets between the transients and their candidate host centers. Four sources (AT~2025xcr, AT~2025amnx, AT~2025amoa, AT~2025amob) exhibit nearly flat light curves spanning $\sim180\text{--}300$ days in the observed frame (Figure~\ref{fig:non_SN}), {lacking the characteristic power-law or exponential decay found in SN light curves}. 

AT~2025amob ($z_{\rm phot}=0.35^{+2.21}_{-0.21}$) is located directly at the center of its host galaxy, while the remaining three sources show substantial spatial offsets from their host centers. Specifically, AT~2025amnx ($z_{\rm spec}=1.485$) is offset by $\sim 0\farcs56 \approx4.8$ kpc (projected), AT~2025amoa ($z_{\rm spec}=2.415$) is offset by $\sim 0\farcs42 \approx 3.4$ kpc (projected), and AT~2025xcr ($z_{\rm phot}=0.57^{+0.03}_{-0.04}$) lies $\sim 2\farcs1 \approx 13.7$ kpc (projected) from its host center (outside the cutout field-of-view in Figure~\ref{fig:non_SN}), appearing adjacent to a satellite of its host galaxy. 

{Although their exact nature remains unknown based solely on light curves, a TDE origin is unlikely given their non-exponential decaying light curves and red colors ($\mathrm{F200W}-\mathrm{F444W}\gtrsim 0$) in most epochs. The nuclear location of AT~2025amob suggests an AGN variability origin, despite its unusually red color. On the other hand, the other three off-nuclear sources are candidates for AGNs residing in dwarf satellite galaxies, or potentially wandering or recoiling black holes given their complicated ambient environment. Among these four sources, PRISM spectroscopy was obtained only for AT~2025amoa in Deep Ep06 (Figure~\ref{fig:Transient_host_spec}). Its spectrum displays prominent blue wings in \Ha, providing evidence for fast, AGN-driven outflows. Unfortunately, AT~2025xcr and AT~2025amnx faded below our NIRCam detection threshold by Deep Ep06, leaving their nature illusive. Spectroscopic follow-up of AT~2025amoa and AT~2025amob from the second year observations in NEXUS is currently underway to constrain their physical origin.}

\begin{deluxetable*}{ccccccccccc}
\tablecaption{Summary of the three JWST-SN programs\label{table:survey_comparison}}
\tabletypesize{\footnotesize}
\tablehead{
\colhead{Survey} & \colhead{Area} & \colhead{Epochs} & \colhead{Filters} & \colhead{Tot Exp. Time} & \colhead{$5\sigma$ Depth} & \multicolumn{3}{c}{SN Counts} & \colhead{Raw SN Rates} & \colhead{Survey Efficiency} \\
\nocolhead{} & \colhead{($\mathrm{arcmin}^2$)} & \nocolhead{} & \nocolhead{} & \colhead{(hr)} & \colhead{(mag)} & \colhead{($z \le 1$)} & \colhead{($1 < z \le 3$)} & \colhead{($z > 3$)} & \colhead{($\mathrm{arcmin}^{-2}\,\mathrm{yr}^{-1}$)} & \colhead{($\mathrm{SN}\, \mathrm{hr}^{-1}$)}  \\
\colhead{(1)} & \colhead{(2)} & \colhead{(3)} & \colhead{(4)} & \colhead{(5)} & \colhead{(6)} & \colhead{(7)} & \colhead{(8)} & \colhead{(9)} & \colhead{(10)} & \colhead{(11)}
}
\startdata
JADES & 25 & 2\tablenotemark{a} & F090W F115W F150W & 155.8 & 30.1 & 11 (11) & 50 (29) & 15 (7) & 1.58\tablenotemark{b} & 0.25\tablenotemark{b}\\
 & & & F200W F277W F335M & & [F277W] & & & & & \\
 & & & F356W F410M F444W & & & & & & &\\
\hline
COSMOS & 133 & 1 & F115W F150W F277W & 26.6 & 28.0--28.3 & 28 (0) & 34 (5) & 6 (2) & 0.51 & 2.56 \\
 & & & F444W & &[F277W] & & & & &\\
\hline
NEXUS & 61.5 & 6 & F200W F444W & 9.7\tablenotemark{c} & 27.2--27.6 & {24 (9)} & {34 (23)} & {6 (3)} & 1.04 & 6.60\\
 & & & & &[F200W] & & & & & \\
\enddata
\tablecomments{
Columns: (1) Survey name; (2) Survey area; (3) Number of science epochs; (4) NIRCam filters used; (5) Total imaging exposure time for the sum of reference epoch and science epoch(s) over all filters and pointings (overhead neglected); (6) $5\sigma$ limiting magnitude in F277W (JADES and COSMOS) or F200W (NEXUS) per epoch  estimated from randomly placed apertures with $r=0\farcs15$ without aperture correction; (7--9) Supernova candidate counts within a 1\,yr baseline across redshift bins $z \le 1$, $1 < z \le 3$, and $z > 3$, with the spectroscopic subset ($z_{\mathrm{spec}}$) shown in parentheses; \textbf{The 4 transients with non-SN origins (Section~\ref{sec:non-sn}) are excluded from the NEXUS-SN sample.} (10) Raw SN candidate rates per unit area per year (uncorrected for detection completeness). Survey properties are adopted from \citet{JADES} and \citet{JADES_transients} for JADES, and \citet{COSMOS-Web} and \citet{COSMOS-Web_transients} for COSMOS.}
\tablenotetext{a}{Although the JADES transient survey contains only one reference epoch and one science epoch, the reported counts include both fading and brightening sources by using each epoch interchangeably as a reference. In contrast, the COSMOS and NEXUS transient surveys include brightening sources only.}
\tablenotetext{b}{The JADES candidate number density is halved to maintain consistency with the single-directional (brightening-only) baseline of the other surveys. Three objects with undetermined redshifts are also included.}
\tablenotetext{c}{NEXUS Wide epochs use NIRCam WFSS as primary mode. F200W imaging is paired with F444W WFSS while F444W imaging is obtained from direct image and out-of-field image for WFSS. We adopt the exposure time of F200W (3.46 hr versus 2.59 hr for F444W) as the exposure time for reference epoch, pretending it was taken using NIRCam imaging mode as primary mode and paired with F444W.}
\end{deluxetable*}

\begin{figure}[t]
    \centering
    \includegraphics[width=\linewidth]{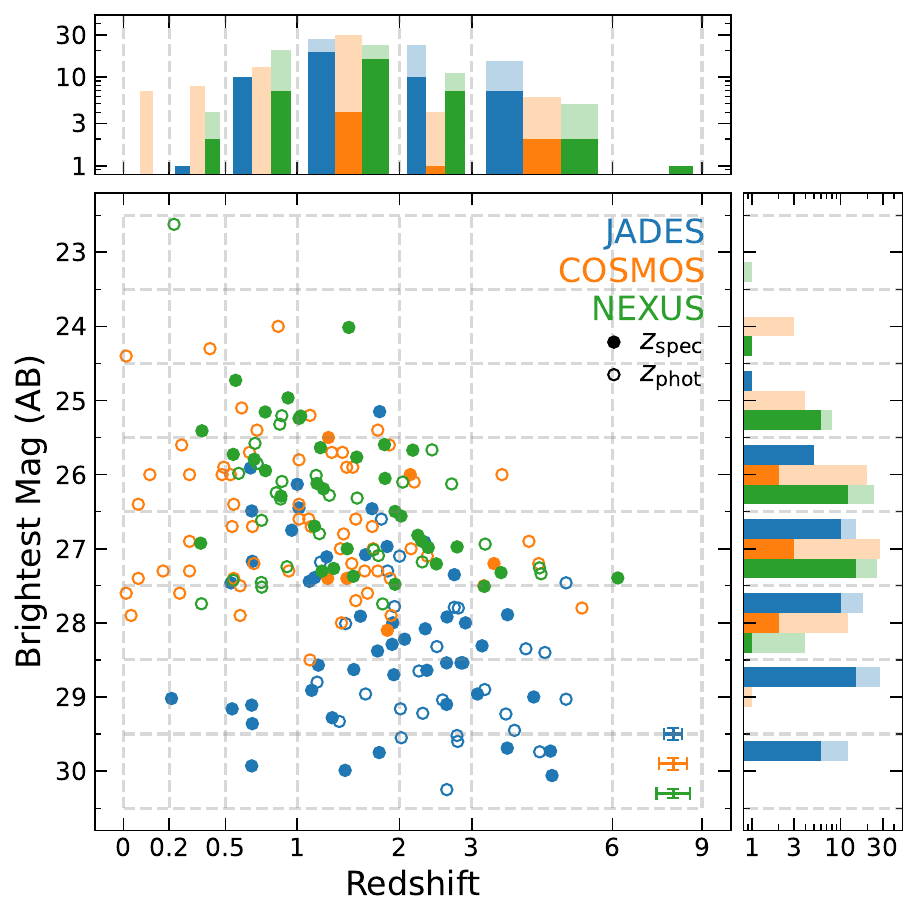}
    \caption{Brightest (on detection) magnitudes versus redshifts for SN candidates discovered in the JADES transient survey (blue), the COSMOS transient survey (orange), and the NEXUS transient survey (green). Solid and open circles indicate objects with $z_{\rm spec}$ and $z_{\rm phot}$, respectively. Top and right side panels show the histograms of redshifts and brightest magnitudes, respectively. Histograms for SN candidates with $z_{\rm spec}$ are shown in solid bars, while those with only $z_{\rm phot}$ are shown in translucent bars. Gray dashed lines indicate the bin edges for the histograms. Typical uncertainties ($\sigma_{z}/(1+z_{\rm phot})$ for $z_{\rm phot}$) are shown in the lower-right corner.}
    \label{fig:SN_mag_z_hist}
\end{figure}

\begin{figure}[t]
    \centering
    \includegraphics[width=\linewidth]{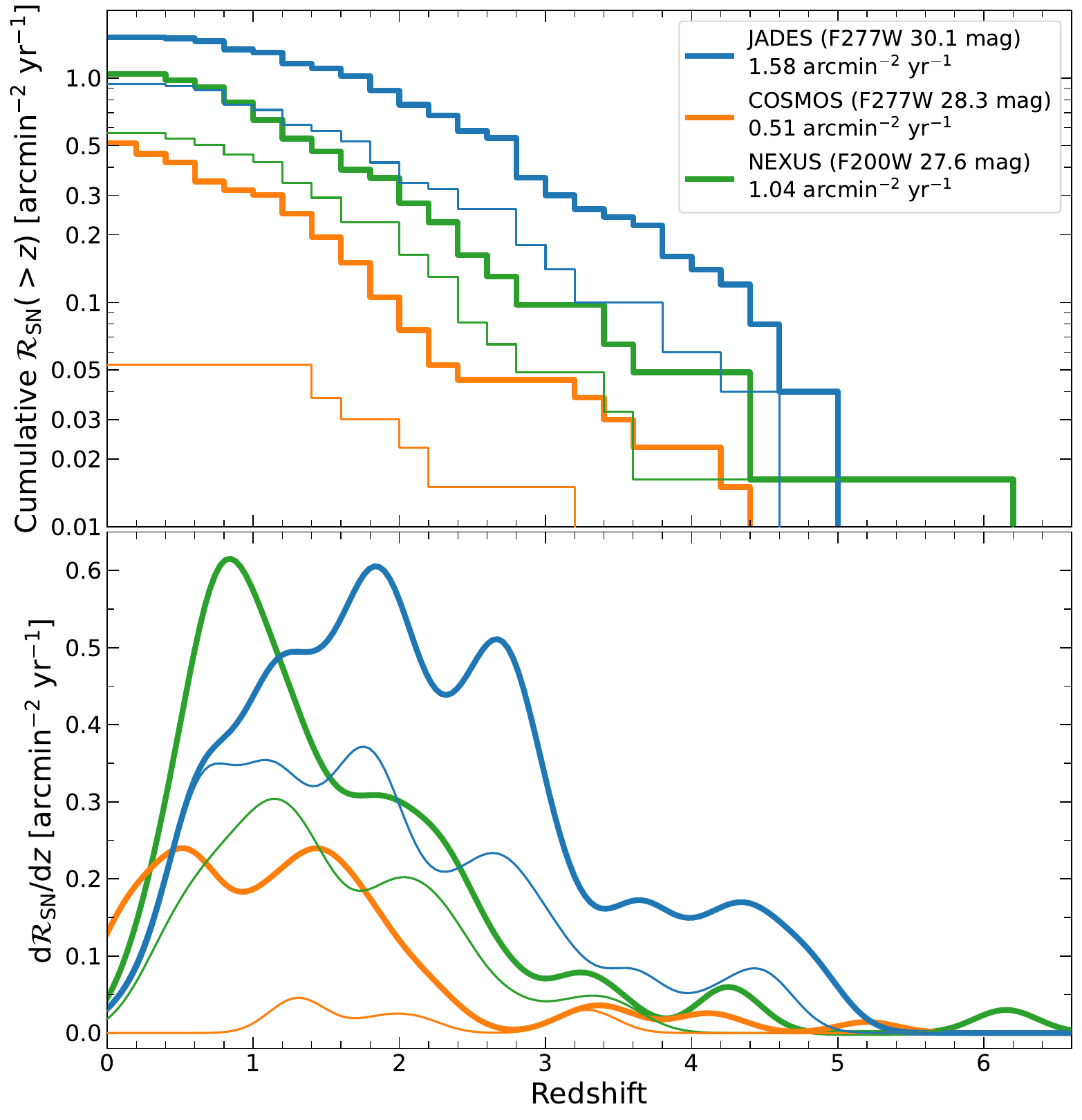}
    \caption{{Raw SN detection rate per unit area per year ($\mathcal{R}_{\rm SN}$; without completeness corrections) as a function of redshift, shown in cumulative form ($\mathcal{R}_{\rm SN} (>z)$; top panel) and differential form per unit redshift(${\rm d}\mathcal{R}_{\rm SN}/{\rm d}z$; bottom panel) for the JADES (blue), COSMOS (orange), and NEXUS (green) transient surveys. In both panels, thick curves represent the full detection sample, while thin curves indicate the subsample with $z_{\rm spec}$. The $5\sigma$ imaging depths and the total integrated detection rates for each survey are shown in the top-left corner. The differential rates are smoothed using a Gaussian kernel density estimator. The greater imaging depth of the JADES survey enables the detection of substantially fainter SNe at higher redshifts, boosting its observed detection rates relative to wider, shallower surveys.}}
    \label{fig:SN_rate}
\end{figure}

\subsection{A wealth of Supernovae at Cosmic Noon and Beyond}

As one of the first JWST programs featuring a dedicated time-domain component, NEXUS has identified 68 robust transients (including 64 SN candidates) within its first year, achieved with short integration times of only $\sim5-10$ minutes exposure per pointing per epoch. The brightest magnitudes of the sample span a broad range of $\sim5.1$~mag from 22.6 to 27.7~mag in either F200W or F444W. Thanks to the rapid built-in NIRSpec/MOS PRISM spectroscopic follow-up, we have successfully obtained $z_{\rm spec}$ for 35 SN candidates. Combined with $z_{\rm phot}$ using host galaxy optical-to-NIR spectral energy distribution (SED), our sample spans $0.22\lesssim z \lesssim 6.15$. Notably, objects at cosmic noon ($1<z<3$) and beyond ($z>3$) constitute $62.5\%$ of the total sample, providing a unique dataset to investigate massive star populations and their environment during the peak of cosmic star formation and into the epoch of reionization. 

\subsubsection{Observational Strategies and Survey Efficiency Across JWST Transient Surveys}

We compare our results with those from the JADES \citep{JADES_transients} and the COSMOS \citep{COSMOS-Web_transients} transient surveys, for a better context of the SN candidates identified in the JWST era (summarized in Table~\ref{table:survey_comparison}). These three programs implement distinct observational strategies driven by primary galaxy science goals, field visibility, and allocation constraints:
\begin{itemize}
    \item \textbf{JADES:} It targets a small area ($25\, \mathrm{arcmin}^2$) with extreme imaging depth ($\rm F277W \approx 30.1$ mag), nine-filter coverage, and low cadence. While JADES only conducted one reference epoch and one science epoch\footnote{The JADES transient survey has several subsequent epochs within 3 months of the science epoch. These later epochs are much shallower and have only partial coverage.} separately by approximately one year, the reported sample includes both fading and brightening sources with each other as reference, effectively doubling the detection, yielding 79 SN candidates up to $z\lesssim4.35$.

    \item \textbf{COSMOS:} It covers a wide area ($133\, \mathrm{arcmin}^2$) with four-filter coverage at moderate depth ($\rm F277W = 28.0\text{--}28.3$ mag) and low cadence. COSMOS identified 68 brightening SN candidates out to $z_{\rm phot}\approx5$ over a 1-year base line.

    \item \textbf{NEXUS:} It targets a medium survey area ($61.5\, \mathrm{arcmin}^2$) with high cadence ($\sim2$ month), shallower imaging depth ($\rm F200W = 27.2\text{--}27.6$ mag), and a two-filter coverage. The dense cadence significantly boosts detection rates by capturing transients near their peak brightness.
\end{itemize}

While JADES yields the largest total number of SN candidates, it requires a substantial investment of telescope time. To quantitatively evaluate the relative transient survey efficiency across different observational strategies, we define a simple metric of \textit{survey efficiency}, which is the number of SN candidates discovered per hour of JWST imaging (overhead ignored). For a fair comparison, the total imaging time includes all constituent science and reference epochs across filters and search area shared between epochs. Based on the APT files of each survey, the total science-only imaging time (reference plus science epochs) amounts to 155.8~hr for JADES (PID 1180, across four pointings) and 26.6~hr for COSMOS (scaled from PRIMER PID 1837 and COSMOS-Web PID 1727 to match the effective search area). In comparison, the NEXUS program (PID 5105, counting one reference epoch and six science epochs) requires a total science exposure time of only 9.7~hr. 

With 64 brightening detections, NEXUS achieves a survey efficiency of $\sim6.60$ SN candidates per hour. This significantly surpasses the efficiency of other JWST transient programs, including COSMOS (2.56 candidates~hr$^{-1}$) and JADES (0.25 candidates~hr$^{-1}$; averaged over brightening and fading sources). We emphasize that this simple survey efficiency metric does not capture the full capacity of a specific transient program. For example, the deep nine-filter coverage in JADES provides significantly better SN classifications than other programs. On the other hand, the dense cadence of NEXUS provides decent light curves for bright transients to trace their evolution. 

\subsubsection{Sample Distributions, Spectroscopic Completeness, and Raw SN Detection Rates}

Figure~\ref{fig:SN_mag_z_hist} shows the brightest observed magnitude and redshift distribution across the three transient surveys. The brightest magnitude here refers to that measured during the detection epochs. Faint SN candidates ($>28$ mag) are predominately from the JADES sample (50/52) due to its extreme imaging depth. On the other hand, low-redshift SN candidates ($z<0.5$) are dominated by the COSMOS sample (15/20) due to its wide sky coverage and short wavelength coverage (F115W+F150W). In all three surveys, the bulk portion of discovered SN candidates reside at cosmic noon ($1<z<3$). We note that there is a large $\sim 5$~mag spread in brightest observed magnitudes at a given redshift in Figure~\ref{fig:SN_mag_z_hist}. This broad dispersion is primarily driven by two key factors: (1) the large intrinsic luminosity diversity of CCSNe and their various subtypes (spanning a peak absolute magnitude range of $\Delta M_B \approx 3\text{--}5$~mag; e.g., \citealt{2014AJ....147..118R}), which dominate the transient population at $z \gtrsim 1$ due to the rising cosmic star formation rate; and (2) observational cadence effects, where programs with long inter-epoch baselines (such as JADES) frequently capture transients well past peak brightness during late-time plateau or exponential decay phases.

As a unique feature of NEXUS, the dedicated NIRSpec/MOS spectroscopic follow-up yields a high $z_{\rm spec}$ fraction: $\sim55\%$ (35/64) for the full sample and $\sim68\%$ (23/34) at cosmic noon. This cosmic-noon spectroscopic fraction is more than four times higher in the $z_{\rm spec}$ fraction than that of COSMOS ($\sim15\%$) and even $\sim10\%$ higher than JADES ($\sim59\%$).

{Figure~\ref{fig:SN_rate} compares the raw SN detection rate per unit area per year ($\mathcal{R}_{\rm SN}$; without completeness corrections) as a function of redshift in both cumulative ($\mathcal{R}_{\rm SN} (>z)$) and differential (${\rm d}\mathcal{R}_{\rm SN}/{\rm d}z$) forms across the three surveys samples. Overall, all three surveys show broadly similar cumulative redshift distribution for the full samples and spectroscopic subsamples. Owing to its significantly deeper imaging sensitivity, the JADES transient survey results in the highest detection rate of $\mathcal{R}_{\rm SN} \approx 1.58\, {\rm arcmin^{-2}\, yr^{-1}}$. In contrast, although the COSMOS transient survey is $\sim0.7$~mag deeper than NEXUS, its sparse cadence yields the lowest observed rate ($\mathcal{R}_{\rm SN} \approx 0.51\, {\rm arcmin^{-2}\, yr^{-1}}$) due to missing SNe in the inter-epoch gaps. Examining the differential rate distribution, while $\mathcal{R}_{\rm SN}$ drops substantially beyond $z\gtrsim2$ for all datasets, the peak detection efficiency occurs at different redshift ranges across the three surveys. $\mathcal{R}_{\rm SN}$ for the two shallower surveys (COSMOS and NEXUS) show peak at lower redshifts ($z\approx1$), whereas JADES exhibit a broad distribution at cosmic noon ($1<z<3$). These differences are primarily driven by survey area, cadence, and depth, as well as potential cosmic variance across different fields. We defer a more careful measurement of volumetric SN rates ($\rm Mpc^{-3}\, yr^{-1}$) to future work, after thoroughly correcting for detection incompleteness, and better classifications and redshift constraints from upcoming NEXUS data.}

\subsubsection{Challenges in Photometric Classification}

{Although NEXUS is by far the most efficient survey in discovering SN candidates, its two-filter strategy poses challenges for photometric classification, in particular in the early phase. During Year 1, only 23 sources (16 with $z_{\rm spec}$ and 7 with only $z_{\rm phot}$) accumulated sufficient photometric points ($\ge 6$) to enable light curve fitting using \texttt{SNCosmo} \citep{sncosmo}. Fits require at lease one degree-of-freedom (i.e., number of data points minus number of free parameters: five parameters for SN~Ia model \texttt{salt3-nir} and four parameters for CCSN models including $z$).}

{Because NIRSpec/MOS target scheduling occurs immediately after each NIRCam imaging visit, the sparsely sampled SED (F200W+F444W) available at initial detection is insufficient to break redshift-model degeneracies, complicating early classification and target prioritization. To optimize follow-up efficiency in future program iterations, incorporating additional filter pairs (e.g., two filter combos per epoch) or adopting a higher cadence ($\sim 1$~month) will be critical to constrain early transient SEDs and rise times.}

\begin{figure}[t]
    \centering
    \includegraphics[width=\linewidth]{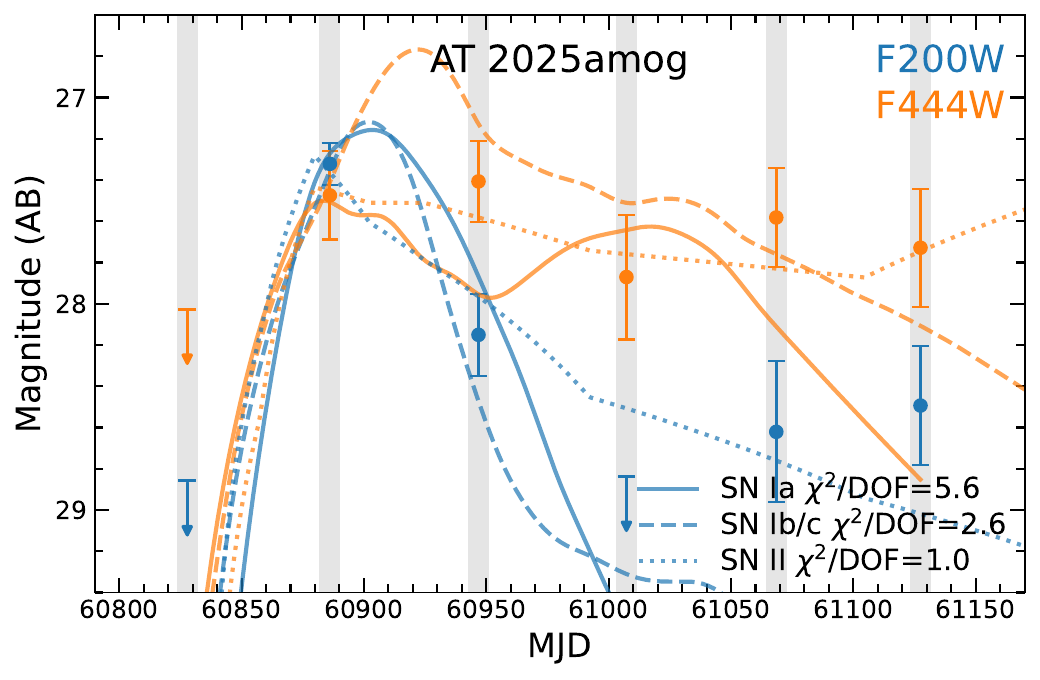}
    \caption{Comparison of best-fit SN~Ia (solid), SN~Ib/c (dashed), and SN~II (dotted) models to the light curve of AT~2025amog. Blue and orange dots indicate data in F200W and F444W bands. Goodness of fitting is described as $\chi^2$ divided by degree-of-freedom (DOF), where DOF is the number of data points minus number of free parameters. }
    \label{fig:2025amog_lc_fit}
\end{figure}

\begin{figure*}[t]
    \centering
    \includegraphics[width=0.8\linewidth]{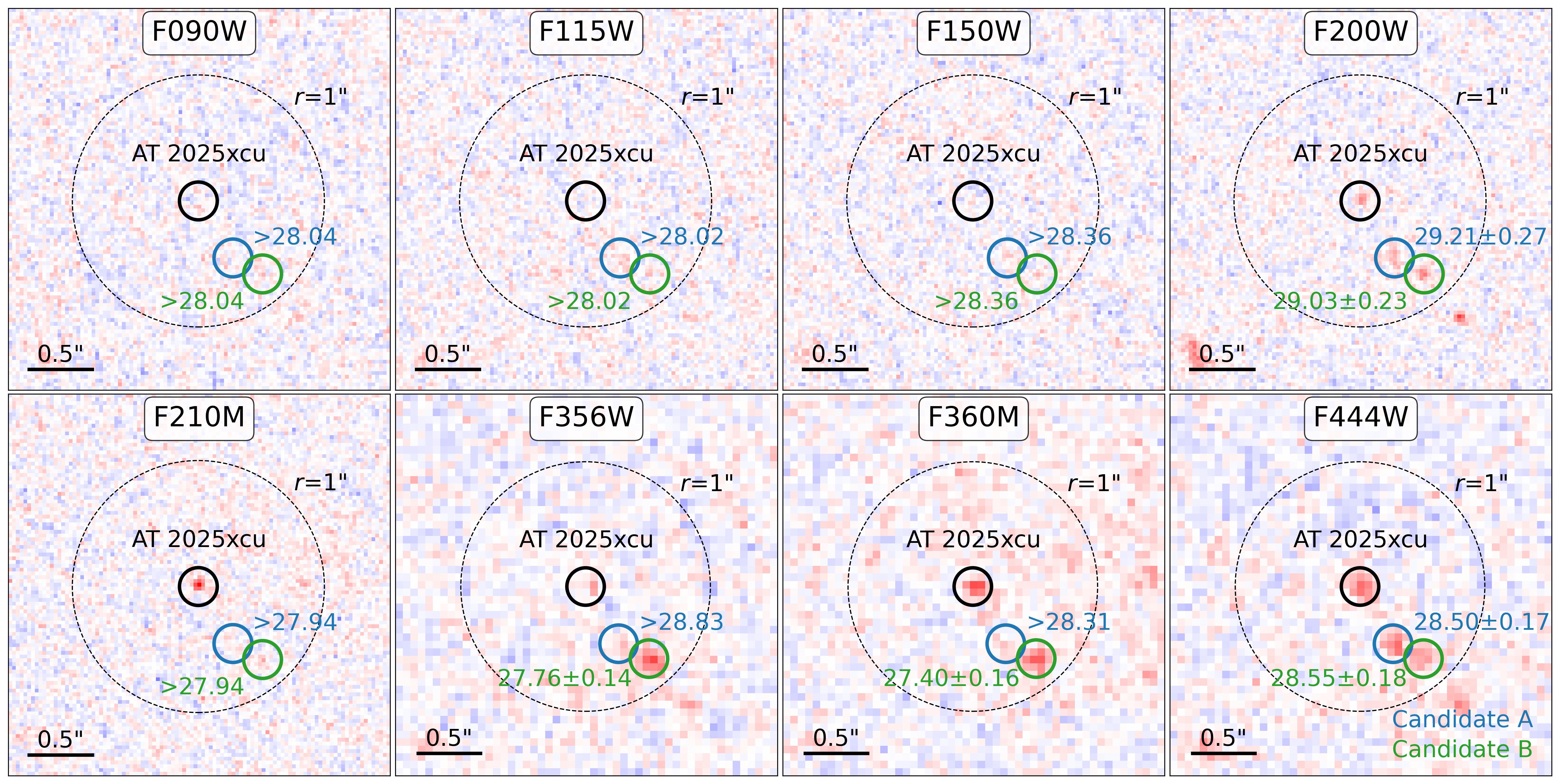}
    \includegraphics[width=0.4\linewidth]{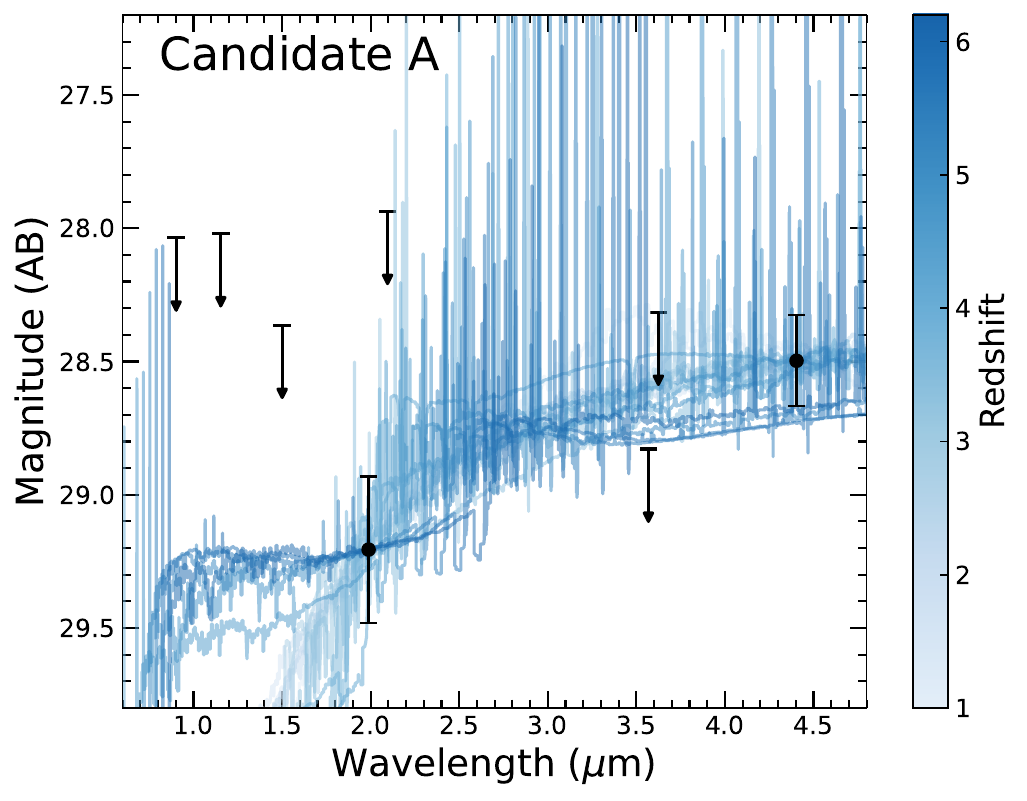}
    \includegraphics[width=0.4\linewidth]{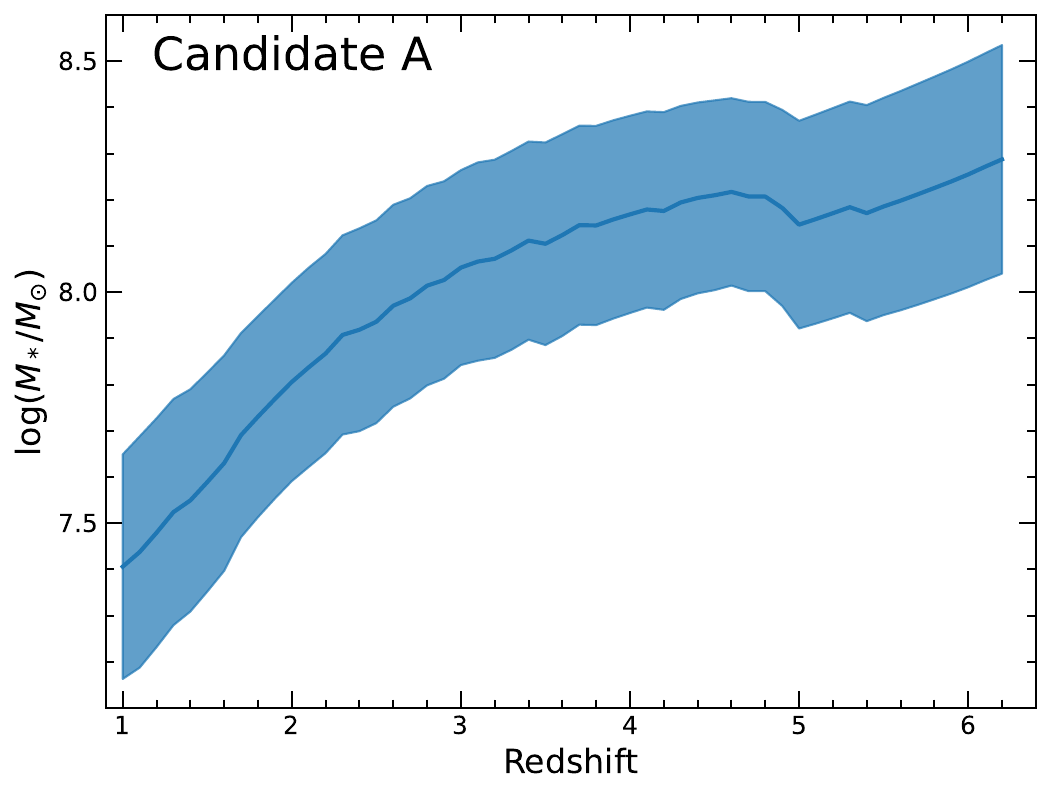}
    \caption{\textbf{Top:} Multi-band NIRCam cutouts of AT~2025xcu (black circle), generated by coadding all available NIRCam observations in the NEXUS field. For potential host association, Candidate~A (blue circle) is closer to the transient (0\farcs55) but faint ($\rm F200W=29.2$ mag; $\rm F444W=28.5$ mag) and detected only in F200W and F444W. Candidate~B at $z_{\rm spec}=6.151$ (green circle) is brighter ($<28$~mag in F356W and F360M) and slightly farther away (0\farcs76). We adopt it as our fiducial host galaxy candidate. Magnitudes for both host candidates are provided in each panel, with 3$\sigma$ upper limits quoted for non-detections. \textbf{Bottom:} SED (black dots) and best-fit galaxy templates at each redshift grid (blue curves) of Candidate~A (left) and stellar mass versus redshift from best-fit templates (right). Error bars indicate 1$\sigma$ uncertainty while downward arrows indicate $3\sigma$ upper limits. Shaded region in the lower right panel represents the $1\sigma$ uncertainty of stellar mass.}
    \label{fig:2025xcu_hosts}
\end{figure*}

\begin{figure*}[t]
    \centering
    \includegraphics[width=0.8\linewidth]{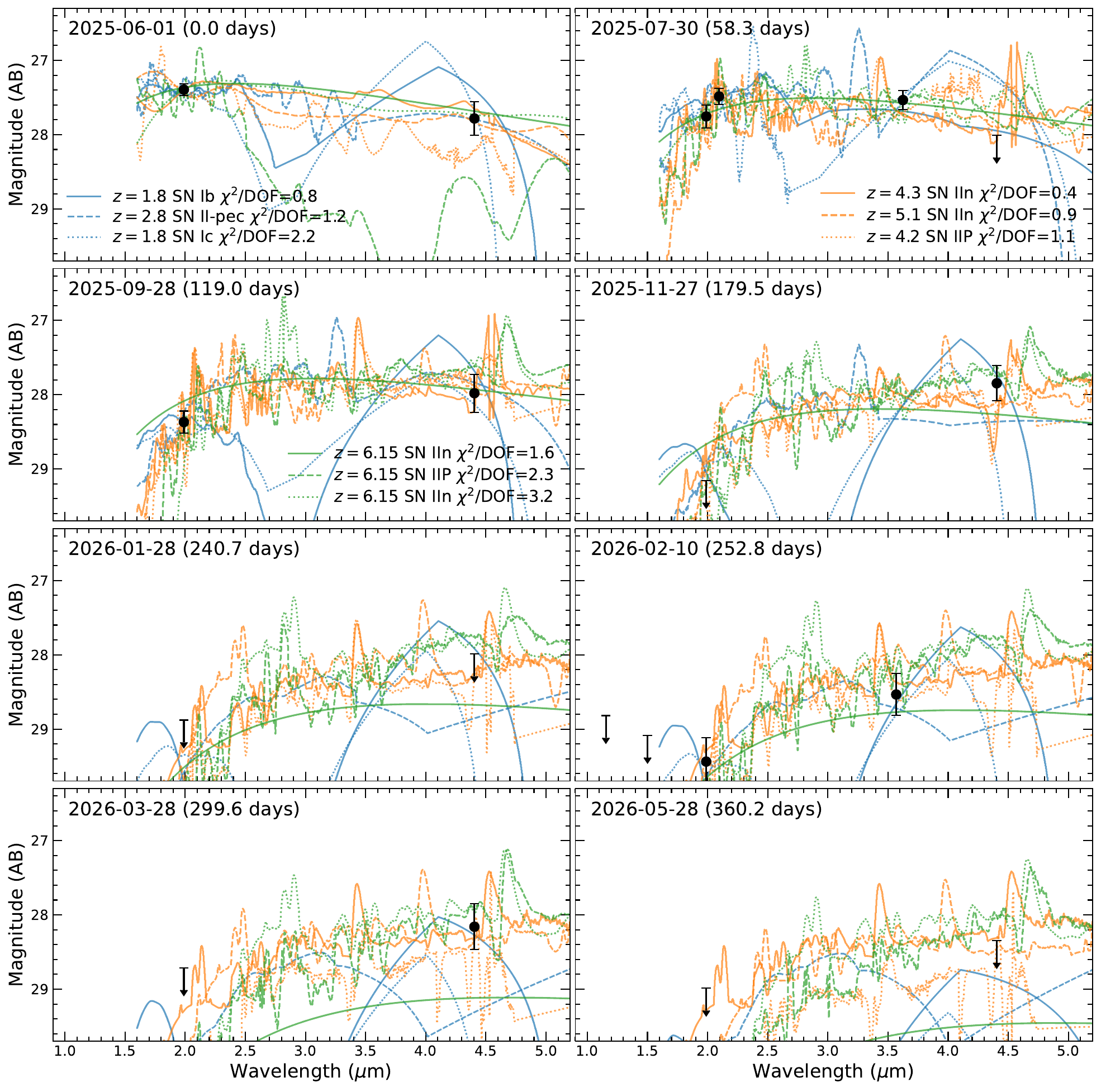}
    \caption{Multi-band photometry of AT~2025xcu (black dots) across eight epochs spanning nearly 1 year, with solid, dashed, and dotted curves representing top three SN templates for the low-$z$ (blue), high-$z$ (orange), and fixed $z=6.151$ (blue) scenarios, respectively. The type of SN templates and goodness of the fits are shown in legend.}
    \label{fig:2025xcu_model_comp}
\end{figure*}

\subsection{Supernova Candidates at $z>3$}\label{sec:highz-SN}

Our high-confidence sample contains two SN candidate with $z_{\rm spec}>3$ (AT~2025amog and AT~2025amot) and one SN candidate at $z_{\rm spec}=6.151$ (AT~2025xcu). We discuss their properties in this subsection.

\subsubsection{A $z_{\rm spec}=3.493$ SN~IIP Candidate AT~2025amog}

AT~2025amog was discovered in Deep Ep02 with a $\sim10\sigma$ detection in F200W and $5\sigma$ detection in F444W. Its F200W flux decreased by $\sim0.8$~mag in Deep Ep03 and remains marginally detected (2--3$\sigma$) in Deep Ep04--Ep06. In contract, its F444W flux stayed relatively stable ever since the discovery epoch with a decrease of $\lesssim0.5$~mag from Deep Ep02 to Ep06. It lies in close proximity to the center of its underlying host galaxy. Follow-up PRISM spectrum was acquired in Deep Ep03, capturing both the host and the transient while AT~2025amog was still active. As shown in Figure~\ref{fig:Transient_host_spec}, the spectrum is host dominated, displaying strong emission lines including \OII$\lambda\lambda3727,3729$, \OIII$\lambda5007$, and \Ha\ pinpointing its $z_{\rm spec}=3.493$. 

The brighter F444W emission over F200W in later epochs and the long-lasting F444W emission suggests an SN~II origin. Figure~\ref{fig:2025amog_lc_fit} shows a comparison of best-fit SN~Ia, SN~Ib/c, and SN~II models using \texttt{SNCosmo} \citep{sncosmo}. Its light curve is best described by an SN~IIP model from \texttt{nugent-sn2p}, with the top five SN models all corresponding to SNe~IIP. The best-fit model yields a peak absolute $B$-band magnitude of $M_B=-18.5$ mag, which exceeds the average brightness of local SN~IIP but remains consistent with the most luminous local analogs \citep{2014AJ....147..118R}. 

AT~2025amog resides in a low-mass galaxy ($M_*\approx 10^{8.7}\, M_{\odot}$) with an exceptionally high \OIII/\OII\ ratio of $\sim10$, suggesting an extremely metal-poor environment with a gas-phase metallicity of $Z\approx0.1\text{--}0.25\,Z_{\odot}$ ($\rm 12+\log(O/H)=7.7\text{--}8.1$) as estimated from $\rm O_3 O_2$ metallicity calibration from \citet{O32_metallicity}. AT~2025amog closely resembles AT~2023adsv, a SN~II candidate at $z_{\rm spec}=3.613$ discovered in the JADES field \citep{2026ApJ..1002...83C}. The light curve of AT~2023adsv is best fitted by an SN~IIP model with peak $M_B=-18.3$~mag. Its host galaxy exhibits similar properties ($M_*\approx10^{8.4}\, M_{\odot}$ and gas-phase metallicity $Z\approx0.3\, Z_{\odot}$). As discussed in \citet{2026ApJ..1002...83C}, the discovery of multiple SN~II candidates in low-mass, metal-poor galaxies may highlight a systematic evolution of SN host galaxy population towards early Universe.

\subsubsection{A $z_{\rm spec}=3.204$ CCSN Candidate AT~2025amot}

AT~2025amot was discovered in Deep Ep03 with a clear F200W detection ($\sim10\sigma$) and a marginal F444W detection ($\sim2.6\sigma$). It remained active $\sim2$ months later in Deep Ep04 with a $\sim4\sigma$ F444W detection. It is located offset northeast to its host galaxy center by $\sim$0\farcs06. Our follow-up PRISM spectrum was acquired in Deep Ep05 after the transient had faded below detection limits. The spectrum shows weak \OII, \Hb, \OIII, and \Ha\ emission lines, indicating a an immediate-age, low-mass host galaxy at $z_{\rm spec}=3.204$. With only two formal photometric detections, no robust properties of the transient can be obtained. Nonetheless, an estimate based on F200W magnitude in Deep Ep03 results to a peak magnitude of $M_B\lesssim-18.1$ mag, reasonable for CCSNe \citep{2014AJ....147..118R}.

\subsubsection{A $z_{\rm spec} = 6.151$ SN~IIP Candidate AT~2025xcu}\label{sec:6.3.3}

AT~2025xcu has the highest $z_{\rm spec}$ in our sample and would be the most distant SN to date if confirmed. It was detected in five Deep epochs (fading below detection only in Deep Ep05). Its F200W flux decreased systematically from Deep Ep01 to Ep03 and fell below detection limits by Deep Ep04, whereas, it remained detected in F444W through Deep Ep03, Ep04, and Ep06. 

The fiducial host galaxy was initially assigned to the only object located within a 1\arcsec\ radius in reference F200W and F444W mosaics: galaxy 102981 (Candidate~B), located $\sim 0\farcs76$ from the transient location. This galaxy is a low-mass line emitter ($M_*\approx10^{8.3}\, M_{\odot}$) with $z_{\rm spec}=6.151$ secured from the presence of strong emission lines including \Hb, \OIII, and \Ha\ in our follow-up PRISM spectrum (see Figure~\ref{fig:Transient_host_spec}). However, detailed investigation of its ambient environment by stacking all the available NIRCam images reveals a very faint ($\rm F200W=29.2$ mag, $\rm F444W=28.5$ mag) object (Candidate~A) located slightly closer ($\sim$0\farcs55) to AT~2025xcu (Figure~\ref{fig:2025xcu_hosts}). Candidate~A shows an elongated morphology in F200W, potentially undergoing interactions. Since Candidate~A is detected only in F200W and F444W in the deepest coadds, its $z_{\rm phot}$ and physical association with candidate B cannot be robustly determined. However, we may still obtain a reasonable stellar mass range from the two-band detection. We perform SED fitting using \texttt{CIGALE} \citep{cigale} across a grid of fixed redshifts ($1\leq z\leq6$ with a step of 0.1). The best-fit stellar template at each redshift grid and the resulting stellar mass distribution are shown in Figure~\ref{fig:2025xcu_hosts}. We find that Candidate~A is a low-mass, dwarf galaxy, with its stellar mass constrained to $10^{7.3\text{--}8.2} \, M_{\odot}$ regardless of redshift. 

PRISM spectroscopy of AT~2025xcu acquired in Deep Ep02 yielded a low SNR spectrum (median $\rm SNR\lesssim1$. We fitted the light curve of AT~2025xcu with a library of SN templates in \texttt{SNCosmo} \citep{sncosmo} using all the available data across three redshift scenarios: 
\begin{itemize}
    \item \textbf{Low-$z$ ($0 < z < 3$):} Favors SN~Ib/c templates at $z \approx 1.8$ (\texttt{snana-2004gv}; reduced $\chi^2 / \rm DOF < 1$). A peculiar SN~II template (\texttt{snana-2007ms}) at $z \approx 2.8$ also provides an acceptable fit ($\chi^2 / \rm DOF  \approx 1.2$).
    
    \item \textbf{High-$z$ ($3 < z < 8$):} Favors SN~IIn models at $4 < z < 5$ ($\chi^2 / \rm DOF < 1$). SN~Ib/c templates at these redshifts are disfavored ($\chi^2 / \rm DOF > 2$).
    
    \item \textbf{Fixed $z = 6.151$ (Candidate B host assumption):} Favors SN~II models (best and third-best fits are SN~IIn; second-best is SN~IIP). Stripped-envelope SN~Ib/c models are strongly disfavored ($\chi^2 / \rm DOF \gtrsim 20$).
\end{itemize}

Figure~\ref{fig:2025xcu_model_comp} compares multi-epoch photometry of AT~2025xcu across eight epochs spanning nearly 1 year against the top three SN templates for each redshift scenario. SN~II templates generally provide good matches to the SED evolution of AT~2025xcu, whereas the low-$z$ SN~Ib fits appear driven by wavelength and temporal extrapolation artifacts within the template. Notably, the detection of F200W and F210M (from parallel NIRCam imaging paired with NIRSpec/MOS observations) on 2025-07-30 (Deep Ep02) likely aligns with the \CaII\ H+K P-Cygni feature, a classic hallmark of SNe~II. Furthermore, stripped-envelope SNe (SNe~Ib/c) are preferentially found in massive, metal-rich galaxies, and the ratio of stripped-envelope to Type~II SNe decreases dramatically towards dwarf, low-metallicity galaxies \citep[e.g.,][]{2010ApJ...721..777A, 2011MNRAS.412.1473L, 2017ApJ...837..121G}. Given the low stellar mass estimated for both host Candidates A and B, an SN~II origin is physically favored. At this point, existing NEXUS data are inconclusive as to whether Candidate~A, Candidate~B, an unseen host galaxy is the true host of AT~2025xcu, or whether it represents a genuinely ``hostless'' transient. Dedicated follow-up deep, multi-band NIRCam imaging will be critical to constraining the properties of Candidate~A and revealing the nature of this $z>6$ SN~II candidate.

\section{Concluding Remarks}\label{sec:con}

In this paper we present an overview of ongoing efforts of transient searches and follow up within the multi-cycle JWST-NEXUS treasury program \citep{NEXUS}. We also report first results from Year-1 observations for a sample of 68 {robust transients (64 SN candidates)}, with detailed studies on individual transient systems to be presented in future work. 

Overall, the NEXUS transient program achieves a high detection efficiency, 6.6 detection per imaging hrs of JWST. This is a factor of 2.6 higher than the COSMOS transient survey, and a factor of 26 higher than the JADES transient survey. {NEXUS is around 0.7~mag shallower than COSMOS in single-epoch depths (27.6 versus 28.3 mag)}, but the much higher cadence of NEXUS (2 months versus 1~year) made it possible to capture substantially more SNe per area {1.04 versus 0.51 ${\rm arcmin^{-2}\, yr^{-1}}$}. The JADE transient survey has substantially deeper ($\sim 30.1$~mag) single-epoch imaging, which allowed the capture of faint SN detection during the fading phase, even with a sparse $\sim$yearly cadence similar to the COSMOS transient survey. However, the detection efficiency per hr of JWST imaging is also the lowest for the JADES transient survey, as such depths require significantly more exposure time. We note that both JADES and COSMOS transient surveys are not dedicated transient programs, as such their program designs are not optimized for transient detection efficiency. 

On the other hand, the results from NEXUS demonstrated that the program is significantly more optimized for transient detection with JWST. The 2-month cadence greatly facilitates the capture of transients during their bright phases, as well as providing decent light curves for bright and gradually evolving transients. This advantage largely hinges on the convenient location of the NEXUS field within the JWST continuous viewing zone (CVZ), as other fields (JADES and COSMOS) have much restricted annual JWST visibility. Being in the JWST CVZ also enables routine spectroscopic follow-up of transients detected therein. The NEXUS program has a built-in NIRSpec MSA component to follow up detected transients with the same 2-month cadence. So far, NEXUS is able to provide spectroscopic redshifts for {$\sim 54\%$} of the photometrically detected transients {($\sim69\%$ at cosmic noon $1<z<3$)}. 

Our initial results on the raw SN rates (uncorrected for survey completeness) are roughly consistent with the two other JWST-SN programs, with a similar decline in rate towards the $z>2$ Universe. The difference in the absolute value of observed rates among these three SN programs is understood by the differences in detection completeness. While NEXUS is more complete than the COSMOS transient survey, the 2-month cadence is still missing a significant number of transients that rapidly fade within 2 observed-frame months. Recognizing that going deeper in epoch photometry is too inefficient, a better solution is to increase the cadence to $\sim$monthly, which will not only recover more transients near peak brightness, but also provide improved light curves for bright transients. 

Another major limitation of NEXUS is that the transient detection imaging is only performed in F200W and F444W. While detecting transients is not an issue, the lack of JWST photometry at other wavelengths severely limits robust identification of $z>3$ candidates. Because of the overwhelming population of $z<3$ SNe that will also be detected in F200W+F444W, the NEXUS program currently does not have an efficient approach to confidently select high-$z$ SNe for deeper JWST spectroscopic follow-up. There are genuine $z>3$ SNe within the full photometric sample for which deep JWST spectroscopy was not acquired. Future transient surveys should deploy at least two NIRCam imaging filter combos (4 wavelengths) to improve the confidence of high-$z$ SN candidate selection.

Nevertheless, the NEXUS program sets a good standard for dedicated JWST-SN programs. We highlight the importance of the north ecliptic pole (NEP) field, which is the ideal location for dedicated SN searches with space facilities and long-range monitoring efforts. As the field is chosen to lie within the Euclid ultra deep field (EUDF), the regular monthly cadenced Euclid photometry in $IYJH$ during 2024--2030 will be valuable to provide additional constraints on NEXUS transients. Furthermore, the NEXUS field is within the Roman eXtreme Deep Field (RXDF), which will provide extremely deep ($\sim 30$ mag) photometry in $RZYJH$. All these photometric data from EUDF, RXDF and NEXUS, in combination with the large number of spectroscopic redshifts from these programs, will provide a comprehensive data set for transient discovery and classification, e.g., with much improved host galaxy association and $z_{\rm phot}$. 

Both the NEXUS and RXDF programs have no proprietary period on data access. The Euclid mission has released first data in the EUDF field \citep{euclid_q1} and is expected to release more data in the coming years. Given the importance of open science and community-driven advances on frontier topics, we highly encourage the community to take advantage of these public data. In addition, the NEXUS team rapidly releases \citep{NEXUS_QDR} the reduced cadenced NIRCam imaging and MSA spectra to aid prompt community investigations and investments.  

\startlongtable
\tabletypesize{\footnotesize}
\begin{deluxetable}{lCCCCcCC}\label{table:SN_host}
\tablecaption{SN Host Information}
\tablehead{
\colhead{TNS ID} & \colhead{Host ID} & \colhead{R.A.} & \colhead{Decl.} & \colhead{$z$} & \colhead{$z$ type} & \colhead{Host Sep.} & \colhead{$d_{\rm Kron}$}
}
\startdata
AT 2025xcm & 98724 & 268.56459 & 65.16535 & 0.70^{+0.03}_{-0.04} & Phot & 0\farcs37 & 0.4 \\
AT 2025xcn & 97025 & 268.43039 & 65.17273 & 0.366 & Spec & 0\farcs49 & 0.3 \\
AT 2025xco & 92994 & 268.46861 & 65.19465 & 1.024 & Spec & 0\farcs95 & 0.8 \\
AT 2025xcp & 93210 & 268.41492 & 65.19166 & 1.189 & Spec & 0\farcs21 & 0.2 \\
AT 2025xcq & 91640 & 268.41784 & 65.20011 & 0.547 & Spec & 1\farcs46 & 1.1 \\
AT 2025xcr & 102010 & 268.43585 & 65.22907 & 0.57^{+0.03}_{-0.04} & Phot & 2\farcs13 & 1.8 \\
AT 2025xcs & 88268 & 268.5173 & 65.21744 & 1.010 & Spec & 0\farcs57 & 0.5 \\
AT 2025xct & 99080 & 268.53253 & 65.16338 & 4.24^{+6.76}_{-3.05} & Phot & 1\farcs58 & 3.1 \\
AT 2025xcu\tablenotemark{a} & 102981 & 268.49541 & 65.22661 & 6.151 & Spec & 0\farcs76 & 1.4 \\
AT 2025xcv & 86532 & 268.35868 & 65.24159 & 0.359 & Spec & 5\farcs92 & 2.6 \\
AT 2025xcw & 94163 & 268.37013 & 65.18705 & 1.946 & Spec & 0\farcs15 & 0.3 \\
AT 2025xcy & 104085 & 268.29117 & 65.24192 & 0.73^{+0.04}_{-0.04} & Phot & 1\farcs63 & 1.0 \\
AT 2025xcz & 31092 & 268.42527 & 65.14893 & 1.308 & Spec & 1\farcs96 & 2.0 \\
AT 2025xda & 96246 & 268.55056 & 65.17682 & 1.831 & Spec & 0\farcs02 & 0.1 \\
AT 2025xdb & 103570 & 268.48706 & 65.2237 & 2.363 & Spec & 0\farcs05 & 0.1 \\
AT 2025xdc & 93508 & 268.46615 & 65.18996 & 0.757 & Spec & 0\farcs08 & 0.1 \\
AT 2025amnx & 91998 & 268.50712 & 65.19855 & 1.485 & Spec & 0\farcs55 & 0.6 \\
AT 2025amny & 91125 & 268.542 & 65.20312 & 1.80^{+0.14}_{-0.35} & Phot & 0\farcs56 & 1.0 \\
AT 2025amnz & 72984 & 268.63543 & 65.21112 & 0.92^{+0.08}_{-0.07} & Phot & 0\farcs39 & 0.7 \\
AT 2025amoa & 105794 & 268.33883 & 65.21346 & 2.415 & Spec & 0\farcs31 & 0.3 \\
AT 2025amob & 150767 & 268.36957 & 65.26635 & 0.35^{+2.41}_{-0.21} & Phot & 0\farcs01 & 0.1 \\
AT 2025amoc & 60109 & 268.31721 & 65.13408 & 0.88^{+0.01}_{-0.07} & Phot & 0\farcs17 & 0.1 \\
AT 2025amod & 31724 & 268.43397 & 65.13588 & 1.53^{+0.34}_{-0.32} & Phot & 0\farcs38 & 0.6 \\
AT 2025amoe & 58739 & 268.34491 & 65.1429 & 0.924 & Spec & 0\farcs05 & 0.2 \\
AT 2025amof & 114748 & 268.2941 & 65.15851 & 2.162 & Spec & 0\farcs11 & 0.2 \\
AT 2025amog & 94237 & 268.51563 & 65.18637 & 3.493 & Spec & 0\farcs01 & 0.1 \\
AT 2025amoh & 90467 & 268.49424 & 65.20627 & 0.871 & Spec & 1\farcs09 & 0.9 \\
AT 2025amoi & 136871 & 268.58171 & 65.25405 & 0.36^{+0.17}_{-0.16} & Phot & 0\farcs18 & 0.3 \\
AT 2025amoj & 152667 & 268.5157 & 65.25644 & 1.159 & Spec & 0\farcs52 & 0.4 \\
AT 2025amok & 152306 & 268.55594 & 65.25832 & 1.215 & Spec & 0\farcs53 & 0.4 \\
AT 2025amol & 151146 & 268.52767 & 65.2643 & 4.27^{+2.33}_{-3.33} & Phot & 0\farcs12 & 0.3 \\
AT 2025amom & 151137 & 268.51117 & 65.26436 & 0.87^{+0.74}_{-0.26} & Phot & 0\farcs11 & 0.2 \\
AT 2025amon & 150175 & 268.55801 & 65.2691 & 1.529 & Spec & 0\farcs29 & 0.4 \\
AT 2025amoo & 32944 & 268.50342 & 65.12731 & 0.58^{+0.06}_{-0.43} & Phot & 0\farcs15 & 0.2 \\
AT 2025amop & 45232 & 268.48277 & 65.13644 & 1.496 & Spec & 0\farcs34 & 0.5 \\
AT 2025amoq & 99119 & 268.50994 & 65.16278 & 0.562 & Spec & 0\farcs37 & 0.5 \\
AT 2025amor & 98065 & 268.56518 & 65.16826 & 1.27^{+0.06}_{-0.07} & Phot & 0\farcs87 & 0.9 \\
AT 2025amos & 80528 & 268.58856 & 65.17162 & 0.682 & Spec & 0\farcs38 & 0.6 \\
AT 2025amot & 94718 & 268.54771 & 65.18416 & 3.204 & Spec & 0\farcs06 & 0.2 \\
AT 2025amou & 109568 & 268.33955 & 65.18886 & 1.825 & Spec & 0\farcs41 & 0.4 \\
AT 2025amov & 107000 & 268.33558 & 65.20495 & 2.772 & Spec & 0\farcs09 & 0.2 \\
AT 2025amow & 151144 & 268.54379 & 65.26427 & 1.451 & Spec & 0\farcs09 & 0.2 \\
AT 2025amox & 148126 & 268.55104 & 65.28009 & 0.22^{+0.05}_{-0.04} & Phot & 0\farcs29 & 0.2 \\
AT 2025amoz & 81855 & 268.62502 & 65.16354 & 0.73^{+0.06}_{-0.04} & Phot & 0\farcs22 & 0.2 \\
AT 2025ampb & 79138 & 268.62813 & 65.17917 & 1.202 & Spec & 0\farcs12 & 0.3 \\
AT 2025ampc & 94614 & 268.44788 & 65.18493 & 2.226 & Spec & 0\farcs19 & 0.2 \\
AT 2025ampe & 109215 & 268.3365 & 65.19166 & 2.017 & Spec & 0\farcs56 & 0.7 \\
AT 2025ampf & 73759 & 268.64324 & 65.20717 & 2.28^{+0.18}_{-0.25} & Phot & 0\farcs14 & 0.2 \\
AT 2025ampg & 89545 & 268.48422 & 65.21129 & 1.435 & Spec & 0\farcs19 & 0.3 \\
AT 2025amph & 102547 & 268.52453 & 65.2281 & 2.41^{+2.51}_{-1.68} & Phot & 0\farcs04 & 0.1 \\
AT 2025ampi & 152270 & 268.35054 & 65.25851 & 1.138 & Spec & 0\farcs05 & 0.1 \\
AT 2026rfz & 44379 & 268.44595 & 65.14233 & 2.274 & Spec & 0\farcs02 & 0.1 \\
AT 2026rga & 98447 & 268.39576 & 65.16596 & 0.759 & Spec & 2\farcs87 & 1.8 \\
AT 2026rgb & 112434 & 268.27231 & 65.17279 & 0.69^{+0.07}_{-0.06} & Phot & 0\farcs51 & 0.5 \\
AT 2026rgc & 96655 & 268.52233 & 65.17499 & 2.69^{+0.61}_{-2.26} & Phot & 2\farcs54 & 5.8 \\
AT 2026rgd & 109709 & 268.25427 & 65.18808 & 2.473 & Spec & 0\farcs31 & 0.2 \\
AT 2026rge & 109350 & 268.23955 & 65.19002 & 0.86^{+0.02}_{-0.01} & Phot & 1\farcs31 & 0.5 \\
AT 2026rgg & 101200 & 268.37917 & 65.23478 & 1.76^{+1.17}_{-1.14} & Phot & 0\farcs48 & 0.8 \\
AT 2026rgh & 151296 & 268.55482 & 65.26334 & 0.84^{+0.41}_{-0.70} & Phot & 0\farcs52 & 0.7 \\
AT 2026rgi & 150358 & 268.47314 & 65.26851 & 1.15^{+0.11}_{-0.32} & Phot & 0\farcs31 & 0.4 \\
AT 2026rgj & 45701 & 268.4881 & 65.13344 & 3.22^{+0.21}_{-0.37} & Phot & 0\farcs06 & 0.1 \\
AT 2026rgk & 44453 & 268.42244 & 65.14092 & 0.73^{+0.04}_{-0.03} & Phot & 1\farcs45 & 1.7 \\
AT 2026rgl & 69392 & 268.61058 & 65.15548 & 0.55^{+0.03}_{-0.07} & Phot & 1\farcs65 & 1.0 \\
AT 2026rgm & 88261 & 268.45507 & 65.21771 & 1.18^{+1.20}_{-0.54} & Phot & 0\farcs13 & 0.2 \\
AT 2026rgo & 165579 & 268.31371 & 65.26223 & 0.88^{+0.07}_{-0.12} & Phot & 0\farcs94 & 1.0 \\
AT 2026rgp & 151363 & 268.41823 & 65.26316 & 2.04^{+0.03}_{-0.01} & Phot & 0\farcs03 & 0.1 \\
AT 2026rgq & 165095 & 268.30049 & 65.26835 & 1.71^{+0.14}_{-0.13} & Phot & 0\farcs89 & 0.6 \\
AT 2026rhb & 94163 & 268.37013 & 65.18705 & 1.946 & Spec & 0\farcs12 & 0.2 \\
\enddata
\tablecomments{Host IDs correspond to source identification numbers in the upcoming NEXUS DR1 catalog (Zhuang et al. in prep). Sources with $z$ type = Spec have spectroscopic redshift derived from NIRSpec PRISM spectra (presented in Figure~\ref{fig:Transient_host_spec}), except for AT~2026rgd, whose $z_{\rm spec}$ is determined from NIRCam WFSS. For sources with only photometric redshift, we report the median and 16th--84th percentile confidence intervals ($\pm1\sigma$) derived from \texttt{EAZY}. The \textit{Host Sep.} column lists the angular distance between the transient and its assigned host galaxy. The $d_{\rm Kron}$ parameter is a dimensionless quantity, representing the host separation normalized by the host size (see Section~\ref{sec:host_assoc}).}
\tablenotetext{a}{The redshift of AT~2025xcu is not robust due to ambiguous host association. See Section~\ref{sec:6.3.3} for details.}
\end{deluxetable}

\begin{acknowledgments}
Based on observations with the NASA/ESA/CSA James Webb Space Telescope obtained from the Barbara A. Mikulski Archive at the Space Telescope Science Institute, which is operated by the Association of Universities for Research in Astronomy, Incorporated, under NASA contract NAS5-03127. Support for Program number JWST-GO-05105 was provided through a grant from the STScI under NASA contract NAS5-03127. L.H. acknowledges support from the STScI grant JWST-AR-05965.
\end{acknowledgments}

\appendix
\section{A Novel MCMC PSF Photometry Code}\label{Appendix_A}

\subsection{Covariance Matrix Estimate}

The correlated noise properties of the drizzled images are characterized empirically using source-free background regions. We first constructed a segmentation-based source mask and identified contiguous square regions with minimal source contamination. Specifically, we searched for the largest square regions with tolerance of handful individual masked pixels, to ensure that the selected regions are dominated by blank sky emission. The covariance structure of the background noise was then estimated directly from these source-free image cutouts. 

For a given background cutout $I(x,y)$, we first replace individual bad pixels by Gaussian random realizations drawn from the measured background distribution in order to avoid artificial edge effects in the Fourier analysis and then subtract any remaining mean background level. The two-dimensional autocorrelation function (ACF) is computed using Fast Fourier Transform (FFT)-based convolution:

\begin{equation}
\mathrm{ACF}(\Delta x,\Delta y)
=
\left( I \star I \right)(\Delta x,\Delta y),
\end{equation}
where $\star$ denotes convolution with the reversed image. The ACF was normalized by its peak value at zero lag to produce the covariance kernel describing the spatial correlation structure introduced by the resampling (drizzling) process. 

\begin{figure*}[t]
    \centering
    \includegraphics[width=\linewidth]{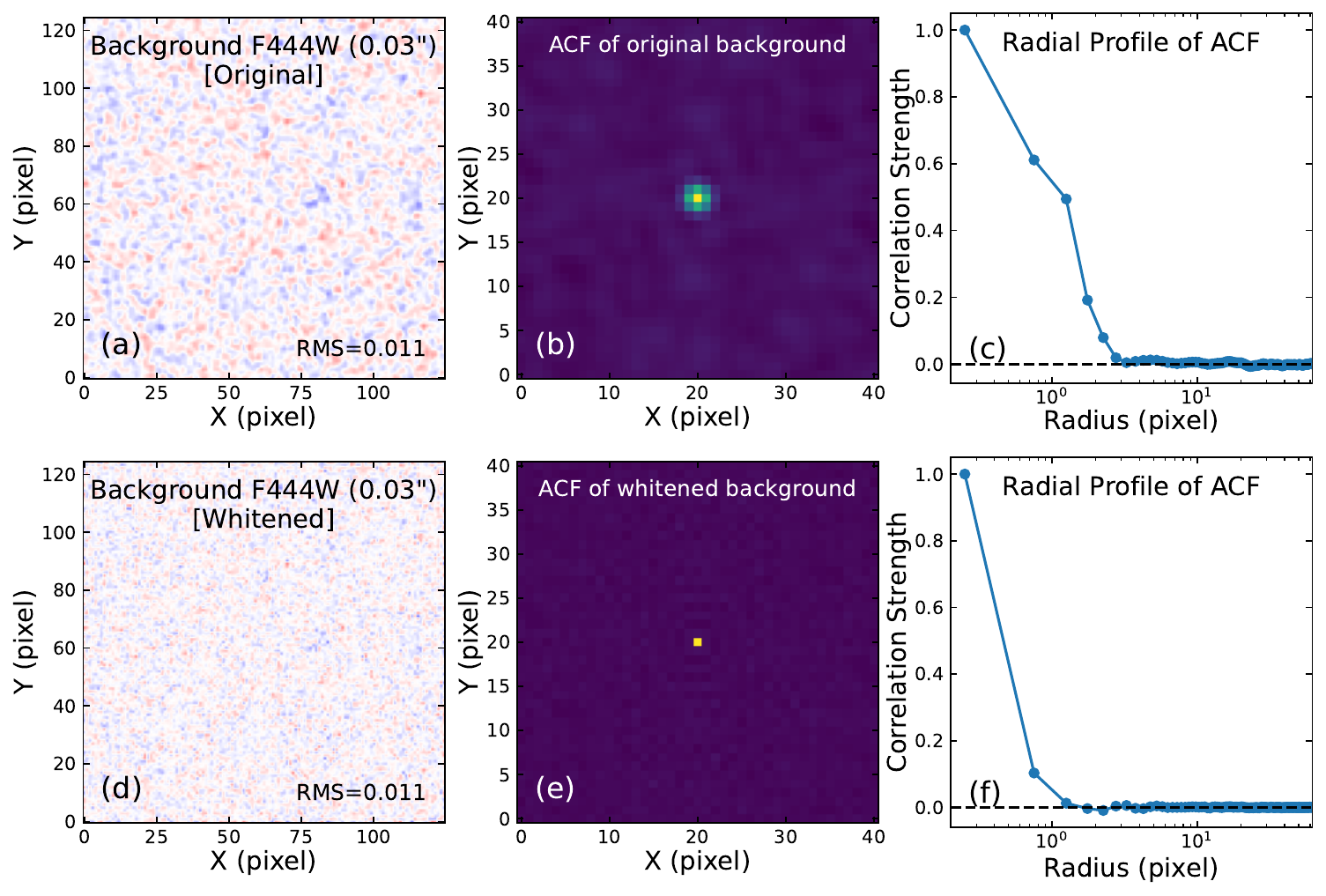}
    \caption{(a) The original background image from difference image in F444W with a pixelscale of 30 mas. The pixel-wise root-mean-square (RMS) value is shown at the lower-right corner. (b) The 2D Autocorrelation Function (ACF) of the original background. (c) The azimuthally averaged-radial profile of the ACF. (d) The same background image as (a) but after noise whitening. Panels (e) and (f) show the same plots as panels (b) and(c) but measured in the whitened background.}
    \label{fig:acf_demo}
\end{figure*}

Figure~\ref{fig:acf_demo} panels (a--c) show an example covariance kernel and its radial profile derived from a representative blank-sky region from a difference image in F444W with a pixel scale of 30 mas. The pixel correlation is evident from blocked patterns in the background image and illustrated as coherent structures surrounding the central peak in panel (b) and a strong excess correlation beyond the central pixel. Specifically, the correlation strength at the adjacent pixel ($r \approx 1$ pixel) is approximately $0.5$, and the correlation extends to roughly 2.5 pixels before falling to the baseline level. 

\subsection{Noise Whitening}

Direct inversion of the full pixel covariance matrix is computationally expensive for MCMC-based PSF fitting. Instead, we exploit the fact that for a spatial homogeneous Gaussian random field, the covariance matrix becomes diagonal in Fourier space. Under this approximation, the correlated noise can be efficiently characterized by its power spectrum.

To suppress high-frequency fluctuations and edge artifacts in the empirically estimated covariance kernel, we apply a split cosine-bell window function with a flat inner region and a cosine taper extending toward larger radii. The corresponding noise power spectrum is then obtained from the Fourier transform of the windowed covariance kernel:

\begin{equation}
P(k_x,k_y)
=
\mathcal{F}\left[\mathrm{ACF}(x,y)\right],
\end{equation}

where $\mathcal{F}$ denotes the two-dimensional Fourier transform, and $k_x$ and $k_y$ denote the spatial-frequency coordinates in Fourier space corresponding to the image axes x and y. The measured power spectrum quantifies the scale-dependent covariance introduced by the drizzling process.

Since the drizzling process introduces spatially correlated noise, the raw Fourier amplitudes at high spatial frequencies are suppressed relative to ideal white noise. To correct for this effect, we whiten the background fluctuations in Fourier space by dividing the Fourier transform of the image by the square root of the measured power spectrum:

\begin{equation}
\widetilde{I}_{\rm white}(k_x,k_y)
=
\frac{\widetilde{I}(k_x,k_y)}
{\sqrt{P(k_x,k_y)}},
\end{equation}
where $\widetilde{I}(k_x,k_y)$ denotes the Fourier transform of the image $I(x,y)$.

To verify the effectiveness of the noise whitening, we reconstruct the whitened image via inverse Fourier transform and recompute the ACF of the whitened background images. As shown in Figure~\ref{fig:acf_demo} panels (d--f), the off-center covariance signals are strongly suppressed after correction, demonstrating that the whitening procedure successfully decorrelates the noise introduced by the drizzling process and restored it to a spatially independent white noise without altering absolute variation of the noise. 

\subsection{Fourier-space Likelihood Function}

The PSF-fitting likelihood is evaluated directly in Fourier space using the empirically measured noise power spectrum. For a model image $M(x,y)$ and observed data $D(x,y)$, we compute the residual image

\begin{equation}
R(x,y)=\frac{D(x,y)-M(x,y)}{\sigma(x,y)},
\end{equation}
where $\sigma(x,y)$ denotes the pixel uncertainty map. In our case, $M(x,y)$ consists of a PSF model $PSF(x,y)$ and a scalar background $B$ to account for any remaining background emission. The Fourier transform of the normalized residual image is then computed:

\begin{equation}
\widetilde{R}(k_x,k_y)=\mathcal{F}[R(x,y)].
\end{equation}

Assuming the residual background fluctuations are described by a spatially homogeneous Gaussian random field, the likelihood can be written as

\begin{equation}
\chi^2=\frac{1}{N_{\rm pix}}
\sum_{k_x,k_y}
\frac{|\widetilde{R}(k_x,k_y)|^2}{P(k_x,k_y)},
\end{equation}
where $N_{\rm pix}$ is the total number of fitted pixels. In this formulation, the correlated noise covariance is naturally incorporated through the Fourier-space weighting by the inverse power spectrum. The posterior parameter distributions are then explored using an MCMC sampler.

\section{Transient Photometry}\label{Appendix_B}

Table~\ref{table:transient_phot} presents the multi-epoch $F200W$ and $F444W$ photometry (mag) across six NEXUS Deep epochs for the 68 robust transient candidates in our Year~1 sample.

\begin{longrotatetable}
\centerwidetable
\begin{deluxetable*}{lCCCCCCCCCCCC}
\tablecaption{Supernova Photometry}\label{table:transient_phot}
\tabletypesize{\scriptsize}
\tablehead{
\colhead{TNS ID} & \colhead{F200W\_Ep01} & \colhead{F444W\_Ep01} & \colhead{F200W\_Ep02} & \colhead{F444W\_Ep02} & \colhead{F200W\_Ep03} & \colhead{F444W\_Ep03} & \colhead{F200W\_Ep04} & \colhead{F444W\_Ep04} & \colhead{F200W\_Ep05} & \colhead{F444W\_Ep05} & \colhead{F200W\_Ep06} & \colhead{F444W\_Ep06}
}
\startdata
AT 2025xcm & 25.85\pm0.04 & 26.71\pm0.14 & 25.94\pm0.05 & 26.77\pm0.16 & 27.12\pm0.13 & 27.31\pm0.24 & 27.59\pm0.18 & >27.73 & 27.83\pm0.23 & >27.66 & \nodata & >27.90 \\
AT 2025xcn & 25.86\pm0.06 & 26.50\pm0.17 & 25.41\pm0.04 & 26.23\pm0.14 & 26.09\pm0.10 & 26.39\pm0.16 & >27.77 & >27.28 & >27.84 & >27.29 & >27.70 & >27.37 \\
AT 2025xco & 25.21\pm0.02 & 26.76\pm0.12 & 26.60\pm0.06 & >27.89 & 27.90\pm0.17 & >27.98 & >28.83 & >27.93 & >28.87 & >27.90 & >28.85 & >27.88 \\
AT 2025xcp & 25.71\pm0.04 & 27.02\pm0.25 & 25.64\pm0.03 & 26.88\pm0.23 & 27.24\pm0.13 & 27.43\pm0.33 & >28.39 & >27.41 & >28.38 & >27.41 & >28.42 & >27.36 \\
AT 2025xcq & 25.73\pm0.04 & 26.52\pm0.11 & 27.24\pm0.12 & >27.80 & >28.59 & >27.86 & >28.50 & >27.73 & >28.47 & >27.76 & >28.63 & >27.78 \\
AT 2025xcr & 27.28\pm0.10 & 27.91\pm0.31 & 27.31\pm0.10 & 27.87\pm0.32 & 27.75\pm0.18 & 27.16\pm0.21 & 27.57\pm0.12 & 27.45\pm0.22 & >28.54 & >28.03 & >28.50 & >27.78 \\
AT 2025xcs & 25.24\pm0.02 & 27.20\pm0.18 & \nodata & 27.12\pm0.22 & 26.27\pm0.05 & >27.99 & 27.84\pm0.16 & >27.98 & \nodata & >27.74 & >28.70 & >27.95 \\
AT 2025xct & 27.58\pm0.15 & 27.26\pm0.21 & >28.81 & 27.90\pm0.31 & >28.58 & >27.92 & >28.51 & >27.85 & >28.82 & >28.02 & >28.66 & >27.80 \\
AT 2025xcu & 27.39\pm0.08 & 27.78\pm0.23 & 27.75\pm0.16 & >28.01 & 28.37\pm0.15 & 27.98\pm0.26 & >29.15 & 27.85\pm0.24 & >28.88 & >27.99 & >28.71 & 28.16\pm0.31 \\
AT 2025xcv & 26.92\pm0.08 & 27.56\pm0.35 & \nodata & \nodata & \nodata & >27.58 & 28.06\pm0.16 & >27.77 & \nodata & \nodata & 28.45\pm0.31 & >27.61 \\
AT 2025xcw & 27.57\pm0.15 & 27.48\pm0.27 & 28.20\pm0.27 & >27.83 & 28.18\pm0.23 & >28.03 & >28.61 & >27.76 & >28.44 & >27.75 & >28.69 & >28.01 \\
AT 2025xcy & 26.62\pm0.08 & 27.58\pm0.28 & \nodata & \nodata & \nodata & \nodata & >28.53 & >27.85 & \nodata & \nodata & \nodata & \nodata \\
AT 2025xcz & 27.27\pm0.12 & 27.70\pm0.34 & 28.04\pm0.20 & 27.49\pm0.28 & >28.78 & >27.79 & >28.57 & >27.77 & >28.54 & >27.72 & >28.73 & >27.83 \\
AT 2025xda & 26.33\pm0.08 & 26.92\pm0.18 & 26.05\pm0.06 & 26.38\pm0.11 & 26.32\pm0.08 & 26.48\pm0.12 & 26.87\pm0.12 & 27.15\pm0.29 & 27.67\pm0.27 & 27.48\pm0.30 & >27.90 & >27.46 \\
AT 2025xdb & 26.98\pm0.09 & 27.39\pm0.21 & 27.65\pm0.17 & 27.71\pm0.30 & 27.78\pm0.16 & 27.78\pm0.33 & 28.34\pm0.26 & 27.76\pm0.33 & >28.67 & >27.93 & >28.75 & >27.94 \\
AT 2025xdc & 25.15\pm0.03 & 26.63\pm0.21 & 26.22\pm0.08 & >27.31 & 27.21\pm0.20 & >27.31 & >27.77 & >27.30 & >27.92 & >27.33 & >27.90 & >27.27 \\
AT 2025amnx & 27.55\pm0.15 & 26.88\pm0.16 & 27.70\pm0.17 & 27.16\pm0.21 & 27.60\pm0.15 & 27.18\pm0.18 & 27.68\pm0.17 & 27.40\pm0.26 & >28.54 & 27.60\pm0.31 & >28.63 & >27.93 \\
AT 2025amny & 27.74\pm0.18 & >27.85 & 27.85\pm0.13 & >28.04 & 28.75\pm0.35 & >28.02 & >28.57 & >27.82 & >28.84 & >27.73 & >28.81 & >28.00 \\
AT 2025amnz & 27.76\pm0.15 & >28.02 & 27.24\pm0.10 & >27.68 & \nodata & \nodata & 28.44\pm0.28 & >27.91 & >28.47 & >27.91 & \nodata & \nodata \\
AT 2025amoa & \nodata & 27.20\pm0.20 & 27.44\pm0.14 & 27.29\pm0.23 & 27.35\pm0.10 & 27.60\pm0.25 & 27.57\pm0.14 & 27.51\pm0.29 & 27.45\pm0.15 & 27.42\pm0.30 & 27.61\pm0.17 & 27.49\pm0.28 \\
AT 2025amob & 27.74\pm0.18 & \nodata & \nodata & \nodata & 27.87\pm0.23 & 26.99\pm0.14 & 28.02\pm0.23 & 27.18\pm0.16 & \nodata & \nodata & 27.83\pm0.21 & 27.37\pm0.21 \\
AT 2025amoc & \nodata & \nodata & 25.21\pm0.05 & 26.61\pm0.31 & \nodata & \nodata & \nodata & \nodata & >27.66 & >26.81 & \nodata & \nodata \\
AT 2025amod & \nodata & \nodata & 26.32\pm0.06 & >27.86 & 27.10\pm0.08 & 27.46\pm0.21 & \nodata & \nodata & 28.20\pm0.26 & >27.79 & 28.62\pm0.30 & >27.99 \\
AT 2025amoe & \nodata & \nodata & 24.96\pm0.02 & 25.80\pm0.05 & \nodata & \nodata & \nodata & \nodata & >28.88 & >28.02 & \nodata & \nodata \\
AT 2025amof & \nodata & \nodata & 25.67\pm0.04 & >27.73 & \nodata & \nodata & \nodata & \nodata & 27.50\pm0.17 & >27.76 & \nodata & \nodata \\
AT 2025amog & >28.86 & >28.03 & 27.32\pm0.10 & 27.47\pm0.21 & 28.15\pm0.20 & 27.41\pm0.20 & >28.84 & 27.87\pm0.30 & 28.62\pm0.34 & 27.58\pm0.24 & 28.49\pm0.29 & 27.73\pm0.29 \\
AT 2025amoh & >28.36 & >27.96 & 26.29\pm0.05 & 27.08\pm0.17 & 26.78\pm0.08 & 27.61\pm0.27 & >28.59 & >27.98 & >28.72 & >27.93 & >28.74 & >27.90 \\
AT 2025amoi & \nodata & \nodata & 27.74\pm0.14 & >28.02 & \nodata & \nodata & \nodata & \nodata & >28.53 & >27.99 & \nodata & \nodata \\
AT 2025amoj & \nodata & \nodata & 26.12\pm0.05 & 27.13\pm0.28 & 28.16\pm0.30 & 27.20\pm0.31 & \nodata & \nodata & >28.39 & >27.42 & >28.35 & >27.38 \\
AT 2025amok & \nodata & \nodata & 27.01\pm0.07 & 27.58\pm0.26 & 26.61\pm0.08 & 26.99\pm0.19 & \nodata & \nodata & 26.19\pm0.05 & 26.45\pm0.10 & 26.73\pm0.08 & 27.00\pm0.20 \\
AT 2025amol & \nodata & \nodata & 27.34\pm0.10 & 27.67\pm0.26 & 28.27\pm0.24 & >28.09 & \nodata & \nodata & 28.75\pm0.35 & >27.96 & >28.68 & >28.02 \\
AT 2025amom & \nodata & \nodata & 26.33\pm0.06 & 26.90\pm0.13 & 27.19\pm0.09 & 27.12\pm0.17 & \nodata & \nodata & 28.07\pm0.20 & >27.92 & 28.02\pm0.19 & 27.77\pm0.32 \\
AT 2025amon & \nodata & \nodata & 25.76\pm0.04 & >27.57 & 25.90\pm0.04 & >27.49 & \nodata & \nodata & 28.12\pm0.27 & >27.52 & >28.58 & >27.24 \\
AT 2025amoo & \nodata & \nodata & \nodata & \nodata & 25.98\pm0.06 & 27.07\pm0.22 & \nodata & \nodata & \nodata & \nodata & >28.26 & >27.64 \\
AT 2025amop & \nodata & \nodata & >28.57 & \nodata & 27.73\pm0.14 & >27.99 & \nodata & \nodata & \nodata & \nodata & 27.37\pm0.11 & 27.49\pm0.23 \\
AT 2025amoq & >28.73 & >27.80 & >28.80 & >28.07 & 24.73\pm0.02 & 25.69\pm0.06 & 24.95\pm0.03 & 25.67\pm0.06 & 25.57\pm0.03 & 26.25\pm0.08 & 26.77\pm0.08 & 26.75\pm0.15 \\
AT 2025amor & >28.68 & >27.89 & >28.88 & >28.09 & 26.28\pm0.06 & 27.47\pm0.27 & 26.82\pm0.08 & 26.84\pm0.16 & 26.95\pm0.07 & 26.97\pm0.14 & 28.39\pm0.33 & >27.76 \\
AT 2025amos & >28.58 & >28.08 & \nodata & \nodata & 25.88\pm0.05 & 26.73\pm0.14 & 25.79\pm0.04 & 26.59\pm0.10 & \nodata & \nodata & 27.35\pm0.14 & 27.30\pm0.25 \\
AT 2025amot & >28.60 & >27.85 & >28.58 & >28.01 & 27.51\pm0.11 & >27.92 & >28.61 & 27.57\pm0.30 & >28.46 & >28.03 & >28.81 & >27.99 \\
AT 2025amou & >28.56 & >27.56 & >28.43 & >27.53 & 25.59\pm0.03 & \nodata & 26.25\pm0.06 & \nodata & 27.30\pm0.14 & \nodata & 28.00\pm0.24 & \nodata \\
AT 2025amov & >28.46 & >27.71 & >28.39 & >27.76 & 27.99\pm0.22 & >27.77 & 27.07\pm0.14 & 27.06\pm0.21 & 27.05\pm0.13 & 26.98\pm0.20 & 27.10\pm0.13 & 27.16\pm0.19 \\
AT 2025amow & \nodata & \nodata & >28.64 & >28.01 & 24.01\pm0.01 & 25.18\pm0.03 & \nodata & \nodata & 25.60\pm0.03 & 25.94\pm0.06 & 26.96\pm0.09 & 26.25\pm0.08 \\
AT 2025amox & \nodata & \nodata & >28.09 & >27.66 & 22.62\pm0.01 & 23.73\pm0.01 & \nodata & \nodata & 26.80\pm0.13 & 25.81\pm0.08 & >28.33 & 26.87\pm0.17 \\
AT 2025amoz & >28.79 & >28.05 & \nodata & \nodata & \nodata & \nodata & 27.52\pm0.12 & >27.85 & \nodata & \nodata & \nodata & \nodata \\
AT 2025ampb & >28.53 & >28.10 & \nodata & \nodata & \nodata & \nodata & 27.30\pm0.10 & 27.54\pm0.25 & \nodata & \nodata & \nodata & \nodata \\
AT 2025ampc & >28.74 & >27.95 & >28.81 & >27.93 & >28.75 & >27.93 & 26.82\pm0.06 & >27.87 & >28.68 & >27.86 & >28.65 & >27.83 \\
AT 2025ampe & >28.72 & >27.92 & >28.77 & >27.90 & >28.76 & >27.96 & 27.55\pm0.15 & >27.73 & 26.56\pm0.06 & 26.98\pm0.18 & 26.87\pm0.09 & 26.92\pm0.15 \\
AT 2025ampf & >28.57 & >27.86 & >28.58 & >27.56 & \nodata & \nodata & 27.43\pm0.18 & 27.35\pm0.25 & 27.82\pm0.20 & 27.18\pm0.21 & \nodata & \nodata \\
AT 2025ampg & >28.63 & >27.59 & >28.51 & >27.57 & >28.16 & >27.59 & 27.00\pm0.11 & >27.57 & 27.90\pm0.24 & 27.02\pm0.22 & >28.16 & >27.58 \\
AT 2025amph & >28.64 & >28.00 & >28.68 & >27.87 & >28.73 & >28.00 & 26.37\pm0.06 & 27.43\pm0.22 & 25.67\pm0.04 & 26.63\pm0.13 & 26.06\pm0.04 & 27.16\pm0.17 \\
AT 2025ampi & >28.12 & >27.25 & \nodata & \nodata & >27.85 & >27.25 & 26.70\pm0.11 & 27.21\pm0.33 & \nodata & \nodata & >27.75 & >27.21 \\
AT 2026rfz & \nodata & \nodata & >28.40 & >27.49 & >28.65 & >27.74 & \nodata & \nodata & 28.09\pm0.23 & >27.88 & 26.90\pm0.10 & 26.97\pm0.20 \\
AT 2026rga & >28.77 & >27.78 & >28.83 & >27.63 & >28.79 & >27.76 & >28.79 & >27.67 & 26.01\pm0.04 & 27.16\pm0.24 & 25.95\pm0.04 & 26.79\pm0.16 \\
AT 2026rgb & \nodata & >27.81 & >28.40 & >27.62 & \nodata & \nodata & >28.60 & >27.67 & 25.58\pm0.04 & 26.25\pm0.11 & \nodata & \nodata \\
AT 2026rgc & >28.69 & >27.94 & >28.70 & >28.12 & >29.00 & >28.14 & >28.73 & >27.85 & 27.76\pm0.19 & 27.78\pm0.30 & 26.13\pm0.04 & 26.54\pm0.09 \\
AT 2026rgd & >28.40 & >27.38 & >28.50 & >27.60 & \nodata & \nodata & >28.45 & >27.39 & 27.53\pm0.14 & 27.20\pm0.26 & \nodata & \nodata \\
AT 2026rge & >28.46 & \nodata & \nodata & \nodata & \nodata & \nodata & >28.46 & >27.59 & 25.32\pm0.03 & 26.46\pm0.14 & \nodata & \nodata \\
AT 2026rgg & >28.72 & >27.86 & >28.93 & >28.09 & >28.96 & >27.89 & >28.82 & >27.85 & 27.09\pm0.09 & 27.44\pm0.22 & 27.37\pm0.10 & 27.16\pm0.17 \\
AT 2026rgh & \nodata & \nodata & >28.84 & >28.12 & >28.83 & >28.09 & \nodata & \nodata & 26.24\pm0.04 & 27.60\pm0.27 & 26.86\pm0.07 & >27.98 \\
AT 2026rgi & \nodata & \nodata & >28.69 & >27.86 & >28.85 & >28.14 & \nodata & \nodata & 26.01\pm0.05 & 27.57\pm0.30 & 27.44\pm0.12 & >28.05 \\
AT 2026rgj & \nodata & \nodata & \nodata & \nodata & >28.83 & >27.94 & \nodata & \nodata & \nodata & \nodata & 26.94\pm0.08 & >27.94 \\
AT 2026rgk & \nodata & \nodata & >28.78 & >27.86 & >29.04 & >27.85 & \nodata & \nodata & >28.58 & >27.85 & 27.45\pm0.14 & >27.86 \\
AT 2026rgl & >28.70 & >27.88 & \nodata & \nodata & >28.44 & >27.69 & >28.28 & >27.85 & \nodata & \nodata & 27.71\pm0.16 & >27.81 \\
AT 2026rgm & >29.01 & >28.08 & >28.46 & >27.96 & >28.76 & >27.89 & >28.66 & >27.94 & >28.84 & >27.87 & 27.20\pm0.12 & 27.45\pm0.24 \\
AT 2026rgo & >28.74 & >27.83 & \nodata & \nodata & >28.55 & >28.04 & >28.47 & >27.86 & \nodata & \nodata & 26.09\pm0.04 & 26.72\pm0.13 \\
AT 2026rgp & \nodata & \nodata & \nodata & \nodata & >28.42 & >27.86 & >28.10 & >27.60 & \nodata & \nodata & 26.96\pm0.11 & 27.23\pm0.22 \\
AT 2026rgq & >28.40 & >27.62 & \nodata & \nodata & \nodata & \nodata & >28.32 & >27.67 & \nodata & \nodata & 27.09\pm0.13 & \nodata \\
AT 2026rhb & >28.37 & >27.74 & >28.34 & >27.73 & >28.54 & >27.93 & >28.32 & >27.67 & >28.32 & >27.68 & 26.50\pm0.07 & 27.14\pm0.18 \\
\enddata
\tablecomments{Multi-epoch $F200W$ and $F444W$ photometry (mag) across six NEXUS Deep epochs for the 68 robust transient candidates in our Year~1 sample. For observations with detection significance below $2\sigma$, we report $3\sigma$ magnitude upper limits.}
\end{deluxetable*}
\end{longrotatetable}

\section{Transient Image Stamp Gallery}\label{sec:stamp_gallery}

Multi-epoch F200W and F444W NIRCam reference, science, and difference image cutouts for all 74 transient candidates identified in this work (Figure~\ref{fig:stamps}). 

\begin{figure*}[t]
    \centering
    \includegraphics[width=\linewidth]{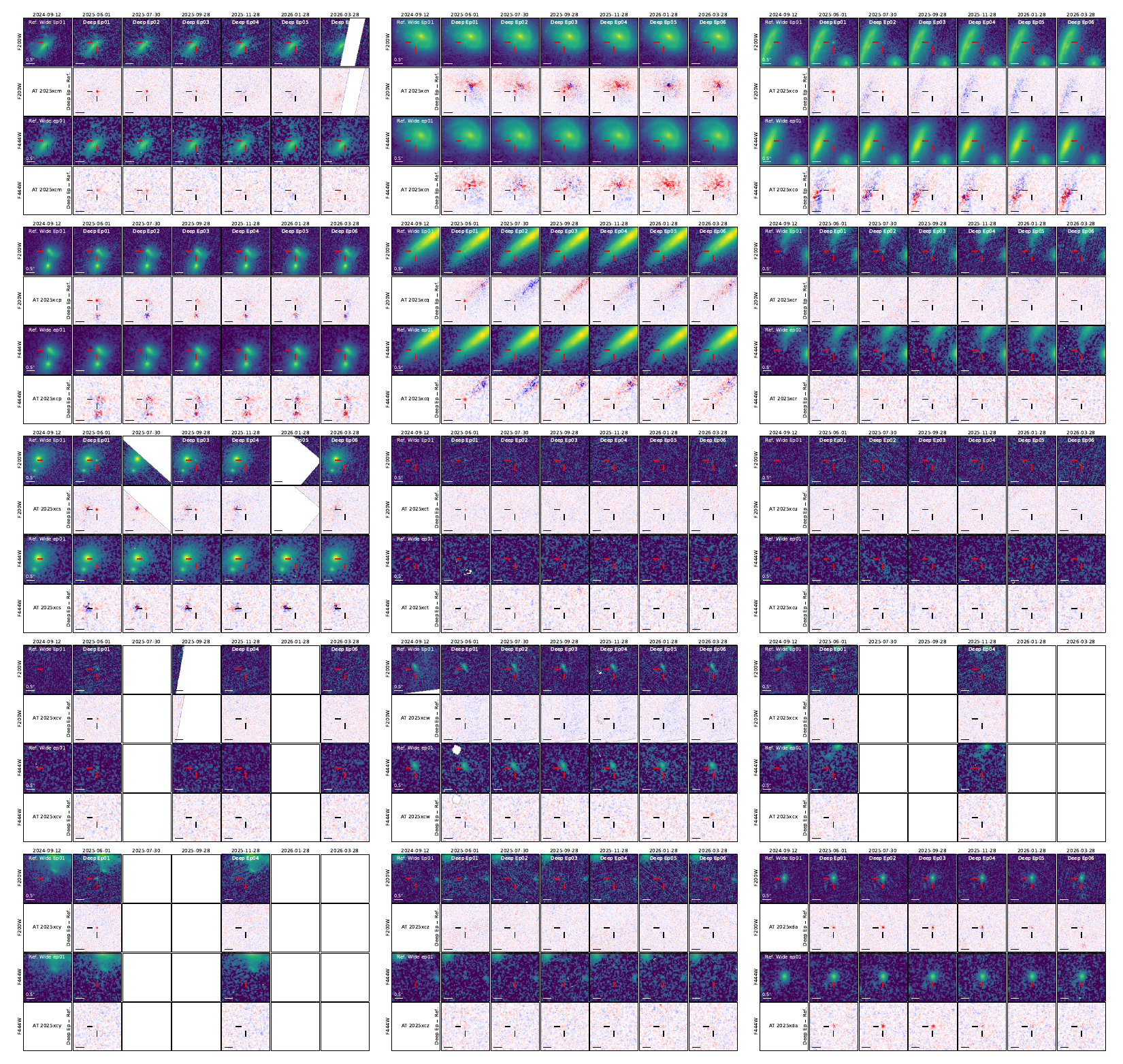}
    \caption{Same as Figure~\ref{fig:SN_diff_img}, but for transients 1--15.}
  \label{fig:stamps}
\end{figure*}

\begin{figure*}[t]
    \ContinuedFloat
    \centering
    \includegraphics[width=\linewidth]{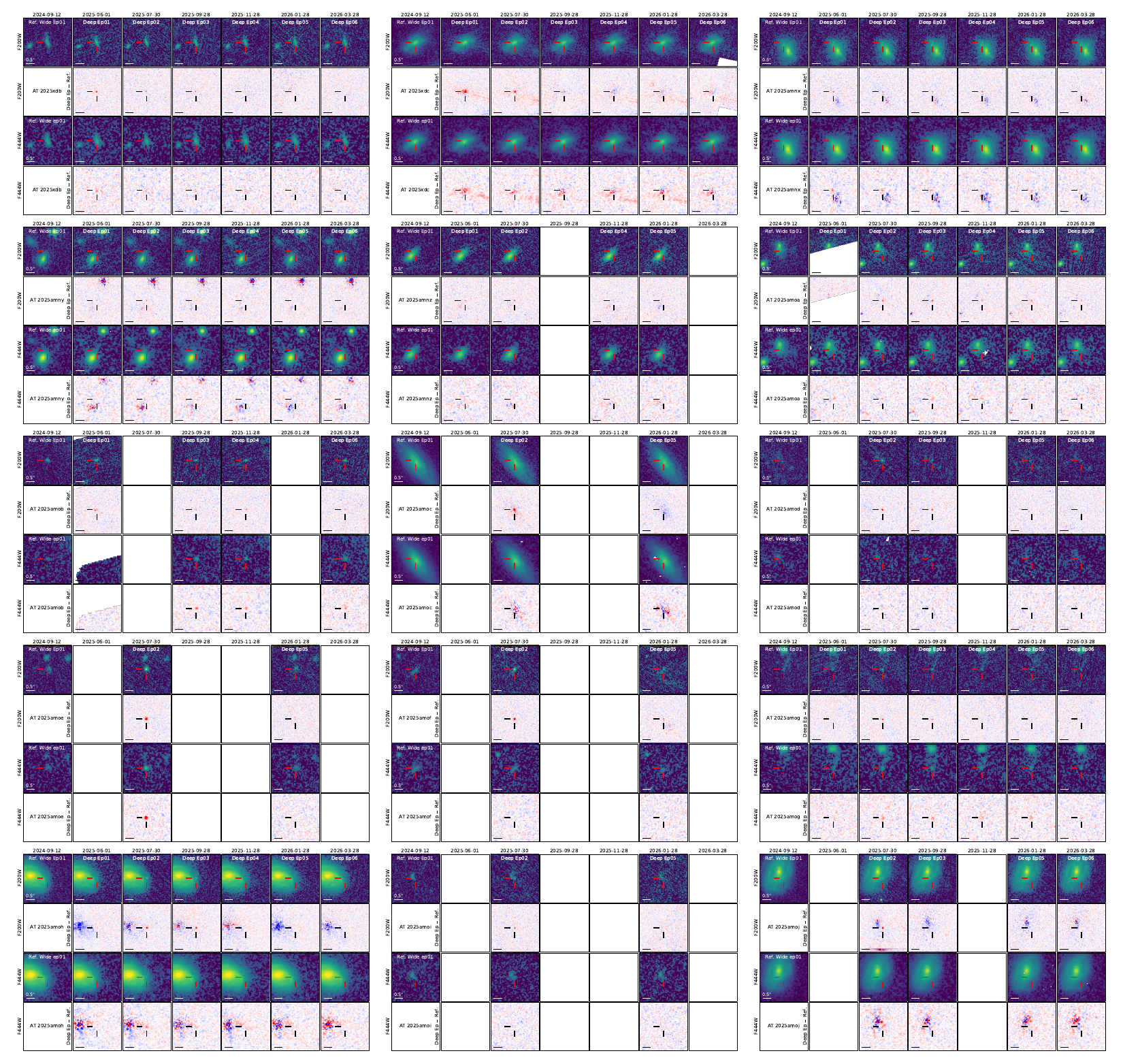}
    \caption{--- \textit{Continued.}}
\end{figure*}

\begin{figure*}[t]
    \ContinuedFloat
    \centering
    \includegraphics[width=\linewidth]{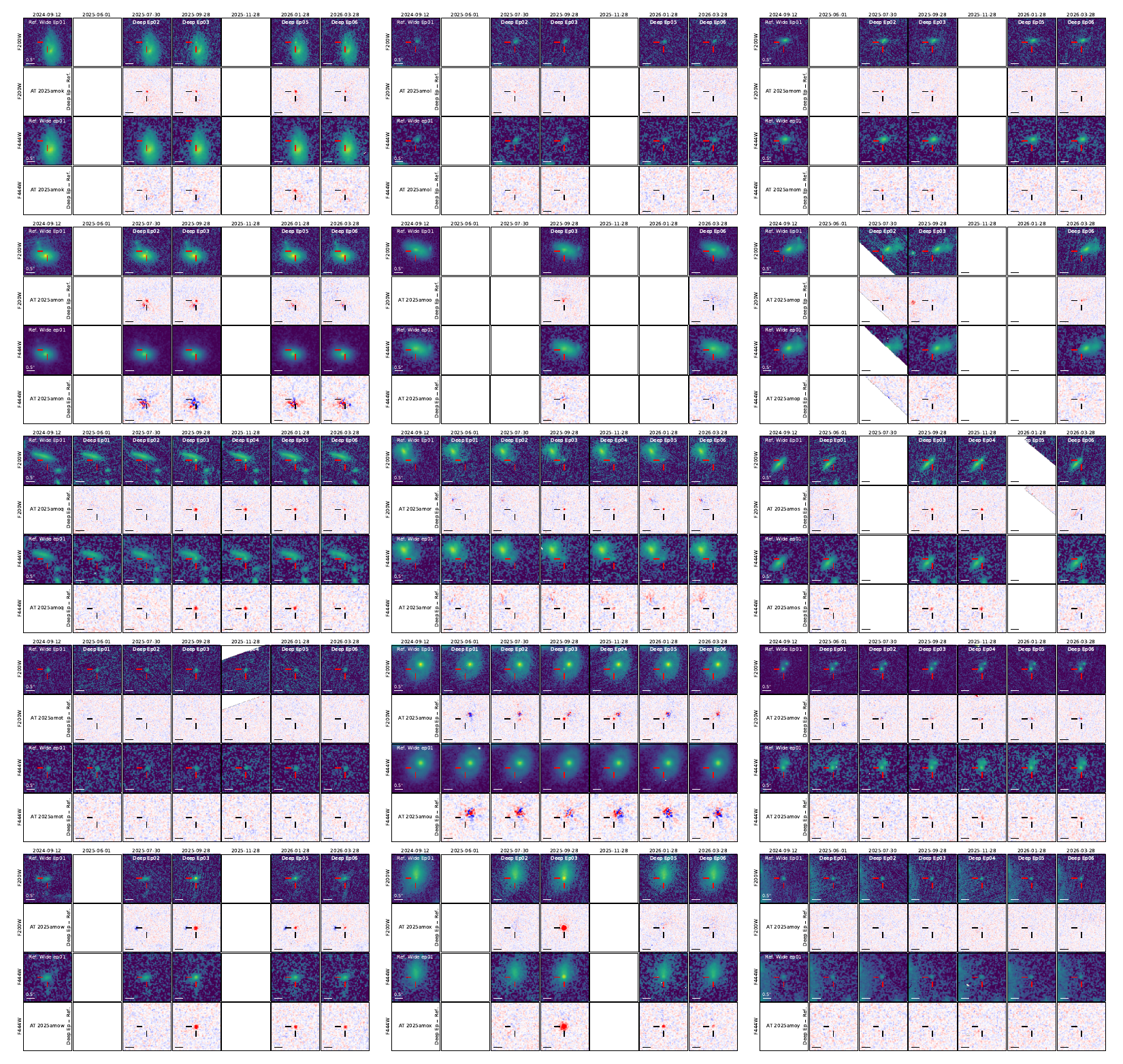}
    \caption{--- \textit{Continued.}}
\end{figure*}

\begin{figure*}[t]
    \ContinuedFloat
    \centering
    \includegraphics[width=\linewidth]{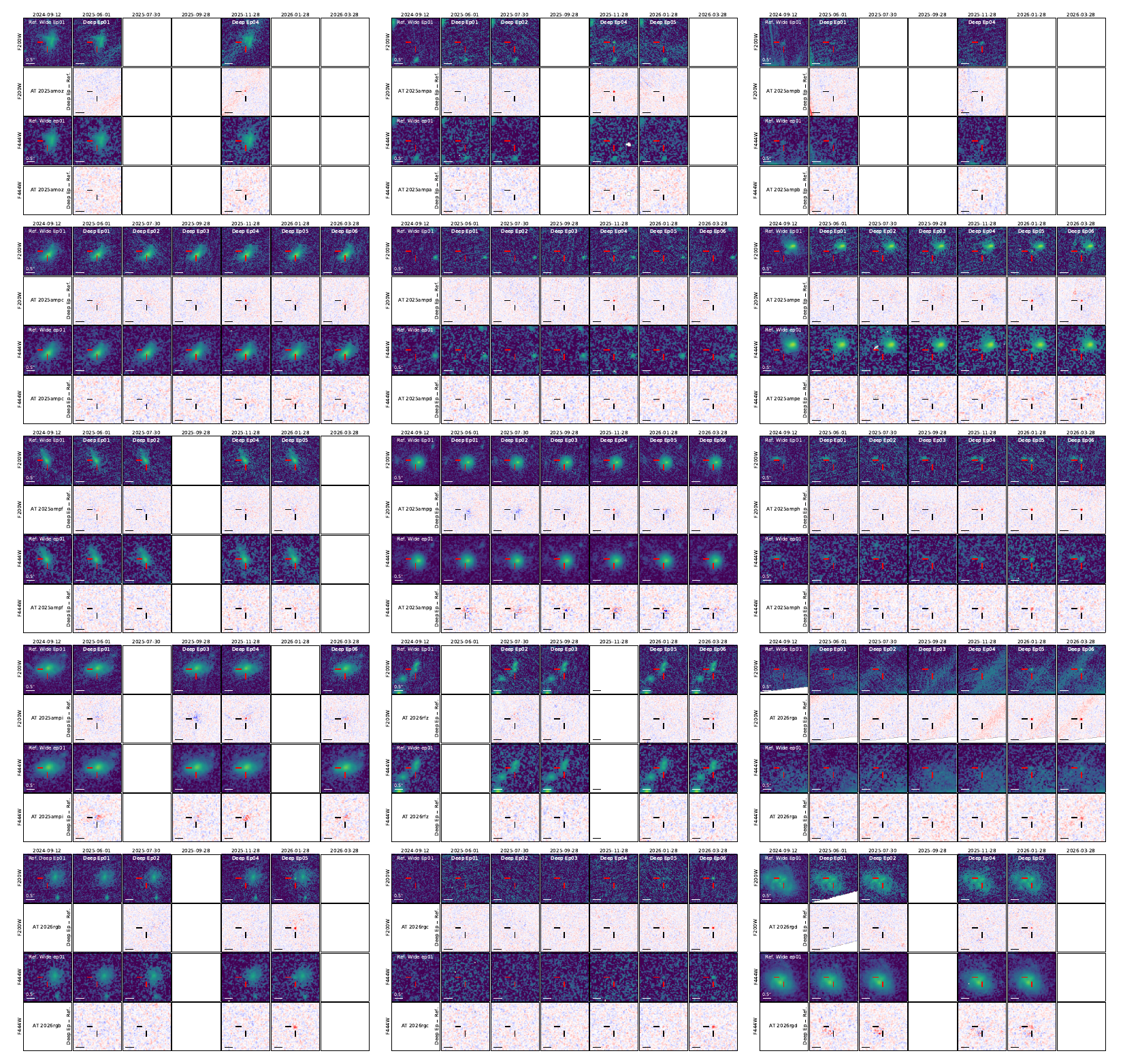}
    \caption{--- \textit{Continued.}}
\end{figure*}

\begin{figure*}[t]
    \ContinuedFloat
    \centering
    \includegraphics[width=\linewidth]{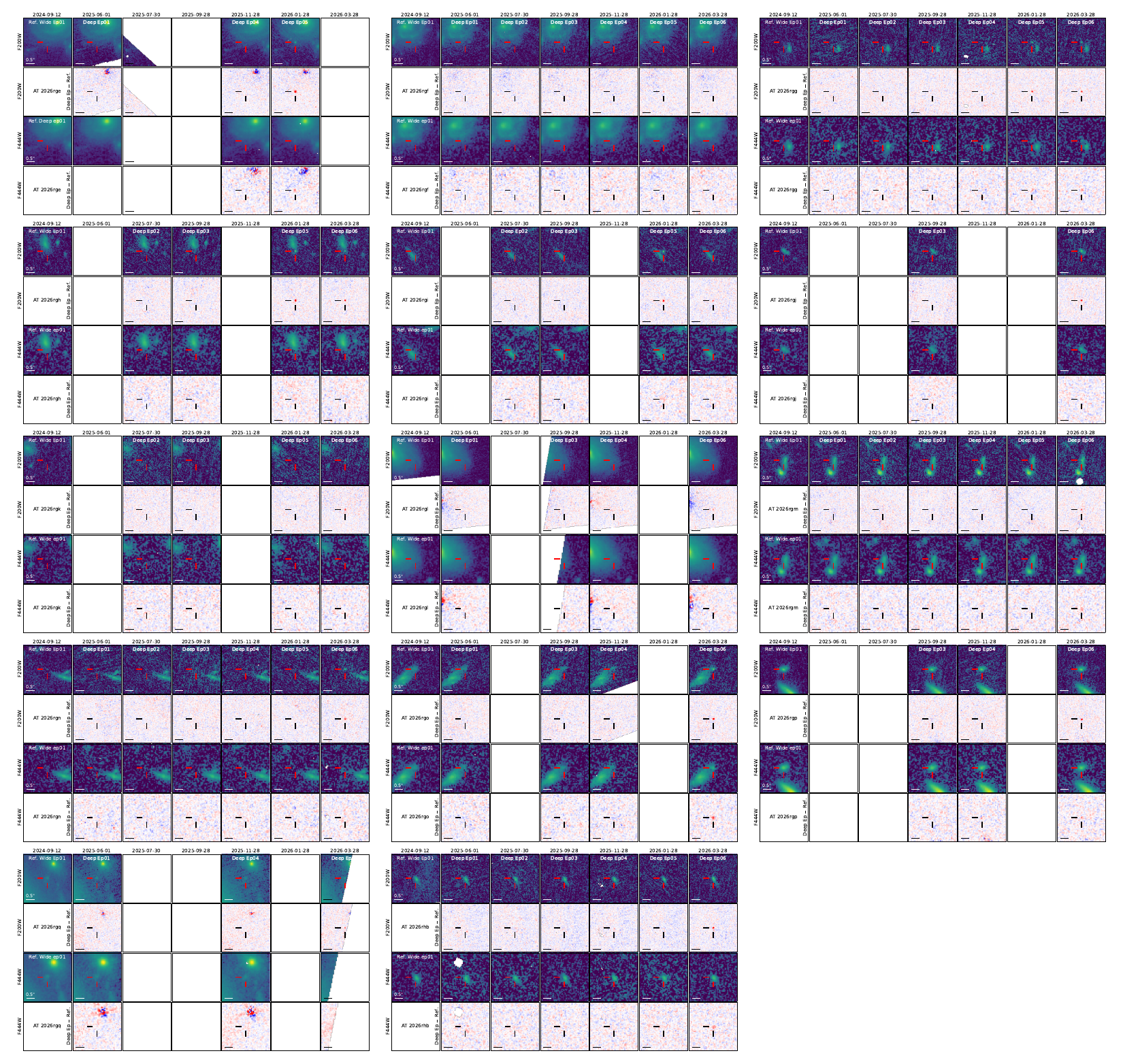}
    \caption{--- \textit{Continued.}}
\end{figure*}

\bibliographystyle{aasjournalv7.1}
\bibliography{refs.bib,sn-bibliography.bib,references.bib}{}

\end{document}